\documentclass{jpp}
\usepackage{graphicx,natbib,setspace}
\usepackage{epstopdf, epsfig}
\usepackage{amsfonts,amsmath,amssymb}
\usepackage[pdftex,colorlinks,citecolor=blue]{hyperref}
\usepackage{bm}
\usepackage{xcolor}
\usepackage{pstool}
\usepackage[]{cleveref}%capitalise, noabbrev
\crefname{section}{\S\!}{\S\S\!}
\crefname{appendix}{appendix}{appendices}
\crefname{equation}{}{}
\Crefname{equation}{Equation}{Equations}
\crefname{figure}{figure}{figures}
\Crefname{figure}{Figure}{Figures}
\crefname{table}{table}{tables}
\Crefname{table}{Table}{Tables}

\usepackage{mathtools,upgreek,pifont}
\usepackage{cuted}

\newcommand{\mockalph}[1]{}	

\newcommand{\imag}{{\rm i}}
\newcommand{\rmd}{{\rm d}}
\newcommand{\rme}{{\rm e}}
\newcommand{\pD}[2]{\frac{\partial #2}{\partial #1}}

\newcommand{\D}[2]{\frac{\rmd #2}{\rmd #1}}

\newcommand\bs[1]{{\boldsymbol{#1}}}
\newcommand\bb[1]{\mbox{\boldmath{$#1$}}}
\newcommand\grad{\bb{\nabla}}
\renewcommand\bcdot{\,\bb{\cdot}\,}
\newcommand{\bscdot}{\,\bs{\cdot}\,}
\newcommand{\bdbldot}{\,\bb{:}\,}
\newcommand\btimes{\,\bb{\times}\,}
\newcommand{\msb}[1]{\mathsfbi{#1}}

\newcommand{\ffold}{.}
\newcommand{\Dp}{\Delta p}
\newcommand{\dpstress}{\grad\bcdot(\bh\bh\Dp)}
\newcommand{\It}{\mathcal{I}}
\newcommand{\bbgu}{\hat{\bm{b}}\hat{\bm{b}}\bdbldot\grad \bm{u} }
\newcommand{\bh}{\hat{\bm{b}} }
\newcommand{\fh}{\Uptheta}
\newcommand{\an}{\mathcal{A}}
\newcommand{\va}{v_{\rm A}}
\newcommand{\omegaa}{\omega_{\rm A}}
\newcommand{\ma}{\mathcal{M}_{\rm A}}
\newcommand{\maint}{\mathcal{M}_{\rm A}^{\rm int}}

\newcommand{\skqs}{S+19}
\newcommand\aksqs{A+22}

\newcommand{\revchng}[1]{#1}

\newcommand{\specialcell}[2][c]{\begin{tabular}[#1]{@{}c@{}}#2\end{tabular}}

\ifCUPmtlplainloaded \else
  \checkfont{eurm10}
  \iffontfound
    \IfFileExists{upmath.sty}
      {\typeout{^^JFound AMS Euler Roman fonts on the system,
                   using the 'upmath' package.^^J}%
       \usepackage{upmath}}
      {\typeout{^^JFound AMS Euler Roman fonts on the system, but you
                   dont seem to have the}%
       \typeout{'upmath' package installed. JPP.cls can take advantage
                 of these fonts, if you use 'upmath' package.^^J}%
       \providecommand\upi{\pi}%
      }
  \else
    \providecommand\upi{\pi}%
  \fi
\fi

\ifCUPmtlplainloaded \else
  \checkfont{msam10}
  \iffontfound
    \IfFileExists{amssymb.sty}
      {\typeout{^^JFound AMS Symbol fonts on the system, using the
                'amssymb' package.^^J}%
       \usepackage{amssymb}%
         \let\leq=\leqslant
         \let\geq=\geqslant
      }{}
  \fi
\fi

\ifCUPmtlplainloaded \else
  \IfFileExists{amsbsy.sty}
    {\typeout{^^JFound the 'amsbsy' package on the system, using it.^^J}%
     \usepackage{amsbsy}}
    {\providecommand\boldsymbol[1]{\mbox{\boldmath $##1$}}}
\fi

\title[Weakly collisional plasma turbulence]{Pressure anisotropy and viscous heating in weakly collisional plasma turbulence}

\author[J.~Squire and others]%
{J.~Squire$^{1}$\thanks{Email address for correspondence: jonathan.squire@otago.ac.nz},
 M.~W.~Kunz$^{2,3}$,
 L.~Arzamasskiy$^{4}$,
 Z.~Johnston$^{1}$,
E.~Quataert$^{2}$,
 A.~A.~Schekochihin$^{5,6}$}

\affiliation{$^1$Physics Department, University of Otago, Dunedin 9010, New Zealand\\[\affilskip]
$^2$Department of Astrophysical Sciences, Princeton University, Peyton Hall, Princeton, NJ 08544, USA\\[\affilskip]
$^3$ Princeton Plasma Physics Laboratory, PO Box 451, Princeton, NJ 08543, USA\\[\affilskip]
$^4$ School of Natural Sciences, Institute for Advanced Study, Princeton, NJ 08540, USA\\[\affilskip]
$^5$The Rudolf Peierls Centre for Theoretical Physics, University of Oxford, 1 Keble Road, Oxford, OX1 3NP, UK\\[\affilskip]
$^6$Merton College, Oxford OX1 4JD, UK
}

\pubyear{2023}
\volume{}
\pagerange{}
\begin{document}
\maketitle

\begin{abstract}
Pressure anisotropy can strongly influence the dynamics of weakly collisional, high-beta plasmas, but its effects are missed by standard  magnetohydrodynamics (MHD). Small changes to the magnetic-field strength generate large pressure-anisotropy forces,
heating the plasma, driving instabilities, and rearranging flows, 
even on scales far above the particles' gyroscales where 
kinetic effects are traditionally considered most important. Here, we study the influence of pressure anisotropy on turbulent plasmas
threaded by a mean magnetic field (Alfv\'enic turbulence). Extending previous results that were concerned with 
Braginskii MHD, we consider a wide range of regimes and parameters using
a simplified fluid model based on drift kinetics with heat fluxes calculated using a Landau-fluid closure. 
We show that viscous (pressure-anisotropy) heating dissipates between a quarter \revchng{(in collisionless regimes)} and half \revchng{(in collisional regimes)} of the turbulent cascade power injected at large scales; \revchng{this does not depend strongly on either plasma beta or the ion-to-electron temperature ratio}.  This will in turn influence the plasma's thermodynamics by regulating energy partition between different dissipation channels (e.g., electron and ion heat).    Due to the pressure anisotropy's rapid dynamical feedback onto the flows that create it -- an effect we term `magneto-immutability' -- the viscous heating is confined to a narrow range of scales near the forcing scale, supporting  a nearly conservative, MHD-like inertial-range cascade, \revchng{via which the rest of the energy is transferred to small scales}.   
Despite the simplified model, our results -- including the viscous heating rate, distributions, 
and turbulent spectra -- compare favourably to recent hybrid-kinetic simulations. This is promising for the more general 
use of extended-fluid (or even MHD) approaches to model weakly collisional plasmas such as the intracluster medium, hot accretion flows, and the solar wind. 
\end{abstract}
%just like isotropic pressure forces rapidly feed back on compressible perturbations to make low-Mach-number flows incompressible. 
%Our study, which also presents extensive background discussion and new numerical methods, will  be helpful for understanding and interpreting the physics of the intracluster medium, hot accretion flows, and the solar wind. 

\section{Introduction}

Many hot, diffuse plasmas in astrophysical environments are weakly collisional, with
Coulomb mean free paths that are comparable to relevant macroscopic scales in the system. 
Canonical examples include the intracluster medium in galaxy clusters \citep[e.g.,][]{Fabian1994,Fabian2006,Kunz2022}, hot accretion flows such 
as those observed by the Event Horizon Telescope \citep{Quataert2001,EHTC2019}, 
the hotter phases of the interstellar and circumgalactic media \citep{Cox2005,Tumlinson2017}, and the solar wind and magnetosphere \citep{Borovsky2018,Marsch2006}. In all of these examples,
the plasma is extremely well magnetised, in that the system scales are much larger than the  ion gyroradius $\rho_{i}$. Coupled with the plasmas' large Coulomb mean free paths, this implies that magnetic fields are crucial in providing the plasma with an internal cohesion
that allows it to behave more or less  like a collisional fluid \citep{Kulsrud1983}. 
However, despite the particle motion's allegiance to the local magnetic field, in many environments the magnetic fields are energetically subdominant as measured by $\beta = 8\upi p/B^{2}$, where $p$ is a thermal pressure and $B^{2}/8\upi$ 
is the magnetic energy density. \revchng{Under such conditions, small relative changes to the magnetic field
can easily cause  particle distributions to become unstable, suggesting that  non-equilibrium kinetic physics should play a key role in the dynamics.}

Nonetheless, such conditions  are commonly modelled using collisional magnetohydrodynamics (MHD), 
usually for reasons of expediency, though with rigorous justification in some circumstances \citep[e.g.,][]{Kulsrud1983,Schekochihin2009,Kunz2015}.
It is the first purpose of this article to  explain why this is usually not appropriate, as a result of the aforementioned kinetic 
physics; it is the second purpose to explain why, in the end, it is  not so bad after all.
The  physics we explore is that of \emph{pressure anisotropy} (equivalently, temperature anisotropy), 
which is the difference between the thermal pressures in the directions perpendicular and parallel to the local magnetic field. 
The relevance of pressure anisotropy stems from it being the only non-isotropic piece of the pressure tensor that can survive 
on scales arbitrarily far above the particles' gyroradii \citep{Braginskii1965},  making it a key physical ingredient 
 in the momentum balance of weakly collisional plasmas (compared to collisional MHD; \citealp{Chew1956,Kulsrud1983}).
At $\beta\gg1$, a very small relative pressure anisotropy can lead to large bulk forces on the plasma.
Furthermore, whenever the magnetic-field strength $B$ changes in time,  $\Dp$ is generated 
as a result of the conservation of single-particle adiabatic invariants.
Put together, these two properties suggest that pressure-anisotropy stresses can vastly overwhelm the forces from the magnetic field or Reynolds stress in many high-$\beta$ environments.

Our study focuses on the role of pressure anisotropy in turbulence, specifically, in turbulence of the `Alfv\'enic' 
variety that occurs when the system is threaded by a large-scale mean magnetic field. A particularly important example of such a system is the solar wind, which is our best natural laboratory for the study of Alfv\'enic turbulence under weakly collisional conditions. Such turbulence may also be relevant quasi-universally at smaller scales in systems without a mean magnetic field \citep{Brandenburg2005,Schekochihin2020}. We explore the structure and dynamics of weakly collisional Alfv\'enic turbulence, with a particular focus on their comparison with MHD and the implications for turbulent plasma heating. Pressure anisotropy provides another channel -- effectively a viscosity -- for damping injected mechanical energy into heat near the outer scales \citep[e.g.,][]{Kunz2010,Yang2017a}. \revchng{(Although this energy transfer can be reversible in some regimes,  
we will often refer to it as `viscous heating' because at higher collisionalities the pressure-anisotropy 
stress increasingly resembles a parallel viscosity, with an associated heating rate that is positive definitive.) }
Because the pressure-anisotropy stress can easily overwhelm the magnetic tension and Reynolds stress in the high-$\beta$ limit \citep{Squire2016},
simple estimates suggest that this viscous heating should completely dominate over  other heating mechanisms. 
However, we show that, because of the dynamical feedback of the pressure anisotropy on the plasma flow, 
such heating is confined to a narrow range of scales near the forcing scales, with a sizeable fraction (typically, the majority) 
of the energy making it through to participate in a nearly conservative turbulent cascade. This 
phenomenon, which  was termed `magneto-immutability' by \cite{Squire2019} (hereafter \skqs) who studied the effect 
 in the context 
of the  simplified `Braginskii MHD' model (see also \citealt{Kempski2019,StOnge2020}),
involves the force from the pressure anisotropy acting  to rearrange the turbulence in order to reduce the 
 influence of itself. This rearrangement constrains the
global variation of $B$, hence the term `magneto-immutability' to describe the effect.
Its physics is analogous to that of incompressibility, in which the pressure force that results from a fluid
compression rapidly drives a flow that counters the compression, thereby eliminating such motions and minimising variations in the density. 
This explains our flippant declaration above about the article's  purpose: 
simple estimates suggest pressure anisotropy should dominate the force balance and heating, modifying
the dynamics compared to MHD; but, the effect of this modification is to minimise the pressure anisotropy's own influence, thus allowing
turbulent dynamics that look broadly similar to MHD below the outer scales (albeit with a few important differences).

These ideas complicate the already complex story of turbulent heating in weakly collisional plasmas \citep[e.g.,][]{Quataert1999,Kunz2010,Howes2010a,Kawazura2020,Meyrand2021,Yang2022,Arzamasskiy2022}.  
A useful concept is the `cascade efficiency': the
fraction of energy available to cascade down to small scales and heat the plasma via different collisionless mechanisms \citep{Schekochihin2009,Chandran2010,Arzamasskiy2019}. In our simulations, between ${\simeq}20\%$ and ${\simeq}45\%$
of the energy is lost through pressure-anisotropy heating at large scales (giving a cascade efficiency of ${\gtrsim}55\%$ \revchng{for our choice of large-scale forcing}), independently of $\beta$ and the electron-to-ion temperature ratio. Magneto-immutability 
ensures that the remainder of the energy cascades nearly conservatively,  presumably eventually heating at small scales in the way predicted 
by gyrokinetics \citep{Howes2008,Schekochihin2009,Kunz2018}. With different heating processes 
implying a different partition of turbulent energy between electron and ion heat, or between perpendicular and parallel heat, 
these apparently esoteric details of the turbulent structure can strongly influence the basic thermodynamics of the plasma. 
For instance, while it is well known that small-scale collisionless processes in high-$\beta$ plasmas generically heat ions more than electrons \citep{Howes2010a,Parashar2018,Kawazura2020,Roy2022},  large-scale pressure anisotropy can also heat collisionless electrons \citep{Sharma2007}, meaning that the cascade efficiency -- and thus magneto-immutability -- could 
directly control  the ion-to-electron heating ratio.
While we do not explicitly study electron physics, our study  provides a useful foundation for distilling and quantifying the important physics, 
particularly through the idea that pressure-anisotropy heating is confined to 
a small range of large scales.
%A key contribution is the understanding that pressure-anisotropy heating is confined to 
%a small range of large scales, with magneto-immutability feeding back on the flow to establish a conservative MHD-like cascade below this range. 

Our approach to the study of pressure anisotropy here is computational and simplified. We use
the so-called CGL-Landau-fluid (hereafter CGL-LF) model \citep{Chew1956,Snyder1997}, which attempts to approximate collisionless heat fluxes in order to obtain 
a closed, simplified fluid model for plasma dynamics on scales well above $\rho_{i}$.
This model, which is implemented using new numerical methods into the \textsc{Athena++} framework \citep{White2016,Stone2020}, is then used to explore a range of conditions as the key parameters of the problem are varied. 
Detailed comparisons to otherwise equivalent MHD simulations  are used to diagnose the influence of the pressure anisotropy on the turbulence and heating. 
A downside of the simplified-fluid-model approach is that various \emph{ad-hoc}, loosely justified approximations are necessary. \revchng{These approximations include those made to derive} the fluid closure and its numerical realisation, 
and perhaps more importantly, the methods used to cope with the fast-growing $\rho_{i}$-scale firehose and mirror instabilities 
that will inevitably arise in real systems \citep{Schekochihin2005,Bale2009,Kunz2014}, but which cannot be captured correctly by drift kinetics \citep[e.g.,][]{Rosin2011,Rincon2015}.
The upside of the approach, by contrast, is that the model effectively assumes infinite scale separation between 
the outer scale and $\rho_{i}$, which is clearly not possible in kinetic models that explicitly resolve $\rho_{i}$, but is the 
 appropriate limit for most astrophysical systems.
With this in mind, a useful subsidiary 
purpose of the article is to assess the successes of this simplified model in reproducing the  physics seen in 
  the hybrid-kinetic simulations of \citet{Arzamasskiy2022} (hereafter \aksqs), which explicitly resolve $\rho_{i}$ and the associated chaos of kinetic-scale instabilities at the expense of a much reduced inertial range.
We find relatively good agreement in general, despite the \emph{ad-hoc} nature of the fluid model and the interpretative difficulties 
associated with limited scale separation in the hybrid-kinetic simulations. In particular, we find similar cascade
efficiencies (\revchng{with a similar forcing scheme}) and similar pressure-anisotropy distributions compared to \aksqs, with some caveats.  
Accordingly, the results of this work are quite promising for the use of  CGL-LF approaches in modelling weakly collisional plasmas.

\subsection{Outline}
Although most of the basic theoretical concepts presented here have appeared in previous literature, 
we feel it is helpful to keep the majority of the discussion and notation self-contained to clarify the approximations and key concepts.
Thus,~\cref{sec: theory}  presents an overview of the physics of pressure anisotropy and magneto-immutability, starting 
from the drift-kinetic model of \citet{Kulsrud1983}. Lacking any quantitative theory of the important effects, we focus 
on qualitative explanations for the behaviour and effect of the pressure anisotropy and heat fluxes in different regimes.
We also define the `interruption number' $\It$, which quantifies the expected strength of pressure-anisotropic effects in Alfv\'enic turbulence, 
analogously  to the Reynolds number or Mach number in fluids. In~\cref{sec: methods} we then outline the numerical methods and 
diagnostics, before presenting detailed simulation results in~\cref{sec: results}. Leveraging  the versatility of the simplified fluid 
model, a focus of the results is the direct comparison to `passive-$\Dp$' simulations. These are identical to the standard 
simulations but with the anisotropic pressure force artificially removed, thereby affording a direct comparison to a counterfactual situation 
where magneto-immutability does not exist. The comparison clearly demonstrates both the similarities and differences between our pressure-anisotropic Landau-fluid model and MHD
in spectra, distributions, and scale-dependent heating functions. We follow with  discussions 
of the key uncertainties related to kinetic instabilities (\cref{subsub: mi and microinstabilities}) and of the distinction between magneto-immutability and the so-called `spherically polarised' states measured in the solar wind (a complicating factor for interpreting spacecraft observations). We also  summarise most salient differences between magneto-immutable and MHD turbulence  (\cref{sub: observations}). A full list of the simulations 
with their important parameters is given in~\cref{tab:sims}. We conclude in~\cref{sec:conclusions}.
An appendix presents various technical results related to the numerical finite-volume implementation in the \textsc{Athena++} code \citep{White2016,Stone2020},
including linear-wave convergence tests and a new Riemann solver for the CGL system (albeit one that we ultimately did not use in the simulations).

\section{Theoretical and phenomenological background}\label{sec: theory}

\subsection{The evolution and effect of pressure anisotropy}

In this section, we introduce the basic concepts necessary to understand the numerical results presented in \S\ref{sec: results}. The goal
is to provide some intuitive understanding of the effects of pressure anisotropy in different plasma regimes, starting
from the basic equations for a collisionless plasma on large scales. We begin by assuming that the
plasma pressure tensor is gyrotropic but anisotropic -- it is invariant under rotations about the magnetic field, but can 
differ in the directions perpendicular and parallel to the field. This is justified for `MHD-range' scales, \emph{viz.,} those 
larger than the ion gyroscale in space and slower than the ion gyrofrequency in time. We also assume a single ion species and isothermal electrons,  which allows us to drop the anisotropic component of the electron pressure and obtain single-fluid equations for 
the dynamics of the ions. Although this can be
formally justified  in the moderately collisional limit using the electron's larger collision frequency and fast thermal speed \citep{Rosin2011}, here the choice is primarily one of simplicity, since we wish to focus on the dynamical effects of pressure anisotropy on fluid motions. Most of our simulations in this work will assume cold electrons anyway, in order to better understand 
and diagnose the basic processes at play. 

With these assumptions, the first three moments of the ion distribution function satisfy \citep{Chew1956,Kulsrud1983} 
\begin{gather} 
 \partial_{t}\rho +\grad \bcdot (\rho \bm{u} ) = 0,\label{eq:KMHD rho} \\[2ex]
\rho \left(\partial_{t}\bm{u} +  \bm{u}\bcdot \grad \bm{u} \right)= -  \grad\left(T_{e}\rho+ p_{\perp} + \frac{B^{2}}{8\upi}\right)+ \grad \bcdot \left[\hat{\bm{b}} \hat{\bm{b}}\left( \Delta p + \frac{B^{2}}{4\upi}\right)\right],\label{eq:KMHD u}\\[2ex]
 \partial_{t} \bm{B} = \grad \btimes (\bm{u} \btimes \bm{B} ) ,\label{eq:KMHD B} \\[2ex]
\partial_{t}p_{\perp} + \grad \bcdot (p_{\perp}  \bm{u}) + p_{\perp} \grad \bcdot \bm{u} + \grad \bcdot (q_{\perp}\hat{\bm{b}}) +q_{\perp} \grad \bcdot \hat{\bm{b}}    = p_{\perp} \bbgu -  \frac{1}{3}\nu_{\rm c}\Delta p,\label{eq:KMHD pp}
 \\[2ex]
\partial_{t}p_{\parallel} + \grad \bcdot (p_{\parallel}  \bm{u}) + \grad \bcdot (q_{\parallel}\hat{\bm{b}}) -2q_{\perp} \grad \bcdot \hat{\bm{b}}     = -2 p_{\parallel} \bbgu +\frac{2}{3}\nu_{\rm c}\Delta p.\label{eq:KMHD pl}
\end{gather}
We use Gaussian-CGS units,  $\bm{B}$ and $\bm{u}$  are the magnetic field and ion flow velocity (also approximately equal to the electron
flow velocity for scales well above the ion gyroscale), $B\doteq \left| \bm{B} \right|$ and $\hat{\bm{b}}=\bm{B}/B$  denote the magnetic-field strength and direction, $\rho$ is the mass density, 
$\nu_{\rm c}$ is the \revchng{ion-ion} collision frequency,\footnote{\revchng{The relevant terms in \cref{eq:KMHD pp,eq:KMHD pl} are derived from a simple BGK collision operator \citep{Gross1956,Snyder1997}.}} $p_{\perp}$ and $p_{\parallel}$ are the components of the ion pressure tensor perpendicular and parallel to the magnetic field, and $q_{\perp}$ and $q_{\parallel}$ are fluxes of perpendicular and 
parallel ion heat in the direction parallel to the magnetic field. The electron temperature $T_{e}$ is constant in time and space, because electrons are assumed isothermal (their pressure 
is $p_{e}=T_{e}\rho/m_{i}$, related to the ion density by quasi-neutrality assuming a single-ion-species plasma, and \revchng{we neglect ion-electron collisions}). Equations \eqref{eq:KMHD rho}--\eqref{eq:KMHD pl} are solved numerically in the conservative form given in~\cref{app: numerical tests }. Importantly, this avoids  explicitly computing  the parallel rate of strain $\bbgu$, which can introduce serious numerical errors in some situations  (\citealp{Sharma2011}; \skqs).  We also define the Alfv\'en speed $\va=B/\sqrt{4\upi\rho}$, the total pressure $p_{0}\doteq2p_{\perp}/3+p_{\|}/3$, the parallel sound speed $c_{\rm s\|}\doteq\sqrt{p_{\|}/\rho }$,  the plasma beta $\beta \doteq 8\upi p_{0}/B^{2}$ (similarly $\beta_{\|}\doteq8\upi p_{\|}/B^{2}$),  \revchng{the pressure anisotropy $\Dp\doteq p_{\perp}-p_{\|}$}, the normalised pressure anisotropy $\Delta \doteq\Dp/p_{0}$, and the `anisotropy parameter,' 
\begin{equation}
    \fh \doteq 1+\frac{4\upi \Dp}{B^{2}},
\end{equation}
which measures the relative change in the propagation speed of linear Alfv\'en waves ({\em viz.,} $\va$) due to the contribution from the pressure-anisotropy stress (see~\cref{subsub: lf waves}). 

%Note that ~\eqref{eq:KMHD pp} and~\eqref{eq:KMHD pl} are not yet closed equations because the heat fluxes remain unspecified.
\revchng{In order to avoid solving a kinetic equation for the heat fluxes, we  close \crefrange{eq:KMHD rho}{eq:KMHD pl} with the expressions
\begin{gather}
q_{\perp} = -\frac{2 c_{\rm s\|}^{2}}{\sqrt{2 \upi }c_{\rm s\|} |k_{\parallel}|+\nu_{\rm c}} \left[ \rho  \nabla_{\parallel} \left(\frac{p_{\perp}}{\rho}\right)  - p_{\perp}\left(1-\frac{p_{\perp}}{p_{\parallel}} \right)\frac{\nabla_{\parallel} B}{B}  \right], \label{eq:GL heat fluxes p}\\ 
q_{\parallel} = - \frac{8 c_{\rm s\|}^{2}}{\sqrt{8 \upi }c_{\rm s\|} |k_{\parallel}|+(3\upi-8)\nu_{\rm c}} \rho \nabla_{\parallel} \left(\frac{p_{\parallel}}{\rho}\right).\label{eq:GL heat fluxes l}
\end{gather} 
These forms, which are  known as a `Landau-fluid' closure,  were devised by \citet{Snyder1997}, in order to 
match the  linear  behaviour of the true kinetic system as closely as possible (\citealp{Hammett1990}; see also \citealt{Passot2012}).  
The $k_{\|}$ and $\nabla_{\|}=\bh\bcdot\grad$ in \crefrange{eq:GL heat fluxes p}{eq:GL heat fluxes l} are the
parallel wavenumber and gradient operator, respectively, with the non-local gradient-like operator $\nabla_{\|}/|k_{\|}|$ 
arising as a result of  approximating the effects of collisionless damping within the fluid model (see \cref{subsub: heat fluxes computational} 
for further discussion). We will use the form \crefrange{eq:GL heat fluxes p}{eq:GL heat fluxes l} both 
computationally and as a useful intuitive guide for understanding the physical effect of the heat fluxes. 
}

\subsubsection{Energy conservation} 

With $T_{e}=0$, equations~\crefrange{eq:KMHD rho}{eq:KMHD pl} conserve the total energy $E_{K}+E_{M}+E_{\rm th}$, where
\begin{equation}
E_{K} \doteq \frac{1}{2}\int \rmd\bm{x}\,\rho |\bm{u}|^{2},\quad E_{M} \doteq \frac{1}{8\upi }\int \rmd\bm{x}\, |\bm{B}|^{2},\quad E_{\rm th} \doteq \int \rmd\bm{x}\,\left(p_{\perp} + \frac{p_{\|}}{2}\right).
\end{equation}
The \revchng{rate of change of the thermal energy is}
\begin{equation}
\D{t}{E_{\rm th}} =  \int \rmd\bm{x}\,(-p_{\perp}\grad\bcdot\bm{u} + \Dp\, \bbgu) = \int \rmd\bm{x}\,\left(p_{\|}\frac{1}{\rho}\frac{\rmd \rho}{\rmd t} + \Dp\frac{1}{B}\frac{\rmd B}{\rmd t}\right),\label{eq: energy conservation}
\end{equation}
\revchng{where $\rmd/\rmd t=\partial_{t}+\bm{u}\bcdot\grad$ is the Lagrangian derivative and we used the continuity and induction equations to derive the second expression (see~\cref{eq: B and bbgu}). We see that a positive correlation between $\Dp$ and $\rmd B/\rmd t$ drives net heating of the plasma. This is not necessarily guaranteed: in essence, the term is similar to the compressive term $p_{\|} \rmd\! \ln\rho/\rmd t$ and can mediate oscillatory transfer between mechanical and thermal energy through pressure anisotropy. 
However, when collisions dominate the evolution of $\Dp$, its effect becomes that of a parallel viscosity and $\Dp\,\rmd \!\ln B/\rmd t$  
is guaranteed to be positive (see~\cref{eq: dp in braginskii} and discussion below). For this reason, 
we will often refer to this term as  `viscous heating,' even in the collisionless regime where it is more general.
Note that, because ion-ion scattering 
must conserve energy, collisions only indirectly influence  the thermal-energy evolution  by changing
the correlation between $\Dp$ and $\rmd B/\rmd t$.} 

The case with $T_{e}\neq0$ is addressed in~\cref{app:sub: cgl equations}.

\subsubsection{Wave behaviour}\label{subsub: lf waves}

\Crefrange{eq:KMHD rho}{eq:KMHD pl} admit five types of linear wave solutions: shear Alfv\'en waves, two modified 
magnetosonic-like waves, and two types of non-propagating entropy modes. These are discussed in more detail in~\cref{appsub: dispersion relation} \citep[see also][]{Hunana2019,Majeski2023}, 
where we give the relevant ideal dispersion relations (computed from \crefrange{eq:KMHD rho}{eq:KMHD pl} with $q_\perp=q_\|=0$ and $\nu_c=0$) and use their nonideal properties  as a test of the numerical solvers. 
Here, we simply note that heat fluxes and/or collisions strongly damp all modes  except for  Alfv\'en waves, which, so long as $4\upi\Dp>-B^2$, propagate undamped  with the modified speed 
\begin{equation}
{v}_{\rm A,eff} = \sqrt{1+\frac{4\upi\Dp}{B^{2}}} \va = \sqrt{{\fh}} \,\va.\label{eq: dp alfven speed}
\end{equation}
Linear Alfv\'enic modes are undamped because they do not involve any perturbation of $B^{2}$, and, therefore, do not create any pressure 
anisotropy. We see that for $4\upi\Dp<-B^{2}$, ${v}_{\rm A, eff}$ becomes imaginary, which is the fluid manifestation of the firehose instability. 

\subsubsection{The collisionless, weakly collisional, and Braginskii-MHD regimes}\label{sub: different regimes}

It is helpful to examine the equation for the evolution of the pressure anisotropy, which may be obtained from~\cref{eq:KMHD pp,eq:KMHD pl}:
\begin{align}
\D{t}{}\Dp  = &\left(p_{\perp}+2p_{\|}\right)\bbgu - \left(2p_{\perp}-p_{\|}\right)\grad\bcdot\bm{u}\nonumber\\ &- 
\left\{\grad\bcdot\bigl[(q_{\perp}-q_{\|})\bh\bigr] + 3q_{\perp}\grad\bcdot\bh \right\} -\nu_{\rm c}\Dp.\label{eq: Dp equation}
\end{align}
Each term on the right-hand side is grouped according to its physical effect, which we  discuss in turn below.
\begin{description}
\item[Changing field strength-- ]The first term on the right-hand side of~\cref{eq: Dp equation} captures the creation of pressure anisotropy through the parallel rate of strain $\bbgu$, 
which is related to changes in the magnetic field strength through 
\begin{equation}
\frac{1}{B}\D{t}{B} = \bbgu - \grad\bcdot\bm{u}\label{eq: B and bbgu}
\end{equation}
(this equation is obtained from~\cref{eq:KMHD B} after dotting it with $\bh$ and rearranging). Due to conservation of the collisionless adiabatic invariants $p_{\perp}/\rho B$ and $p_{\|}B^{2}/\rho^{3}$, positive $\Dp$
is created in regions of increasing field strength, while negative $\Dp$ is created in regions of decreasing field strength. 
\item[Compressions-- ]The second term is similar to the first but relates to changes in the plasma density. It is less important here, because of our focus on 
high-$\beta$ plasmas with low compressibility. 
\item[Heat fluxes-- ]The third term involves the heat fluxes, which, although neglected in the so-called double-adiabatic model, are always 
large in the $\beta\gtrsim1 $ regime where $\Dp$ has a dynamically important effect \citep[e.g.,][]{Hunana2019}. The effect of this term can 
be understood by examining the `Landau fluid' form of the heat fluxes given in \crefrange{eq:GL heat fluxes p}{eq:GL heat fluxes l}.
When $\Dp\ll p_{0}$ and the spatial variation of density and $\bh$ is small compared to that of the temperature, these take the approximate forms 
\begin{gather}
-\grad\bcdot(q_{\perp}\bh) \approx -\bh\bcdot\grad q_{\perp} \approx \sqrt{\frac{2}{\upi}} \nabla_{\|}\left[\frac{c_{\rm s\|}^{2}}{c_{\rm s\|}|k_{\|}|+a_{\perp}\nu_{\rm c}}\nabla_{\|}p_{\perp}\right], \label{eq: approx hf prp}\\
-\grad\bcdot(q_{\|}\bh) \approx -\bh\bcdot\grad q_{\|} \approx   \sqrt{\frac{8}{\upi}}\nabla_{\|}\left[\frac{c_{\rm s\|}^{2}}{c_{\rm s\|}|k_{\|}|+a_{\|}\nu_{\rm c}}\nabla_{\|}p_{\|}\right], \label{eq: approx hf prl}
\end{gather}
where $a_{\perp}$ and $a_{\|}$ are order-unity numerical factors (see~\cref{eq:GL heat fluxes p,eq:GL heat fluxes l}). We see that, broadly speaking, the heat fluxes act like a parallel diffusion operator on $p_{\perp}$ and $p_{\|}$ (albeit a nonlocal one
if $c_{\rm s\|}|k_{\|}|\gtrsim \nu_{\rm c}$), thus reducing the spatial variation of $\Dp$ along the magnetic field.
\item[Collisions-- ]The final term in~\cref{eq: Dp equation} reduces the pressure anisotropy due to collisions, and, if large enough, causes~\crefrange{eq:KMHD rho}{eq:KMHD pl} to reduce to standard, adiabatic MHD.
\end{description}

From this discussion, we see that the first two terms on the right-hand side of~\cref{eq: Dp equation} are responsible for creating a pressure
anisotropy from the motions of the plasma.
By comparing the relative sizes of the other terms -- $\rmd\Dp/\rmd t$, the heat fluxes, and the collisional term -- one finds three regimes each with qualitatively different 
$\Dp$ evolution. To distinguish these, it is helpful to consider $\Dp$ structures of parallel scale $k_{\|}^{-1}$
and assume that $\nabla_{\|} p_{\perp}\sim \nabla_{\|} p_{\|}\sim \nabla_{\|}\Dp$, \emph{viz.,} that the relative variation in $p_{\perp}$, $p_{\|}$ and $\Dp$ is of similar magnitudes (but note that it is not true that $p_{\perp}\sim p_{\|}\sim \Dp$ because the variation in $p_{\perp}$ and $p_{\|}$ is small compared to 
their mean). Assuming Alfv\'enic motions, the time derivative scales as $\rmd\Dp/\rmd t\sim \omegaa\Dp$ where $\omegaa=k_{\|}\va$ is the Alfv\'en frequency, 
 the heat-flux term scales as $\sim \! [(k_{\|}c_{\rm s})^{2}/(|k_{\|}|c_{\rm s}+\nu_{\rm c})] \Dp \sim [\beta \omegaa/(\beta^{1/2} + \nu_{\rm c}/\omegaa)] \Dp$, 
 and the collisional term is just $\nu_{\rm c}\Dp$. Restricting our discussion to $\beta\gtrsim 1$, the three regimes are:
 \begin{description}
\item[Collisionless, $\bm{\nu_{\rm c}\ll \omegaa}$-- ]If $\nu_{\rm c}\ll \omegaa$, the collisional term cannot compete with $\rmd\Dp/\rmd t\sim \omegaa\Dp$ to  
reduce $|\Dp|$ significantly and so can be neglected. Because $\nu_{\rm c}/\omegaa\lesssim 1$, the heat flux term scales as $\sim\!\beta^{1/2}\omegaa\Dp$, implying that heat fluxes
are always important in this regime, smoothing $\Dp$ in space rapidly compared to the Alfv\'enic motions that drive it. 
\item[Weakly Collisional, $\bm{\omegaa\ll \nu_{\rm c}\lesssim \beta^{1/2}\omegaa}$-- ]For $\nu_{\rm c}\gtrsim \omegaa$, we can neglect $\rmd\Dp/\rmd t$ in the balance of terms because collisions  isotropize the pressure faster than $\Dp$ can be created. However, if $\nu_{\rm c}$ also satisfies $\nu_{\rm c}\lesssim \beta^{1/2}\omegaa$, the heat fluxes remain collisionless: they retain a similar form to  the collisionless regime \revchng{and thus have a similar effect on the dynamics}, scaling as  $\sim\!\beta^{1/2}\omegaa\Dp$,
which remains larger than the collisional draining of $\Dp$ (unless the parallel scales self-adjust; see \citealt{Squire2017}). This weakly collisional regime is thus a hybrid collisional--collisionless one: although motions in the plasma are slower than the collision timescales, heat fluxes remain strong and are governed
by collisionless physics \citep{Mikhailovskii1971}.
\item[Braginskii MHD, $\bm{\beta^{1/2}\omegaa\ll \nu_{\rm c}}$-- ]Once $\nu_{\rm c}\gg \beta^{1/2}\omegaa$, in addition 
to neglecting $\rmd\Dp/\rmd t$ compared to $\nu_{\rm c}\Dp$, we see that the heat fluxes take the collisional
form, scaling as $\sim\!(\beta \omegaa^{2}/\nu_{\rm c})\Dp$. Further, unlike in the weakly collisional regime, the heat fluxes become subdominant  to $\nu_{\rm c}\Dp$, 
since $\nu_{\rm c}\gg\beta \omegaa^{2}/\nu_{\rm c}$ for $\nu_{\rm c}\gg\beta^{1/2}\omegaa$. Thus at the same point that the heat fluxes take their collisional 
form, they become subdominant overall. In this regime, assuming incompressibility and $\Dp\ll p_{\perp}\sim p_{\|}$, equation~\cref{eq: Dp equation} takes the simple form 
\begin{equation}
\Dp \approx \frac{3p_{0}}{\nu_{\rm c}}\bbgu,\label{eq: dp in braginskii}
\end{equation}
which, when inserted into~\cref{eq:KMHD u}, yields a parallel viscous stress. Given this is effectively the parallel viscosity of \cite{Braginskii1965}, we refer to this regime as `Braginskii MHD.'
\end{description}

In previous work (\skqs), we explored the effect of pressure anisotropy in the Braginskii-MHD regime, because of the
simplicity of its physics and computational implementation. However, when combined with  the condition
that the pressure anisotropy has a dynamically important influence on the turbulence, $\nu_{\rm c}/\omegaa\lesssim \beta\, \delta B_{\perp}^{2}/B_{0}^{2}$ (see~\cref{sub: magneto-immutability}), the 
constraint $\nu_{\rm c}/\omegaa\gg \beta^{1/2}$  requires  $\beta$ to be 
very large \revchng{in order for there to be sufficient range between $\beta^{1/2}$ and $\beta\,\delta B_{\perp}^{2}/B_{0}^{2}$}. 
Thus, for application to the solar wind and other hot astrophysical plasmas with $\beta\lesssim 100$, the collisionless and weakly collisional
regimes are more  relevant.

A detailed derivation of the effects described above for a single shear-Alfv\'en wave, including simplified asymptotic equations valid in
each regime, is given in \cite{Squire2017}.

\subsection{Magneto-immutability}\label{sub: magneto-immutability}

In previous work \citep{Squire2016,Squire2017,Squire2017a}, we explored the idea that Alfv\'enically polarised magnetic-field or flow perturbations are `interrupted'
if their amplitude satisfies %, $\delta B_{\perp}/B_{0}\sim \delta u_{\perp}/\va$, is larger than the limit 
\begin{equation}
\frac{\delta u_{\perp}}{\va}\sim \frac{\delta B_{\perp}}{B_{0}} \gtrsim2 \beta^{-1/2} 
\begin{cases} 
1, & \omegaa\gg \nu_{\rm c} \\ 
\sqrt{{\nu_{\rm c}}/{\omegaa}}, & \omegaa\ll \nu_{\rm c} 
\end{cases}
\label{eq: interruption limit}
\end{equation}
(the first limit applies in the collisionless regime; the second applies in the weakly collisional and Braginskii-MHD regimes).  The 
origin of the effect is straightforward: above the limit~\eqref{eq: interruption limit}, the (nonlinear) perturbation of the  magnetic-field strength 
drives the pressure anisotropy to the fluid  firehose limit, $\Delta p = -B^{2}/4\upi$, or $\Delta = -2/\beta$. As can be seen from the final term of~\cref{eq:KMHD u}, 
this nullifies the plasma's magnetic tension (indeed this is the cause of the firehose instability), which thus robs the Alfv\'en wave of its 
restoring force (see~\cref{eq: dp alfven speed}). The consequence, for a single linearly polarised Alfv\'en wave, is  that the perturbation dumps most of its energy into plasma heating and/or magnetic-field
perturbations that cease to evolve in time, rather than oscillating in the
usual way \citep{Squire2016,Squire2017a}. Below the limit \cref{eq: interruption limit}, waves slowly damp due to  nonlinear Landau damping \citep[e.g.,][]{Hollweg1971} and/or nonlinear viscous damping \citep{Nocera1986,Russell2023}.

\subsubsection{Alfv\'enic turbulence}

Given that  Alfv\'enic motions underlie magnetised plasma turbulence \citep{Chen2016a,Schekochihin2020}, a natural question 
that arises is: what happens when large-scale random perturbations to the plasma are driven past the interruption limit~\eqref{eq: interruption limit}?
Na\"{i}vely, one might expect that only motions below the limit~\eqref{eq: interruption limit} would be allowed, which would imply that the large-scale fluctuation energy would be limited to less than $\delta u_{\perp}^{2}\sim \va^{2}/\beta$, with the remainder  of the  energy injected at large scales directly heating 
the plasma without creating smaller-scale motions and a turbulent cascade. 
Instead, we showed in \skqs\ that the turbulence rearranges itself  mostly to avoid the motions that would
create large pressure anisotropies in the first place, allowing the turbulent cascade to proceed in a way rather similar 
to MHD. It does this by reducing the variations in $B$ that would have driven large $\Dp$, creating a turbulence
in which $B$ varies significantly less than in turbulence where the pressure anisotropy does not back-react on the plasma motions. This  reduces the spread of $\Dp$ produced by the turbulence, thus reducing 
the plasma heating done by these motions and increasing the `cascade efficiency' (the proportion of  energy that cascades to small scales).

That the plasma does this is not altogether surprising. Indeed, it is well 
known that collisionless effects damp out motions that involve variations in the magnetic-field strength, and 
that this damping is  fast compared to Alfv\'enic timescales at high~$\beta$ \citep{Barnes1966,Foote1979}\footnote{\revchng{An exception 
is the gyrokinetic ($k_{\|}\ll k_{\perp}$) ``non-propagating'' mode, which is damped more slowly than the Alfv\'en frequency at high $\beta$. Its 
structure requires a specific $\Dp$ perturbation with  $\delta p_{\perp}> \delta p_{\|}$, meaning it is also strongly damped in the presence of modest collisions \citep{Schekochihin2009,Majeski2023}. }}. However, 
 the effect does not simply involve a selective damping of those motions that involve 
variations in $B$, thereby leaving behind those motions that do not. Rather, there is a direct force on the plasma  due to 
the final term $\grad\bcdot(\bh\bh\Dp)$ in~\cref{eq:KMHD u}, and this force opposes motions that would be strongly damped 
(those involving variation in $B$). The heating is thus significantly reduced compared to 
what would occur in the absence of this force. The origin of this behaviour is best understood by 
 analogy to compressive motions and  density fluctuations.
It is well known that isotropic pressure forces in a fluid resist compressional flows with $\grad\bcdot\bm{u}\neq0$: such 
a flow will generate a local pressure perturbation, which then (through the $-\grad p$ term) generates a 
force on the fluid that opposes the compressional motion.  In a fluid where 
the thermal energy is large compared to other energies (such as a plasma at high $\beta$), this process 
is rapid compared to the timescales on which the flow or magnetic field change, thus rendering the
system effectively incompressible. Magneto-immutability involves a  similar process, but with `{magneto-dilational}'
flows that have $\bbgu\neq0$. Such flows generate a local pressure-anisotropy perturbation (see~\cref{eq: Dp equation}), the feedback of which -- through the 
pressure anisotropy force $\grad\bcdot(\bh\bh\Dp)$ -- opposes the motion that created  $\bbgu$ in the first place. If the 
generation of pressure anisotropy is fast compared to the Alfv\'enic timescales of the turbulence, 
these forces will render the system `magneto-immutable', since motions with small $\bbgu$ are those that
minimise changes to the magnetic-field strength (see~\cref{eq: B and bbgu}).\footnote{A complication to this story 
involves the difference between a dissipative reaction, such as a Braginskii viscosity ${\propto}\grad\bcdot[\bh\bh(\bbgu)]$ or a bulk 
viscosity ${\propto}\grad(\grad\bcdot\bm{u})$, and a non-dissipative one, such as an isothermal pressure response ${\propto} \grad\rho$.
The pressure-anisotropy response in weakly collisional plasmas spans  both regimes: in the Braginskii-MHD regime, it is purely 
dissipative; in pure CGL without heat fluxes, it is non-dissipative; and in the collisionless and weakly collisional regimes, it lies between these two extremes. But, these different regimes seem to make less difference than one might expect.  Although this is rarely studied, standard neutral fluids 
are rendered incompressible by a large bulk viscosity \citep{Pan2017}, even in the absence of  pressure forces. 
Similarly, we find little obvious dependence of magneto-immutability on the collisionality regime, which controls 
both the level of dissipation caused by different types of motions, and the phase offset between $\Dp$ fluctuations and magneto-dilations $\bbgu$ (see~\cref{eq: Dp equation}). Fundamentally, all that  is needed is a large back-reaction force that inhibits motions of a particular 
form (compressions or magneto-dilations).  }

\subsubsection{A reduced model for magneto-immutable turbulence? }\label{subsub: a reduced model?}

The analogies between incompressibility and magneto-immutability lead 
one to speculate whether there could exist a reduced model -- similar to incompressible hydrodynamics -- that describes magneto-immutable turbulence. Incompressible fluid models are formulated by stipulating 
that, because the compressible back-reaction happens so rapidly, the pressure force is just what it needs to be in order 
to enforce $\partial_{t}(\grad\bcdot \bm{u} )= 0 $. This lets one solve for $\grad^{2}p$ in terms of $\bm{u}$, thus closing the system. 
By analogy, for a magneto-immutable fluid model, we should solve for the $\Dp$ in $\grad\bcdot(\bh\bh\Dp)$ that enforces $\partial_{t}(\bbgu) = 0 $, 
which will ensure that the flow cannot generate magneto-dilations as it evolves.
This immediately reveals a complex technical problem: while $\grad\bcdot$ is a linear operator, thus enabling straightforward solution of $p$, the combination $\bh\bh\bdbldot\grad$ is nonlinear, and solving for $\Dp$ (which must be 
achieved at every time step for a numerical algorithm) becomes complex and expensive. 

 More generally, there is another key difference with incompressibility that argues against
 the utility of formulating a magneto-immutable fluid model: regions with large pressure anisotropies 
 become unstable. The strong back-reaction force needed
 to suppress a large $|\bbgu|$ in some region will  require a large $|\Dp|$, 
 which will then grow small-scale instabilities, presumably tempering its influence. This effect is
 unavoidable because such instabilities are always triggered when $|\Dp|\gtrsim B^{2}/4\upi$  (see~\cref{sub: theory microinstabilities}), which
 is also the pressure anisotropy needed to start feeding back significantly on the flow. 
 In contrast, in a large compression or rarefaction, the distribution function can remain  isotropic and thus stable, 
 so there is no similar effect for incompressibility. This subtle, but important, difference
 between incompressibility and magneto-immutability is discussed in more detail in \cref{sub: sec 2 summary} and~\cref{subsub: mi and microinstabilities}.

\subsubsection{Interruption number}\label{subsub: interruption number}

It will prove helpful to have a simple dimensionless number that 
can be used to quantify the expected influence of the pressure anisotropy on the flow. To construct this, we return to the idea 
that individual shear-Alfv\'en waves are `interrupted' -- meaning that they dissipate a large fraction of their energy into thermal energy within ${\lesssim} \tau_{A}$ -- if their amplitude satisfies \cref{eq: interruption limit}, or
\begin{equation}
\frac{\delta B_{\perp}}{B_{0}}\sim \frac{\delta u_{\perp}}{\va}\gtrsim \maint \doteq \left[\max\left(1,\frac{\nu_{\rm c}}{\omegaa}\right) \frac{1}{\beta}\right]^{1/2}.
\end{equation}
%
%Here $\maint$ is simply the amplitude of a perturbation that will cause $|\Delta|\sim 2/\beta$. 
Applying this limit to a random turbulent collection of fluctuations with root-mean-square (rms) amplitudes $\ma\doteq \langle \delta u_{\perp}^{2}\rangle^{1/2}/v_{\rm A0} \approx \langle \delta B_{\perp}^{2}\rangle^{1/2}/B_{0}$, we define the `interruption number' to be 
\begin{equation}
\It \doteq \left(\frac{\maint}{\ma}\right)^{2}=\frac{\max\left(1,\dfrac{\nu_{\rm c}}{\omegaa}\right)}{\beta\ma^{2}}.\label{eq:interruption number}
\end{equation}
If $\It\lesssim1$, the turbulence (unless otherwise restricted) should be of sufficiently large amplitude  to generate $|\Delta|\gtrsim 2/\beta$, which would be na\"{i}vely expected to damp out the energy faster than it cascades to small scales. In this sense, $\It$ can be interpreted similarly to the Reynolds number, capturing viscous-like 
effects due to the pressure anisotropy, with $\It\lesssim1$ suggesting they dominate over the Alfv\'enic forces in the flow.

However, as described above, this expression ignores the feedback of the pressure anisotropy (magneto-immutability), which reduces the production 
of $\Dp$ below the estimate used to derive~\eqref{eq:interruption number}. Thus, as shown by \skqs, the turbulent cascade can in general proceed when $\It\lesssim 1$, 
meaning $\It$ is better thought of as an estimate of the importance of pressure-anisotropy effects, rather than whether the viscous damping dominates the dynamics (in other words, 
when $\It\lesssim1$, the flow can rearrange itself so as to avoid the motions that would be strongly viscously damped).
The purpose of our study is to understand the properties of turbulence 
in this $\It\lesssim 1 $ regime and characterise how it heats the plasma. 

Another way to think of the interruption number is by analogy to the Mach number of a compressible neutral fluid. 
Start by considering pressure-anisotropy production with the balance $\omegaa \Dp  \sim p_{0}\omegaa \delta B_{\perp}^{2}/B^{2}+ \nu_{\rm c}\Dp$ obtained from~\cref{eq: Dp equation} (ignoring the heat fluxes) and  giving the estimate $ \Dp \sim p_{0}\ma^{2}/\max(1,\nu_{\rm c}/\omegaa)$ for the size of $\Dp$ fluctuations that result from Alfv\'enic magnetic-field fluctuations. Because pressure anisotropy will 
feed back strongly on the plasma once $\Dp\gtrsim B^{2}$, and the estimate above gives $\Dp/B^{2}\sim \It^{-1} $, we see  that $\It^{-1} $ is
the ratio between the $\Dp$ that is driven and the $\Dp$ that would substantially change the flow. Analogously, 
in compressible hydrodynamics, isotropic pressure fluctuations $\delta p$ are related to density fluctuations $\delta \rho$ by $\delta p\sim c_{\rm s}^{2} \delta \rho$, where $c_{\rm s}$ is the sound speed. Pressure fluctuations will feed back strongly on the flow once they are comparable to the ram pressure $\delta p\sim\rho u^{2}$. \revchng{Therefore, the ratio of the naturally generated pressure fluctuations to those needed to feed back on the flow} is $\delta p/\rho u^{2}\sim \mathcal{M}^{-2}\delta \rho/\rho$, where $\mathcal{M}=u/c_{\rm s}$ is the Mach number. 
Thus, the isotropic hydrodynamic equivalent of $\It$ is $\mathcal{M}^{2}/(\delta \rho/\rho)$. But, in a completely unrestricted flow, $\delta \rho/\rho\sim 1$ (because 
the turnover rate ${\sim} ku$ at scale $k$ scales in the same way as $\grad\bcdot\bm{u}$), showing that 
$\mathcal{M}^{2}$ itself provides the relevant analogy with $\It$. This makes intuitive sense: $\mathcal{M}\sim 1$ marks the boundary between supersonic turbulence, which approaches the limit of the pressure-free Burgers equation, and incompressible turbulence, where $\grad\bcdot\bm{u}$ and $\delta \rho/\rho$ become strongly restricted by the feedback of the pressure force on the flow.

We can also define a local interruption number with respect to the scale-dependent amplitude of the eddies, assuming that the
turbulence follows a standard Goldreich--Sridhar cascade \citep{Goldreich1995}. This assumption is clearly  highly questionable 
if the effects of pressure anisotropy are strong, but our simulations will show it to be nevertheless reasonable  and it serves a useful purpose for basic estimates. Turbulent amplitudes scale 
as $\delta B_{\perp}^{2}\propto \delta u_{\perp}^{2}\propto k_{\perp}^{-2/3}$, where $k_{\perp}$ is the inverse scale 
of an eddy perpendicular to the local magnetic field. This implies that, in the collisionless regime where $\omegaa \Delta \propto \omegaa \delta B_{\perp}^{2}/B_{0}^{2}$ (see~\cref{eq: Dp equation,eq: B and bbgu}), the interruption number should be smallest 
at the outer scale and grow as $\It\propto k_{\perp}^{2/3}$. By contrast, in the Braginskii regime where $\nu_{\rm c}\Delta\propto \omegaa \delta B_{\perp}^{2}$, the critical balance scaling $\omegaa\propto k_{\|}\propto k_{\perp}^{2/3}$ implies that the interruption number is effectively constant across all scales above the collisionless transition where $\omegaa\gtrsim\nu_{\rm c}$.

\subsection{Microinstabilities, limiters, and the choice of collisionality}\label{sub: theory microinstabilities}

A key physical effect that has been omitted in the discussion above is the influence of kinetic microinstabilities. Most important at high $\beta$
are the firehose  and mirror instabilities \citep{Rosenbluth1956,Vedenov1958,Chandrasekhar1958a,Parker1958a,Hasegawa1969}, which are thought to be the fastest growing with 
the strongest back-reaction on the large-scale plasma dynamics. In their simplest forms, they are triggered for $\Delta\lesssim -2/\beta$ (firehose) or $\Delta\gtrsim 1/\beta$ (mirror), 
with minor modifications to these limits from fundamentally kinetic effects (resonances and finite-Larmor-radius physics), particularly at moderate $\beta$ \citep{Yoon1993,Hellinger2006}. Although most features of the linear instability thresholds and 
growth rates are captured by~\cref{eq:KMHD rho,eq:KMHD pl}  with the Landau-fluid heat fluxes~\cref{eq:GL heat fluxes p,eq:GL heat fluxes l} (see \citealt{Snyder1997}),
the detailed nonlinear saturation mechanisms, which involve particle scattering and trapping \citep{Schekochihin2008b,Kunz2014,Rincon2015,Melville2016}, are certainly not. 
Even more important is the 
separation of time scales that is inherent in how the microinstabilities feed back on the plasma: they grow and evolve on time scales comparable to
the ion gyro-frequency, which is far faster than any motions related to the outer scale for any astrophysical system of interest. Thus, 
as far as the large-scale ($\ell\gg\rho_{i}$) plasma dynamics are concerned, microinstabilities should saturate and feed back on the 
plasma effectively instantaneously. This poses an extreme difficulty for fully kinetic simulations, which must attempt to 
determine which observed features are dependent on the necessarily modest scale separation, and which are robust to asymptotically large scale separations
(e.g., \citealp{Kunz2020}; \aksqs). 

While the details of firehose and mirror saturation are rather complex \citep{Schekochihin2008b,Hellinger2008,Kunz2014,Riquelme2015,Sironi2015a,Melville2016},
roughly, the process involves the instabilities' magnetic fluctuations scattering particles at the rate needed to maintain $\Dp$ at its marginal level.\footnote{Scattering 
from the mirror instability seems to  reach this level only after a macroscopic shear time once $\delta B/B\sim 1$, with particle trapping 
alone able to maintain marginality over shorter times. However,
for the purposes of this discussion, the exact mechanism through which $\Dp$ is limited is actually not of great importance, so we  refer the reader 
to \citet{Schekochihin2008b,Kunz2014,Rincon2015,Melville2016} for details, rather than discussing these issues here.} This implies an additional microinstability-induced collisionality, $\nu_{\rm c}\sim S\beta$, where $S\sim B^{-1}\rmd B/\rmd t\approx \bbgu$ is the shearing rate,
which operates in regions where $\Dp$ is being driven beyond the mirror or firehose thresholds. 
Broadly speaking, such a picture has been commonly invoked to understand the  solar-wind `Brazil plots' \citep{Kasper2002,Hellinger2006,Bale2009},
which show how  the measured $\Dp$ appears to be limited between the instability threshold values ($|\Delta|\lesssim1/\beta$; see~\cref{fig: Brazil plot}): plasma that
strays beyond the boundaries will be rapidly pushed back via scattering, thus maintaining only a small deviation from $\Dp\approx0$ at high~$\beta$.\footnote{\revchng{In its  simplest form, this idea suggests that plasma should end up clustered near the firehose threshold, where it is driven 
by expansion. Instead, it is observed to have a rather broad distribution centred near $\Dp=0$. There are a number 
of plausible explanations for this difference, including anisotropic heating, Coulomb collisions \citep{Bale2009}, scattering with a `memory' (scattering sites that persist even 
as the plasma becomes stable; see \cref{subsub: mi and microinstabilities}), compressive oscillations that carry the plasma beyond the  thresholds and back \citep{Verscharen2016}, and, indeed,   magneto-immutability (see \cref{sub: sec 2 summary})}.}
Since the nominal effect of this scattering is simply to 
maintain $\Dp$ at marginality,  a simple phenomenological method to capture such effects in a fluid simulation is via the inclusion of `limiters', which 
 halt the growth of $\Dp$ locally in space whenever it is driven past the firehose or mirror thresholds \citep{Sharma2006}.
Physically, the approach assumes that (i) microinstabilities act quasi-instantaneously to return the plasma to marginal stability\footnote{This assumption is 
easy to relax via the implementation of a limiter collisionality; see~\cref{subsub: microinst computational}.}, and (ii)
that the microinstabilities do not directly influence the plasma's evolution outside of the regions that are being driven unstable, either in 
space or time. Despite its clear shortcomings, which will be discussed in detail in~\cref{subsub: mi and microinstabilities} after we present our computational results, the method is at least simple and well controlled, and we will use it
throughout this work.

\subsection{The expected impacts of magneto-immutability}\label{sub: sec 2 summary}

In this subsection  we summarise the basic impact of pressure-anisotropy feedback (magneto-immutability) by comparison to the counterfactual situation where  it does not exist. Because 
these effects tend to cause the plasma to revert to behaviour that more closely resembles the collisional (MHD) limit, they can be
 somewhat subtle and not easily diagnosed. Nonetheless, their influence on the heating processes and turbulent statistics can be strong and  appreciation of it is needed to understand the behaviour of turbulent  collisional  high-$\beta$ plasmas. 

As explained in~\cref{sub: magneto-immutability}, the basic picture involves the plasma rapidly reacting to suppress `magneto-dilational' flows
with large $\bbgu$, which are those that
would create large pressure anisotropies. The effect is very similar to incompressibility, if we substitute $\grad\bcdot\bm{u}$ with $\bbgu$, isotropic pressure $p$ with $\Dp$, 
and the $-\grad p$ force with that from $\grad\bcdot(\bh\bh \Dp)$.
As a consequence:
\begin{enumerate}
\item The standard deviation of $\Dp$ will be suppressed, \emph{viz.,} it will be lower than if 
$\Dp$ were driven by a turbulent flow with similar $\mathcal{M}_{\rm A}$ but that did not feel the force $\grad\bcdot(\bh\bh \Dp)$ (e.g., MHD turbulence).
\item The standard deviation of $B=|\bm{B}|$ will be suppressed in the same way as (i) (e.g., compared to a similar-amplitude MHD case). This is because suppressing $\bbgu$ also suppresses changes of $B$ (fundamentally, it is the changing $B$
that drives $\Dp$). 
\item The primary influence on the flow statistics will be the suppression of magneto-dilations ($\bbgu$) compared to MHD.
\item As a consequence, the net viscous-like heating of the plasma through pressure anisotropy will be suppressed by  magneto-immutability (compared to 
the counterfactual situation where $\Dp$ did not influence the flow). Since such heating is dominated by the outer scales in Alfv\'enic turbulence (see~\cref{subsub: interruption number}), there will thus 
be more vigorous turbulence with a larger cascade efficiency, and therefore a larger fraction of heating will occur via kinetic processes at the smallest scales \citep{Schekochihin2009}. This 
may influence bulk thermodynamical properties such as the ion-to-electron heating ratio or parallel-to-perpendicular heating ratios \cite[cf.][]{Sharma2007,Howes2008,Kawazura2019,Kawazura2020} or even the plasma's thermal stability \citep{Kunz2010}.
\end{enumerate}

While we provide solid numerical evidence in support of each of these points (i)--(iv) in~\cref{sec: results}, it is worth 
clarifying that these effects can never dominate all other dynamics, because there is no physical limit in which any of the aforementioned `suppressions' becomes complete. 
The reason for this is discussed in~\cref{sub: theory microinstabilities}, with more detail to arrive  in~\cref{subsub: mi and microinstabilities}: when pressure anisotropies become large, they cause plasma microinstabilities, which
act to limit $\Dp$ through scattering, trapping, and/or microscale fields. Such effects always occur together with the direct $\grad\bcdot(\bh\bh\Dp)$ feedback of $\Dp$ on 
the flow.\footnote{One exception to this is when there exists a mean pressure anisotropy, which can be driven, e.g., by 
plasma expansion \citep{Hellinger2008,Bott2021} or turbulent heating~(\aksqs). In this case, there are microinstabilities without a corresponding bulk force because 
there are no turbulent pressure-anisotropy gradients.} This
implies that even in the limit of extremely high $\beta$ and large amplitudes (small $\It$), $\bbgu$ can never be arbitrarily strongly suppressed,
because the $\grad\bcdot(\bh\bh\Dp)$ force that causes this suppression  is also attenuated by the microinstabilities. 
This is the most important difference between magneto-immutability and incompressibility; the latter does not suffer the same fate because 
isotropic pressure changes (driven by $\grad\bcdot\bm{u}$) are neither kinetically unstable nor attenuated by particle scattering (see~\cref{subsub: a reduced model?}). 
Similarly, related to point (iv) above, we still expect (and measure) significant  viscous heating at the outer scale (a cascade efficiency below unity); but, the cascade efficiency is independent of $\It$ and 
larger than what would be expected without the $\Dp$ feedback (in which case it would decrease continuously with $\It$, with turbulence amplitudes  limited to ${\ll}\maint$; see~\cref{eq: interruption limit}).

%%%%%%%%%%%%%%%%%%%%%%%%%%%%%%%%%%
\begin{figure}
\begin{center}
\includegraphics[width=0.485\columnwidth]{\ffold/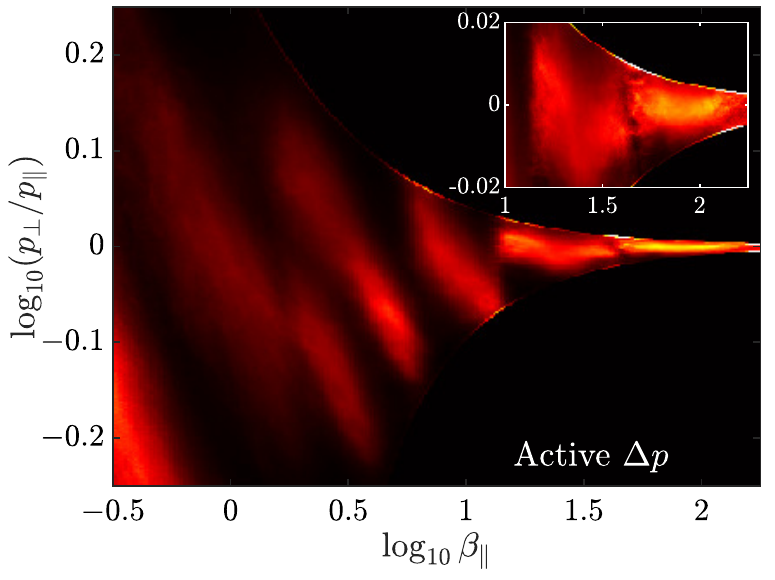}\includegraphics[width=0.515\columnwidth]{\ffold/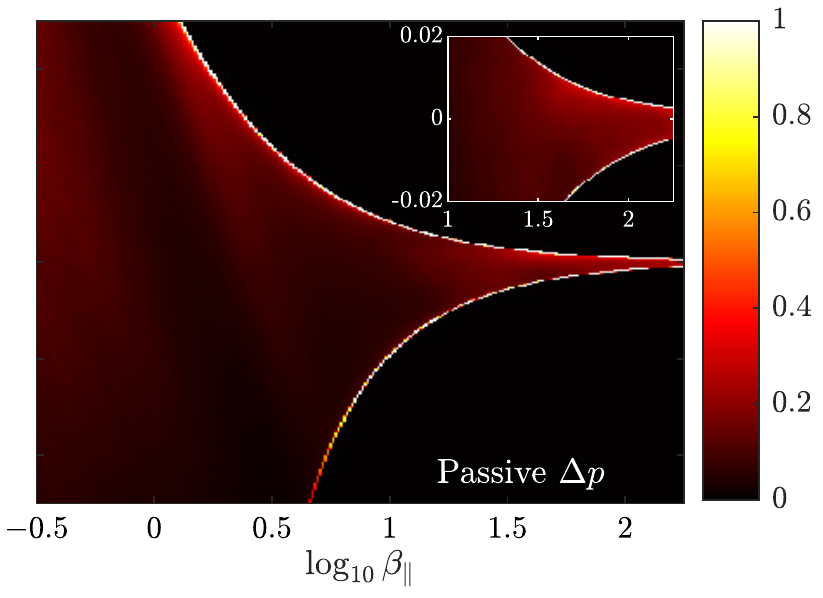}
\caption{`Brazil plots' -- histograms of (local) $\beta$ versus $p_{\perp}/p_{\|}$ -- formed by combining the data from \revchng{all runs that have both  active-$\Dp$ and passive-$\Dp$ simulations with the same parameters, as indicated by the `Passive?' column of \cref{tab:sims} (for passive-$\Dp$ simualations, the $\grad\bcdot(\bbgu)$ force in~\cref{eq:KMHD u} was artificially removed). The left panel combines  all such active-$\Dp$ simulations; the right panel combines all such passive-$\Dp$ simulations.}
The colour (probability) scale is normalised differently for each $\beta$ to better show the main features, but is done so identically for the left and right panels. The insets zoom in on the right-hand regions of each plot. The clear differences between
the active- and passive-$\Delta p$ simulations highlight the  effect of the 
 pressure-anisotropy feedback on the turbulence:  
 active-$\Dp$ cases maintain themselves primarily within the microinstability boundaries with small $|\Dp|$, while passive-$\Dp$ simulations (which have identical parameters otherwise, and similar turbulence amplitudes) have most of their volume artificially constrained by the hard-wall limiters. The basic conclusion is that the plasma employs dynamical pressure-anisotropy feedback, in addition to particle scattering (as more commonly discussed; \citealp{Kasper2002,Hellinger2006,Bale2009}), 
in order  to limit its deviations from local thermodynamic equilibrium and stay within 
 the kinetically stable region of parameter space.}
\label{fig: Brazil plot}
\end{center}
\end{figure}
%%%%%%%%%%%%%%%%%%%%%%%%%%%%%%%%%%

A simple, familiar visual illustration of the effects of magneto-immutability is shown in~\cref{fig: Brazil plot}, where 
we combine most of the simulations run as part of this work into two `Brazil' plots. These are 2-D probability distribution functions (PDFs) of $\beta$ and temperature (pressure) 
anisotropy, illustrating the approximate magnitude of the deviation from pressure isotropy as a function of $\beta$. As mentioned earlier, its
eponymous shape has usually been attributed to the action of microinstabilities scattering particles once $|\Delta|\gtrsim \beta^{-1}$ (the mirror and firehose thresholds;  \citealp{Kasper2002,Hellinger2006,Bale2009}). 
The purpose of~\cref{fig: Brazil plot} is to compare simulations of~\cref{eq:KMHD rho}--\cref{eq:KMHD pl} (left panel) to  equivalent
simulations  where the dynamical feedback of $\Dp$ has been artificially eliminated by removing $\grad\bcdot(\bh\bh\Dp)$ in~\cref{eq:KMHD u} and assuming an
isothermal ion-pressure response (`passive-$\Dp$' simulations; see~\cref{subsub: passive dp}). 
Both sets of simulations  include the effect of microinstabilities via hard-wall limiters at the firehose and mirror boundaries, which
prevent $\Dp$ from straying beyond the relevant instability thresholds. We see a clear difference between the two cases, with 
$\Dp$ distributions strongly dominated by regions at the microinstability boundaries (the artificial limiters) in the passive cases, but not when 
$\Dp$ feeds back on the flow. 
This demonstrates that, in its efforts to remain
close to local thermodynamic equilibrium, the plasma has two 
separate methods at its disposal. 
The first -- particle scattering through microinstabilities -- 
has been discussed by \citet{Kasper2002}, \citet{Hellinger2006}, \citet{Bale2009} and other subsequent works. The second -- the dynamical feedback from the pressure anisotropy -- 
should  be of similar importance for maintaining near-isotropy in most plasmas (see~\cref{subsub: mi and microinstabilities}), but 
has been largely ignored thus far.

% the two separate ways that the plasma
% maintains itself near local thermodynamic equilibrium. 
% In the first, as  discussed by \citet{Kasper2002} and subsequent authors, particle scattering maintains the isotropy. I 
% This demonstrates that the classic results of \citet{Kasper2002} and subsequent works can
% be equally well explained through magneto-immutability as through the scattering induced by kinetic microinstabilities. Indeed, 
% both effects are likely to be of approximately equal relevance  to the suppression of the pressure-anisotropy variation (see~\cref{subsub: mi and microinstabilities}).

%Put together, we thus expect that in a weakly collisional plasma, variation in $B$ should be minimized across all $\beta$, albeit
%with different mechanisms at low and high $\beta$. It is important to distinguish these
%effects theoretically, even though, in practice,  their observational consequences may be similar. In this study, they can be easily distinguished by a direct comparison of the active-$\Dp$ and passive-$\Dp$ simulations (see~\cref{subsub: passive dp}).

%
%
%
\section{Methods}\label{sec: methods}

\subsection{CGL Landau-fluid model}\label{sub: methods cgl lf}

Our computational study is based on the `CGL-Landau-fluid' (CGL-LF) model of \citet{Snyder1997}.
This allows us to probe the effects discussed above without the complications of a true kinetic 
model (see~\cref{sub: theory microinstabilities}).
The model solves~\cref{eq:KMHD rho}--\cref{eq:KMHD pl} supplemented by the Landau-fluid closure for heat fluxes (described in~\cref{subsub: heat fluxes computational}) and  simple `hard-wall' limiters on  $\Dp$ to approximate the effect of kinetic microinstabilities (\cref{subsub: microinst computational}).

We use the finite-volume \textsc{Athena++} code \citep{White2016,Stone2020}, modified to solve~\crefrange{eq:KMHD rho}{eq:KMHD pl} in the conservative form detailed 
in~\cref{app: numerical tests }. Briefly, this uses the total energy $E = p_{\perp}+p_{\|}/2+ B^{2}/8\upi + \rho |\bm{u}|^{2}/2$ and the
`anisotropy' $\an \doteq \rho \ln(p_{\perp} \rho^{2}/p_{\|}B^{3})$ as conserved variables, the latter following \cite{Santos-Lima2014}. We found 
through extensive numerical tests that using
$\an$ leads to a solver that is more numerically robust than if $p_{\perp}/B$ or $p_{\|}B^{2}/\rho^{2}$ is used as the second conserved variable, particularly when there exists significant variation in $B$. Given that $\an$, and the equations themselves, become ill defined for $B\rightarrow0$, we implement a numerical floor on  $B$, below which 
the equations revert to standard adiabatic MHD. We use the piecewise parabolic method with an HLL Riemann solver \citep{Toro2009}; although we developed and tested 
various other HLLD-inspired solvers \citep{Miyoshi2005,Mignone2007}, these were found to be insufficiently 
robust for this study (see~\cref{app:subsub: hlled solver}). 

The Riemann solver solves only the conservative part of~\crefrange{eq:KMHD rho}{eq:KMHD pl}, \emph{viz.,} that with $q_{\perp}=q_{\|}=0$ and $\nu_{\rm c}=0$.
We evaluate the heat fluxes in the form described below (see~\cref{eq: final computational heat fluxes}) using operator splitting with
slope limiters to ensure numerical stability and monotonicity of the solutions \citep{Sharma2007a,Dong2009}. We use the RKL1 super-time-stepping
algorithm of \cite{Meyer2014} to allow for much larger time steps by stepping over the CFL condition that results from the small-scale diffusive form of the
computational heat fluxes (although heat-flux-related time-step constraints are still relatively severe at high $\beta$ and high resolution).
Collisional terms are  evaluated at the end of each global time step $\delta t$ from the exact solution
of $\partial_{t}p_{\perp} = -\nu_{\rm c}\Dp/3$, $\partial_{t}p_{\|} = 2\nu_{\rm c}\Dp/3$ from $t$ to $t+\delta t$ (see~\cref{eq: pprp pprl coll soln}). This method is implicit and numerically stable for any time step, and so has the property 
that adiabatic MHD can be easily recovered from the CGL-LF system by setting $\nu_{\rm c}\gg \delta t^{-1}$ (which also causes the heat
fluxes to become negligible). Similarly, if we choose $\nu_{\rm c}\gtrsim \beta^{1/2}\omegaa$ (see~\cref{sub: different regimes}), the method becomes a convenient, stable, and computationally  efficient way to include a Braginskii-viscous 
stress in the standard MHD system. 

Turbulence is driven via a large-scale incompressible forcing term $\bm{F}$ added to the right-hand side of~\cref{eq:KMHD u}. This applies only in the directions perpendicular to the mean magnetic
field in order to drive primarily Alfv\'enically polarised fluctuations, and consists of the eight largest-scale modes in the box evolved in time as an Ornstein--Uhlenbeck process with correlation
time $t_{\rm corr}=\tau_{\rm A}$ (where $\tau_{\rm A}$ is the box-scale Alfv\'en time). At each time step, its amplitude is adjusted to enforce a constant energy-injection rate $\varepsilon = \langle \rho \bm{u}\bcdot\bm{F}\rangle$ (see, e.g., \citealp{Lynn2012}).

\subsubsection{Heat fluxes}\label{subsub: heat fluxes computational}

I\revchng{n the `Landau-fluid' heat fluxes \crefrange{eq:GL heat fluxes p}{eq:GL heat fluxes l}, the $k_{\|}$ and $\nabla_{\|}=\bh\bcdot\grad$  contain $\bh$, which varies in space along with  $c_{\rm s\|}^{2}=p_{\|}/\rho$. This implies that}
 that $2 c_{\rm s\|}^{2}({\sqrt{2 \upi }c_{\rm s\|} |k_{\parallel}|+\nu_{\rm c}})^{-1}$ and $8 c_{\rm s\|}^{2}[\sqrt{8 \upi }c_{\rm s\|} |k_{\parallel}|+(3\upi-8)\nu_{\rm c}]^{-1}$ should 
rightly be considered operators that are not diagonal in either Fourier space or in real space, making  them complex and  computationally expensive to evaluate (see \citealt{Snyder1997}).
 For this reason, we use a simplified form, motivated by \citet{Sharma2006}, in which  $|k_{\|}|$ in the denominators of~\cref{eq:GL heat fluxes p,eq:GL heat fluxes l} is replaced by a constant  $k_{L}$, which we   take to be $2\times 2\upi/L_{\|}$ to approximate $k_{\|}$ of  larger-scale motions.\footnote{This choice 
 is justified by the fact that, in all cases considered, pressure anisotropy has the strongest influence on the largest scales in the system (due to magneto-immutability
 in weakly collisional and Braginskii-MHD cases; see \cref{subsub: interruption number}). We have tested the dependence on $k_{L}$
in lower-resolution simulations and noticed no significant differences for reasonable choices.}
However, this substitution also leads to the undesirable property that, due to large parallel gradients at small scales,  the heat flux, which is now diffusive, can become ${\gg}\rho c_{\rm s\|}^{3}$, the maximum possible heat flux 
\citep{Hollweg1974a,Cowie1977}. To mitigate this, we additionally limit $q_{\perp}$ and $\,q_{\|}$ to their maximum possible value deduced from~\cref{eq:GL heat fluxes p,eq:GL heat fluxes l}, \emph{viz.,}
\begin{equation}
q_{\perp,{\rm max}} \doteq \sqrt{\frac{2}{\upi}} c_{\rm s\|} p_{\perp},\quad q_{\|,{\rm max}} \doteq \sqrt{\frac{8}{\upi}} c_{\rm s\|} p_{\|}
\end{equation}
(the second term in~\cref{eq:GL heat fluxes p} is ${\sim}\Delta\, \delta B_{\|}/B\ll1$ times smaller than the first, so we ignore its contribution). The heat fluxes are thus computed 
as
 \begin{equation}
\widetilde{q}_{\perp} = q_{\perp{L}}\frac{q_{\perp,{\rm max}}}{q_{\perp,{\rm max}}+|q_{\perp{L}}|},\quad \widetilde{q}_{\|} = q_{\|{L}}\frac{q_{\|,{\rm max}}}{q_{\|,{\rm max}}+|q_{\|{L}}|},\label{eq: final computational heat fluxes}
\end{equation}
where $q_{\perp{L}}$ and $q_{\|{L}}$ are evaluated using~\cref{eq:GL heat fluxes p,eq:GL heat fluxes l} with $|k_{\|}|=k_{L}$. We note 
that this approach effectively generalises the practice of  limiting an MHD heat flux to be $|\bm{q}|\leq5\phi \rho c_{\rm s}^{3}$
with $\phi\approx0.3$ \citep{Cowie1977},\footnote{Note that this standard multiplier $5\phi\approx1.5$ is numerically 
similar to that used here for $q_{\|}$.} which is commonly used in MHD simulations (e.g., \citealp{Vaidya2017}).

The method~\eqref{eq: final computational heat fluxes} is somewhat \emph{ad hoc}, which led us to explore various  other approaches in some detail. One  possibility for some regimes is to evaluate $|k_{\|}|$ along the constant mean field using Fourier 
transforms \citep{Passot2014,Finelli2021}, thus effectively approximating $|k_{\|}|=|\bh_{0}\bcdot\grad|$, rather than $|k_{\|}|=k_{L}$, in~\cref{eq:GL heat fluxes p,eq:GL heat fluxes l}. However, with extensive testing, we found this approach to be more prone to numerical instability and more computationally 
expensive than the simpler method described above, when the fluctuations are  large
compared to the background magnetic field (as is explored here). 
In any case, because heat fluxes will be strongly modified by scattering from microinstabilities (see~\cref{subsub: microinst computational} below),
an exact evaluation of the Landau-fluid form~\eqref{eq:GL heat fluxes p}--\eqref{eq:GL heat fluxes l} is likely irrelevant for 
detailed agreement to nonlinear collisionless physics, so long as the method captures their general influence on the flow (see~\cref{sub: different regimes}).

\subsubsection{Microinstability limiters}\label{subsub: microinst computational}

As discussed in~\cref{sub: theory microinstabilities}, although aspects of the firehose and mirror instabilities are captured by~\crefrange{eq:KMHD rho}{eq:KMHD pl} with the closure~\crefrange{eq:GL heat fluxes p}{eq: final computational heat fluxes}, their nonlinear saturation, which involves particle scattering and trapping, is not. 
We therefore artificially limit the pressure anisotropy to 
\begin{equation}
-\Lambda_{\rm FH} \frac{B^{2}}{8\upi}  < \Dp < \Lambda_{\rm M} \frac{B^{2}}{8\upi}, \label{eq: Dp boundary}
\end{equation}
where $\Lambda_{\rm FH}$ and $\Lambda_{\rm FH}$ define the firehose and mirror instability thresholds, respectively.\footnote{Another limiter
could be used to include the ion-cyclotron instability if desired, but this is only more important than mirror at lower $\beta$ \citep{Hellinger2006} and unimportant
for our overall results anyway.}
By default, we set $\Lambda_{\rm FH}=2$ and $\Lambda_{\rm M}=1$, which describe the canonical versions of these instabilities, but our results
do not  depend strongly on this choice.\footnote{At high $\beta$, the kinetic oblique firehose is destabilised at $\Lambda_{\rm FH}\approx 1.4$, which,  for sufficiently slow motions,
can scatter particles  fast enough to limit $\Dp$ \citep{Bott2021}.} Computationally, 
the   limiters work by applying a large scattering rate ($\nu^{{\rm lim}}_{c}=10^{10}\tau_{\rm A}^{-1}$ in all simulations) to any region outside of~\cref{eq: Dp boundary}, 
which quickly 
(within one time step) reduces $\Dp$ to lie on the relevant instability boundary  (see~\cref{app: heat fluxes and collisions}). 
We also apply this enhanced collisionality to the heat fluxes~\crefrange{eq:GL heat fluxes p}{eq:GL heat fluxes l}, thus strongly suppressing 
them in limited regions. This may be appropriate for mirror-limited regions, which show strong heat-flux suppression in simulations \citep{Kunz2020},
but it is likely much too strong 
in firehose-limited regions, which seem to be well described by the Braginskii estimate \citep{Kunz2020}, meaning it might be more 
appropriate to take $\nu_{\rm c}^{\rm lim}\sim \beta \bbgu$ (though this would be numerically complicated).

\subsubsection{Passive-pressure-anisotropy simulations}\label{subsub: passive dp}

For most simulations, in addition to the standard CGL Landau-fluid model (termed `active-$\Dp$' below), we have run an otherwise equivalent `passive-$\Dp$' simulation. The latter is identical to the former, except that the feedback of the  pressure anisotropy into the momentum equation is artificially removed. Instead, we use an isothermal equation of state with the sound speed chosen to give the same value of $\beta$ as in the active-$\Dp$ run. Note that these passive-$\Dp$ runs still evolve the pressure anisotropy and include heat fluxes in an identical way to the standard CGL Landau-fluid model. In this way, the $p_{\|}$ and $p_{\perp}$ statistics can be compared directly to understand the feedback of the pressure-anisotropy stress on the flow. 

In summary, a `passive-$\Dp$' simulation solves~\crefrange{eq:KMHD rho}{eq:KMHD pl} using the same forcing, initial conditions, and parameters as a standard (`active-$\Dp$') run, but, in~\cref{eq:KMHD u} we remove the $\Dp$ term and set $p_{\perp}=T_{i0}\rho$ with $8\upi\rho T_{i0}/B_{0}^{2} =\beta $ (a chosen initial parameter).
This implies that the velocity and magnetic-field evolution in such a simulation is  described by the isothermal MHD equations.

\subsection{Study design}\label{sub:study design}

All simulations are run in a $L_{x}=L_{\|}=2L_{\perp}$ aspect ratio box of volume $V$, which has mean density $\rho=\rho_0$ and is
threaded by a mean magnetic field $\bm{B}_{0}=B_0\hat{\bm{x}}$. The energy injection (forcing) level  $\varepsilon$
is set to $\varepsilon = 0.16 \va^{3}/L_{\perp}$, which is chosen empirically to give $\ma\sim \delta u_{\perp}/\va\sim \delta B_{\perp}/B_{0}\sim 1/2$ in steady state for MHD, as needed for critical balance at the outer scale (from hereon, $\va$ will refer to the mean-field Alfv\'en speed $\va=B_0/\sqrt{4\pi\rho_0}$). We initialise 
with isotropic pressure $p_{0}$, chosen to yield the desired initial $\beta$, $\beta_{0} = 8\upi p_{0}/B_{0}^{2}$. Recall that here and throughout, $\beta$ and $p_{0}$ refer 
to the ion contribution, and most simulations use $T_{e}=0$ by default in order to  diagnose more easily the influence of pressure anisotropy. 
We do not use an explicit isotropic  viscosity or resistivity in any simulations, relying on the grid to dissipate energy that reaches the smallest scales. Most simulations have a matching passive-$\Dp$ run, which is set up as explained in~\cref{subsub: passive dp}. With these default parameters, each simulation is specified by its initial ion $\beta$ and collisionality $\nu_{\rm c}$ (in units of $\va/L_{\perp}$; these are sometimes omitted for conciseness). These parameters then fix the expected interruption number from~\cref{eq:interruption number}, assuming
$\ma$ is not strongly modified by the effects of pressure anisotropy. 
The full set of simulations is listed in~\cref{tab:sims}.

We focus particularly on three simulations to probe both the collisionless and weakly collisional regimes with $\It\lesssim 1$ (meaning that
pressure-anisotropy feedback is a significant effect).
These use $\beta_{0}=10,\,\nu_{\rm c}=0$ (labelled \emph{CL10}); $\beta_{0}=100,\,\nu_{\rm c}=33\va/L_{\perp}$ (\emph{B100});
and $\beta_{0}=100,\,\nu_{\rm c}=0$ (\emph{CL100}). \emph{CL10} and \emph{B100}, which have a numerical resolution of $1120\times560^{2}$, are chosen to have similar $\It\simeq 0.4$ but 
explore two different,  collisionless and Braginskii, regimes laid out in~\cref{sub: different regimes}.  \emph{CL100} (with resolution $560\times280^{2}$) is chosen to examine a situation with stronger pressure-anisotropy feedback,
$\It\approx 0.04$. The extremely small 
time steps required for the stable evaluation of the heat-flux terms make high-resolution simulations quite costly in wall-clock time, and so  
 \emph{CL10} and \emph{B100} are initialised from the saturated state of lower-resolution simulations (see below). We then run them for only a relatively short time of $t\approx L_{\perp}/\va$, which is  shorter than the outer-scale turnover time ${\simeq}2 L_{\perp}/\va$, but longer than  the time, $t\approx 0.1L_{\perp}/\va$, that it takes the new smaller scales  to reach turbulent steady state (we observe the evolution of the energy spectrum to ensure that this is the case). This means that these simulations are useful 
for exploring detailed properties of the turbulence (e.g., turbulence structure and energy transfers), but the largest scales (those above around a quarter of the box scale) are not properly statistically averaged.

In addition to these simulations, we have 
a large array of other ones  at low resolution ($400\times 200^{2}$). These are designed to explore how the physics of magneto-immutability varies 
with plasma parameters.  `\emph{lrCL}' simulations are all collisionless, changing $\beta_{0}$ from ${\lesssim}1$ to ${\approx}100$ in order to probe the dependence
of the turbulence on the interruption number from $\It\approx 0.04$ to $\It\gtrsim1$. In simulations with $\beta_{0}\lesssim1$,  the plasma is heated,  thus increasing $\beta$, rather rapidly, limiting the time they sit near their initial parameters.  The `\emph{lrB}' simulations probe the collisionless, weakly collisional, and Braginskii-MHD regimes discussed in~\cref{sub: different regimes}
by keeping $\It\simeq0.4$ approximately constant while changing $\beta_{0}$ and $\nu_{\rm c}$, with $\beta_{0}$ ranging from $30$ to $600$ and $\nu_{\rm c}\approx0.33\beta_{0}\,L_\perp/\va$ (see~\cref{eq:interruption number}). The `$\beta16$' simulations, which 
do not have a matching set of passive runs,  all have $\beta_{0}=16$ and
scan from the collisionless to the MHD regime in order to probe the approach to collisional MHD and  compare more directly to the solar wind and/or kinetic simulations of \aksqs. Finally, 
we have carried out a number of simulations to test additional physical effects, including the effect of finite electron temperatures and different microinstability limiters.

%%%% A table of the simulations %%%%%%%%%%%%%%%%%%%%%%%%%%%%%%%%%%%%%%%%%%%%%%%%%%%%%%%%%%%%
\begin{table*}
\begin{center}
\scalebox{0.75}{
 \begin{tabular}{c c c c c c c c c }  Name  & $\beta_{0} $ & $\nu_{\rm c}\frac{L_{\perp} }{\va}$ & $N_{\|}\times N_{\perp}^{2}$ & $\It$ & $\frac{\ell_{\perp}^{\rm WC}}{L_{\perp}}$ & $\frac{\ell_{\perp}^{\rm CL}}{L_{\perp}}$ & Passive?& Notes  \\ [0.5ex] 
 \hline\hline
\emph{CL10} & 10 & 0 & $1120\times560^{2}$ & $\simeq0.4$  & N/A & $>1$ &\checkmark &   \specialcell{Refined from \emph{lrCL}10;\\run for $t\approx L_{\perp}/\va$} \\
\hline
\emph{B100} & 100 & 33 & $1120\times560^{2}$ & $\simeq0.4$ &$ \simeq1$ & $ \simeq0.03$ & \checkmark &  \specialcell{Refined from \emph{lrB}100;\\run for $t\approx L_{\perp}/\va$} \\
\hline
\emph{CL100} & 100 & 0 & $560\times280^{2}$ & $\simeq0.04$ & N/A & $>1$ & \checkmark   &   \\
\hline\hline
\emph{lrB}30 & 30 & 10 & $400\times200^{2}$ & $\simeq 0.4$ & $>1$ &$\simeq 0.2$ & \checkmark   &   \\
\hline
\emph{lrB}100 & 100 & 33 & $400\times200^{2}$ & $\simeq 0.4$ & $ \simeq1$ &$\simeq 0.03$ & \checkmark   &   \\
\hline
\emph{lrB}600 & 600 & 200 & $400\times200^{2}$ & $\simeq 0.4$ & $ \simeq0.2$ &$\simeq 0.002$ & \checkmark   &   \\
\hline\hline
\emph{lrCL}0.2 & 0.2 & 0 & $400\times200^{2}$ & $\simeq 10$ & N/A &$>1$ & \checkmark   &   \\
\hline
\emph{lrCL}1 & 1 & 0 & $400\times200^{2}$ & $\simeq 4$ & N/A &$>1$ & \checkmark   &   \\
\hline
\emph{lrCL}3 & 3 & 0 & $400\times200^{2}$ & $\simeq 1$ & N/A &$>1$ & \checkmark   &   \\
\hline
\emph{lrCL}10 & 10 & 0 & $400\times200^{2}$ & $\simeq 0.4$ & N/A &$>1$ & \checkmark   &   \\
\hline
\emph{lrCL}30 & 30 & 0 & $400\times200^{2}$ & $\simeq 0.1$ & N/A &$>1$ & \checkmark   &   \\
\hline
\emph{lrCL}100 & 100 & 0 & $400\times200^{2}$ & $\simeq 0.04$ & N/A &$>1$ & \checkmark   &   \\
\hline\hline
CL10fh1.4 & 10 & 0 & $400\times200^{2}$ & $\simeq 0.4$ & N/A &$>1$ & \ding{55}  &  Reduced firehose limit $\Lambda_{\rm FH}=1.4$ \\
\hline
CL10Te3 & 10 & 0 & $400\times200^{2}$ & $\simeq 0.4$ & N/A &$>1$ &  \ding{55}  &  Isothermal electrons, $\frac{T_{e}}{T_{i0}}=0.6$  \\
\hline
CL10Te5 & 10 & 0 & $400\times200^{2}$ & $\simeq 0.4$ & N/A &$>1$ & \ding{55}  & Isothermal electrons,  $\frac{T_{e}}{T_{i0}}=1.0$  \\
\hline\hline
$\beta$16$\nu$0 & 16 & 0 & $400\times200^{2}$ & $\simeq 0.25$ & N/A &$>1$ & \ding{55}     &   \\
\hline
$\beta$16$\nu$3 & 16 & 3 & $400\times200^{2}$ & $\simeq 0.25$ & $>1$ &$\simeq 1$ & \ding{55}     &   \\
\hline
$\beta$16$\nu$6 & 16 & 6 & $400\times200^{2}$ & $\simeq 0.5$ & $>1$ &$\simeq 0.4$ & \ding{55}     &   \\
\hline
$\beta$16$\nu$12 & 16 & 12 & $400\times200^{2}$ & $\simeq 1$ & $\simeq1$ &$\simeq 0.13$ & \ding{55}     &   \\
\hline
$\beta$16$\nu$24 & 16 & 24 & $400\times200^{2}$ & $\simeq 2$ & $\simeq0.4$ &$\simeq 0.05$ & \ding{55}     &   \\
\hline
$\beta$16$\nu$50 & 16 & 50 & $400\times200^{2}$ & $\simeq 4$ & $\simeq0.13$ &$\simeq 0.016$ & \ding{55}     &   \\
\hline
$\beta$16$\nu$100 & 16 & 100 & $400\times200^{2}$ & $\simeq 8$ & $\simeq0.04$ &$\simeq 0.006$ & \ding{55}     &   \\
\hline
$\beta$16$\nu$200 & 16 & 200 & $400\times200^{2}$ & $\simeq 16$ & $\simeq0.016$ &$\simeq 0.002$ & \ding{55}     &   \\
\hline
$\beta$16$\nu$400 & 16 & 400 & $400\times200^{2}$ & $\simeq 32$ & $\simeq0.006$ &$\simeq 0.0007$ & \ding{55}     &   \\
\hline\hline
\end{tabular}
}
\end{center}\caption{A list of all simulations used in this article. Key input 
parameters are $\beta_{0} = 8\upi p_{0}/B_{0}^{2}$ and $L_{\perp}\nu_{\rm c}/\va$, where the subscript `$0$' refers to 
an initial value ($\beta$ decreases as $B$ grows  and the turbulence causes ion heating). The interruption number $\It$ is computed 
from these initial parameters using~\cref{eq:interruption number}. The collisionality regime is specified via $\ell_{\perp}^{\rm WC}$ and $\ell_{\perp}^{\rm CL}$, which are
the approximate scales below which motions transition into the weakly collisional regime and collisionless regime, respectively. These are computed 
by equating $\nu_{\rm c}/\omegaa(\ell_{\perp})=\beta^{1/2}$ and $\nu_{\rm c}/\omegaa(\ell_{\perp})=1$, where  $\omegaa$ is estimated by $(2\upi\va/L_{\|})(\ell_{\perp}/L_{\perp})^{-2/3}$, as for a non-aligned  critically
balanced cascade  \citep{Goldreich1995,Schekochihin2020}. A turbulent eddy of scale $\ell_{\perp}$ is in the Braginskii-MHD regime for $\ell_{\perp}\gtrsim \ell_{\perp}^{\rm WC} $, in the weakly collisional 
regime for $\ell_{\perp}^{\rm WC} \gtrsim\ell_{\perp}\gtrsim \ell_{\perp}^{\rm CL} $, and in the collisionless regime for $\ell_{\perp}\lesssim  \ell_{\perp}^{\rm CL} $ (see~\cref{sub: different regimes}). The `Passive?' column indicates whether an otherwise identical passive-$\Dp$ simulation was run for comparison purposes.  All simulations
have $L_{\|}=2 L_{\perp}$ and the energy injection rate $\varepsilon = 0.16 \va^{3}/L_{\perp}$.}
\label{tab:sims}
\end{table*}

\subsection{Diagnostics}

\subsubsection{Energy and rate-of-strain spectra}\label{subsub: ros spectra defs}

The spectrum of a field $ \phi$ is defined as
\begin{equation}
\mathcal{E}_{ \phi}(k) = \frac{1}{\delta k}\sum_{|\bm{k}|= k}|\hat{\phi}(\bm{k})|^{2},
\end{equation}
where $\hat{\phi}(\bm{k})$ is the Fourier transform of $\phi$,  $|\bm{k}|= k$ indicates a sum over all modes that fall in the relevant bin of  $k= \sqrt{k_{x}^{2}+k_{y}^{2}+k_{z}^{2}}$ or $k_{\perp} = \sqrt{k_{y}^{2}+k_{z}^{2}}$, and $\delta k$ is the width of the bin. 
The spectra are computed using a fine, logarithmically spaced $k$ or $k_{\perp}$ grid, removing the large-scale bins that 
do not contain any modes.  Wavenumbers and spectra and are plotted in units of $L_\perp^{-1}$ and $\phi^2 L_\perp$ (for various $\phi$), respectively, which will be omitted for conciseness in most figures.

Useful measures of turbulence structure are the spectra of the parallel and perpendicular rates of strain. These are formed via
\begin{gather}
\nabla_{\|}u_{\|} \doteq \hat{b}_{i}\hat{b}_{j}\nabla_{i}u_{j},\nonumber \\
(\nabla_{\|}\bm{u}_{\perp})_{l} \doteq \hat{b}_{i}(\delta_{lj}- \hat{b}_{j} \hat{b}_{l})\nabla_{i}u_{j},\nonumber \\
(\grad_{\perp}u_{\|})_{l} \doteq (\delta_{li}- \hat{b}_{i} \hat{b}_{l})\hat{b}_{j}\nabla_{i}u_{j},\nonumber \\
(\grad_{\perp}\bm{u}_{\perp})_{kl} \doteq(\delta_{ki}- \hat{b}_{i} \hat{b}_{k})(\delta_{lj}- \hat{b}_{j} \hat{b}_{l})\nabla_{i}u_{j},\label{eq: grad prp prl defs} 
\end{gather}
with the spectrum of a vector or tensor (e.g., $\nabla_{\|}\bm{u}_{\perp}$) computed by summing the spectra of each component. This  gives a $\grad_{\perp}\bm{u}_{\perp}$ spectrum that is nearly identical to the dissipation spectrum ($\grad \bm{u}$ spectrum).
Note that $\nabla_{\|}{u}_{\|}$ with this definition is exactly what must be minimised to minimise the generation of pressure anisotropy.

We also consider pressure anisotropy gradients defined in a similar way: 
\begin{gather}
\nabla_{\|}\Dp \doteq \bh\bcdot\grad \Dp\nonumber, \\
\grad_{\perp}\Dp \doteq (\msb{I}-\bh\bh)\bcdot\grad\Dp\label{eq: grad prp prl Dp}.
\end{gather}
These will prove useful to quantify the reduction in the
pressure-anisotropy stress $\grad\bcdot (\bh\bh\Dp)$ compared to $\Dp$ itself.

\subsubsection{Structure functions}\label{subsub: structure functions}

To diagnose the 3-D structure of turbulent eddies, we use three-point second-order structure functions, conditioned on the angle between the point-separation vector and the local field and/or the local perturbation. 
For any field $\phi$, these are defined as 
\begin{equation}
S_{2}[ \phi](\bm{\ell}) =  \langle | \phi(\bm{x}+\bm{\ell}) - 2 \phi(\bm{x}) +  \phi(\bm{x}-\bm{\ell})|^{2}\rangle,
\end{equation}
with the average taken over all $\bm{x}$.\footnote{\revchng{Note that the results from  three-point structure-function measurements can be 
interpreted in  the same way as those of the more common  two-point structure functions. But higher-point measurements have the ability to capture  steeper power-law scalings, similar in many ways to using wavelets} \citep{Cho2009,Lovejoy2012}.} The separation vector $\bm{\ell}$ is conditioned on its angle to $\bm{B}_{\ell}=[\bm{B}(\bm{x}+\bm{\ell}) + \bm{B}(\bm{x}) + \bm{B}(\bm{x}-\bm{\ell})]/3$ to obtain parallel and perpendicular structure functions ($\ell_{\|}=\hat{\bm{b}}_{\ell}\bcdot\bm{\ell}$ and $\bm{\ell}_{\perp} = \bm{\ell}-\ell_{\|}\hat{\bm{b}}_{\ell}$ with $\hat{\bm{b}}_{\ell}\doteq \bm{B}_{\ell}/|\bm{B}_{\ell}|)$ \citep{Cho2009,StOnge2020}. We also  condition $\bm{\ell}$ on 
its direction with respect to the local field and flow perturbations to study the alignment of the turbulent fluctuations \citep{Boldyrev2006,Chen2012,Mallet2015}.

Following the computation of $S_{2}$, a useful diagnostic of the turbulence structure is the scale-dependent anisotropy $\ell_{\|}^{\phi}(\ell_{\perp})$ for
a given field $ \phi$. 
This is computed by solving the equation $S_{2}[ \phi](\ell_{\perp})=S_{2}[ \phi](\ell_{\|})$ using numerical interpolation.

\subsubsection{Energy transfer functions}\label{subsub: energy transfers}

Energy-transfer functions are defined as in \cite{Grete2017} and \aksqs. The transfer function 
$\mathcal{T}_{q\rightarrow k}^{\rm AB}$
measures the average transfer  of energy from $k$-shell $q$ to shell $k$ due to the interaction labelled AB. 
Here, $k$ shells are defined by the Fourier-space filtering operation,
\begin{equation}
 \phi_{k}(\bm{x}) = \sum_{|\bm{k}|= k}\,\hat{\phi}(\bm{k})\rme^{\imag\bm{k}\bscdot \bm{x}},
\end{equation}
where $\hat{\phi}(\bm{k})$ is the Fourier transform of some field $\phi(\bm{x})$, and $|\bm{k}|= k$ represents those wavenumbers inside the logarithmic
shell centred around $k$ (i.e., $\ln k - \tfrac12 \rmd\ln k \leq |\bm{k}| <\ln k+\tfrac12 \rmd\ln k $). Such a definition clearly satisfies the property $\phi(\bm{x}) = \sum_{k}\phi_{k}(\bm{x})$ and represents the part of $\phi$ centred around wavenumber $k$. The label AB relates to the influence of different terms in~\crefrange{eq:KMHD rho}{eq:KMHD pl}:
e.g., kinetic energy can be transferred between shells through the Reynolds stress in the momentum equation $\rho \bm{u}\bcdot\grad\bm{u}$, with 
\begin{equation}
\mathcal{T}_{q\rightarrow k}^{\rm UU} = -\int \rmd \bm{x}\, \langle \sqrt{\rho}\bm{u}\rangle_{k}\bcdot\left[\bm{u}\bcdot\grad \langle \sqrt{\rho}\bm{u}\rangle_{q}\right],\label{eq: UU  transfer}
\end{equation}
or magnetic energy
can be transferred to kinetic energy through  $\bm{B}\bcdot\grad\bm{B}$.  Further details of the specific transfer terms are given in~\cref{sub: heating}.

The full 2D transfer functions can give useful information on the locality of the cascade, but can be difficult to interpret quantitatively. Two useful reductions 
are the \emph{\revchng{net energy transfer}}
\begin{equation}
{\rm T}^{\rm AB}(k) = \sum_{q}{\mathcal{T}}^{\rm AB}_{q\rightarrow k},\label{eq: total energy transfer}
\end{equation}
and the \emph{flux} 
\begin{equation}
\Pi^{\rm AB}(k) = \sum_{q\leq k}\sum_{p>k}{\mathcal{T}}^{\rm AB}_{q\rightarrow p}.\label{eq: energy fluxes}
\end{equation}
\revchng{The net transfer ${\rm T}^{\rm AB}(k)$ is the contribution of a given term ${\rm AB}$ in \crefrange{eq:KMHD rho}{eq:KMHD pl} to the rate of change of the energy spectrum at a particular $k$ (for example ${\rm T}^{\rm UU}(k)$ is the net contribution to the kinetic-energy spectrum at $k$ from the term $\rho \bm{u}\bcdot\grad\bm{u}$)}. 
In the inertial range in steady state, all terms should be
zero because $\partial_{t}\mathcal{E}(k)=0$, unless there is a continual transfer of energy into or out of the particular shell or between terms (for example, a damping). 
The flux $\Pi^{\rm AB}(k)$
quantifies the transfer of energy across a particular $k$, such that the sum over all contributions AB measures  the cascade rate. The interpretation of individual terms is less obvious, but gives interesting information about the dominant energy-transfer processes in the cascade (e.g., whether energy proceeds to smaller scales via transfers between $\bm{u}$ and $\bm{B}$, or due to transfers from larger scale $\bm{u}$ to smaller scale $\bm{u}$ directly).

\section{Results}\label{sec: results}

%%%%%%%%%%%%%%%%%%%%%%%%%%%%%%%%%%
\begin{figure}
\begin{center}
%\!\includegraphics[height=3.452cm]{\ffold/pic-uz.pdf}\includegraphics[height=3.49cm]{\ffold/pic-uz-p.pdf}\\
%\includegraphics[height=3.15cm]{\ffold/pic-dp.pdf}\includegraphics[height=3.26cm]{\ffold/pic-dp-p.pdf}\\
%\includegraphics[height=3.15cm]{\ffold/pic-B.pdf}\includegraphics[height=3.26cm]{\ffold/pic-B-p.pdf}
\includegraphics[width=\columnwidth]{\ffold/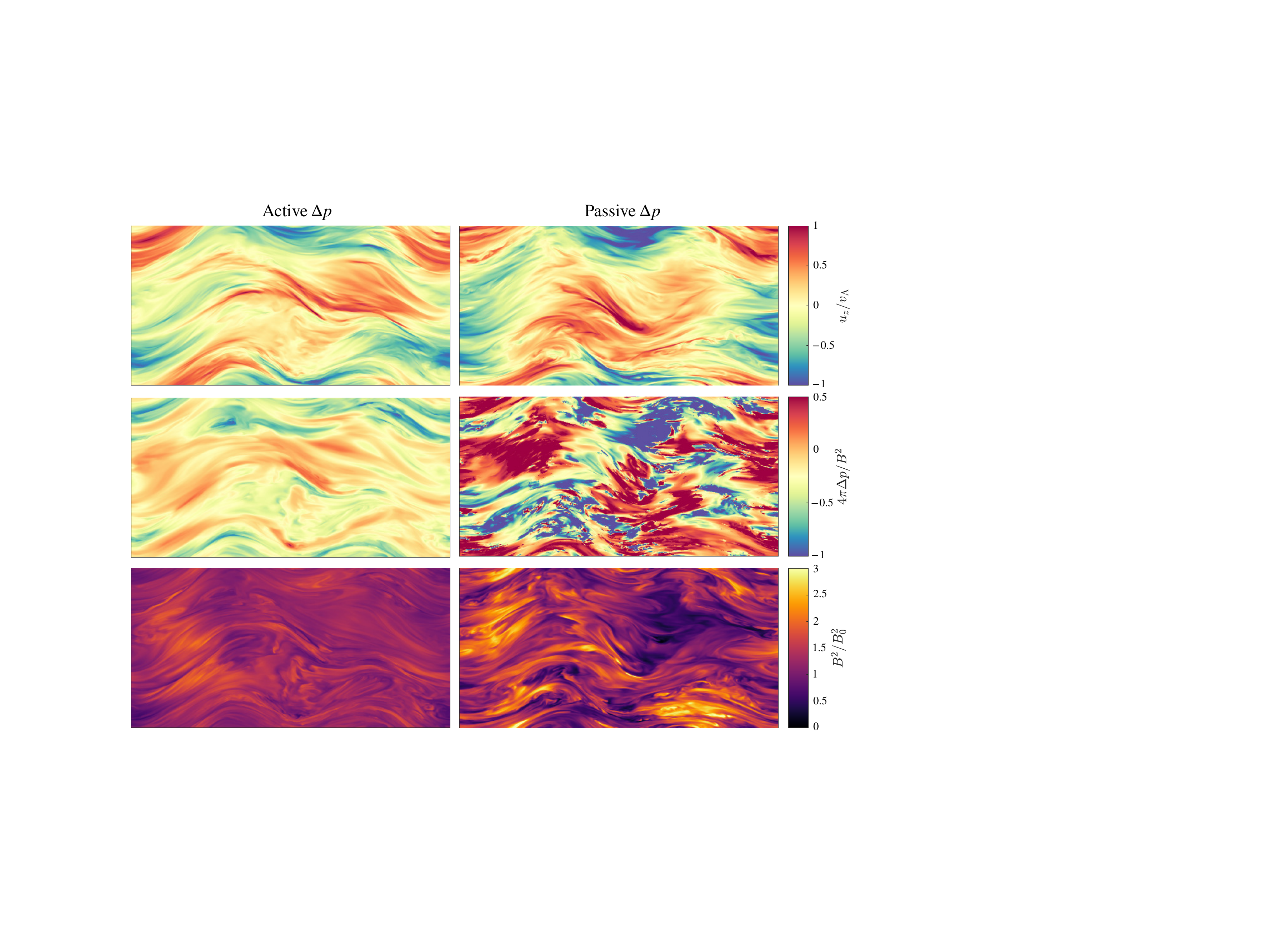}
\caption{Visualization of $u_{z}/v_{\rm A}$ (top panels), $4\upi\Dp/B^{2}$ (middle panels) and  $B^{2}=|\bm{B}|^{2}$ (bottom panels) through an $x$-$y$ slice of the \emph{CL10}  simulation
with $\beta_{0}=10$ and $\nu_{\rm c}=0$ ($\It\approx 0.4$). Left panels show the active-$\Dp$ run (which solves~\crefrange{eq:KMHD rho}{eq:KMHD pl}), and right panels show the passive-$\Dp$ run, in which the $\grad\bcdot(\bh\bh\Dp)$ force was artificially removed. While the fluctuations in $u_{z}$ look rather similar (i.e., the turbulence has a similar 
amplitude), magneto-immutability significantly reduces the variation in $B$ and $\Dp$ in the active-$\Dp$ run. Note that $-1<4\upi \Dp/B^{2}<1/2$ is enforced by the limiters in both runs. }
\label{fig: visualization}
\end{center}
\end{figure}
%%%%%%%%%%%%%%%%%%%%%%%%%%%%%%%%%%

While  the most astrophysically relevant consequences of magneto-immutability concern turbulent heating, it is necessary first to understand the details of how the field and flow self-organize to be magneto-immutable and how this characteristic manifests in the turbulence statistics.
We thus start by diagnosing the basic effect of pressure anisotropy feedback by comparing probability distribution functions (PDFs) and Fourier spectra of various fluctuations in the active-$\Dp$ simulations with those obtained in the passive-$\Dp$ simulations (\crefrange{sub: results 1}{sub: results 2}). 
We then consider the changes to the flow structure that are necessary in order to enable these effects in~\cref{sub: flow rearrangement results}, before diagnosing the viscous heating and cascade efficiency using energy transfer functions in~\cref{sub: heating}. 
Overall, the results show that, aside from a small range at the outer scale, magneto-immutability allows the system to set up a vigorous, nearly
conservative cascade that is in most respects similar to standard MHD.

\subsection{Reduction of the pressure anisotropy}\label{sub: results 1}

\begin{figure}
\begin{center}
\includegraphics[width=0.495\columnwidth]{\ffold/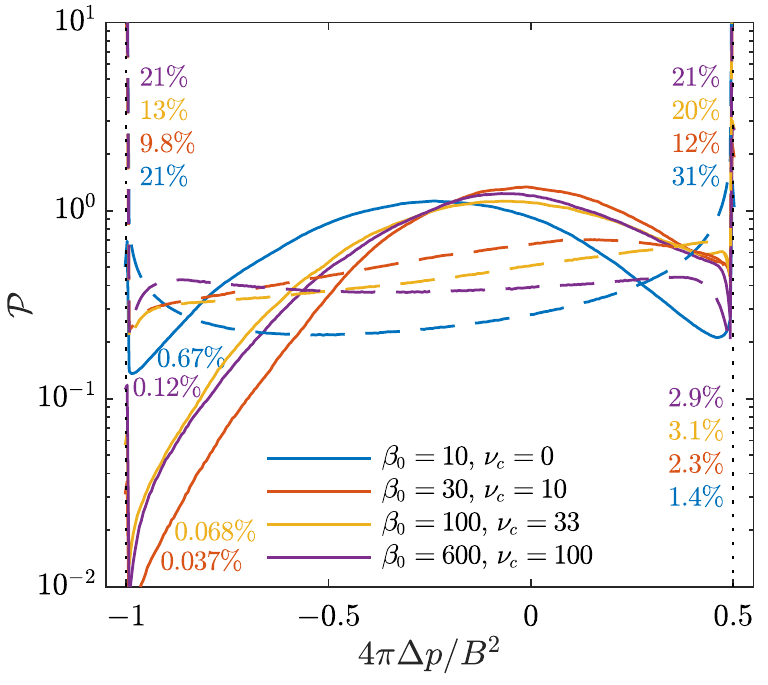}\includegraphics[width=0.505\columnwidth]{\ffold/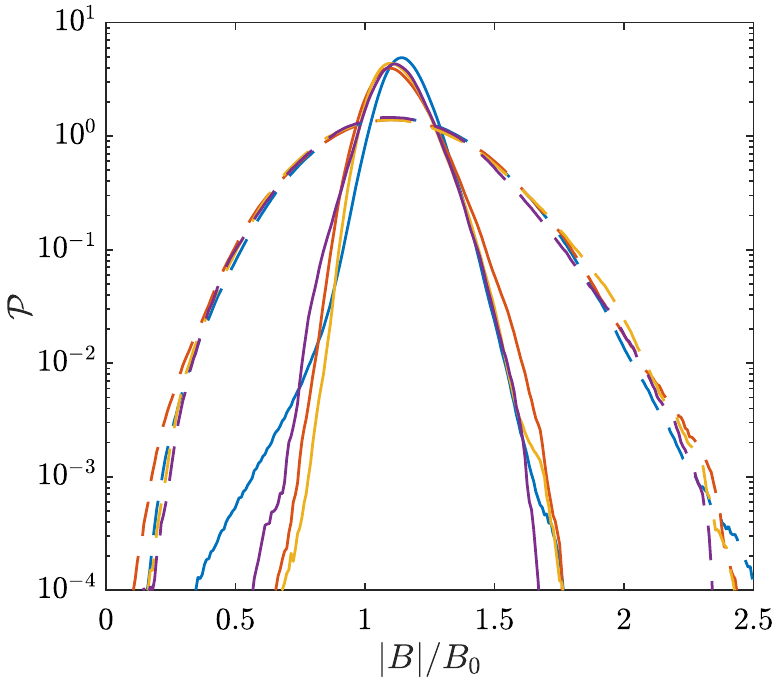}
\caption{Left panel: PDFs of $4\upi \Dp/B^{2}$ in the \emph{lrB} simulation set, which explores the collisionless, weakly collisional, and Braginskii regimes at  fixed $\It\approx 0.4$ (see~\cref{sub:study design}).   Active-$\Dp$ simulations are shown with solid lines  and passive-$\Dp$ simulations with dashed lines. The percentages shown on the left and right sides of the panel indicate the proportions (in volume) of each simulation that lie at the firehose and mirror thresholds, respectively (colours match the line styles; the larger numbers in the upper part of the plot are for the passive simulations). The effect of pressure anisotropy is relatively strong; only a few percent of the box lies at the mirror/firehose thresholds in the active-$\Dp$ cases, while ${\sim}50\%$ of the box lies beyond these thresholds when $\Dp$ has no effect on the flow. Right panel: PDFs of $B=|\bm{B}|$ 
for the same simulations.  In active-$\Dp$ cases, we  see a factor-of-a-few decrease in the variance of $B$ (hence,  `magneto-immutability'), which does not  depend strongly on the regime of collisionality.}
\label{fig: Dp PDFs brag}
\end{center}
\end{figure}

\begin{figure}
\begin{center}
\includegraphics[width=0.495\columnwidth]{\ffold/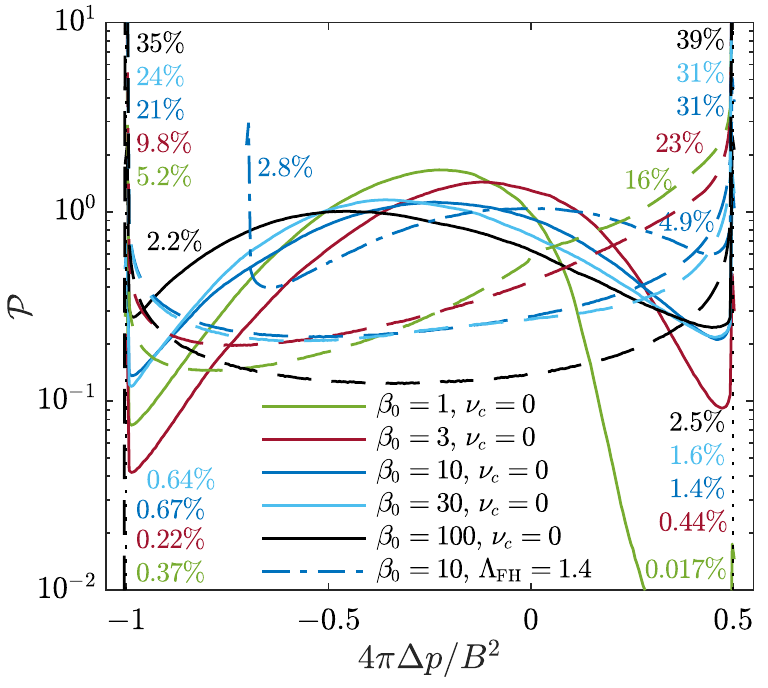}\includegraphics[width=0.505\columnwidth]{\ffold/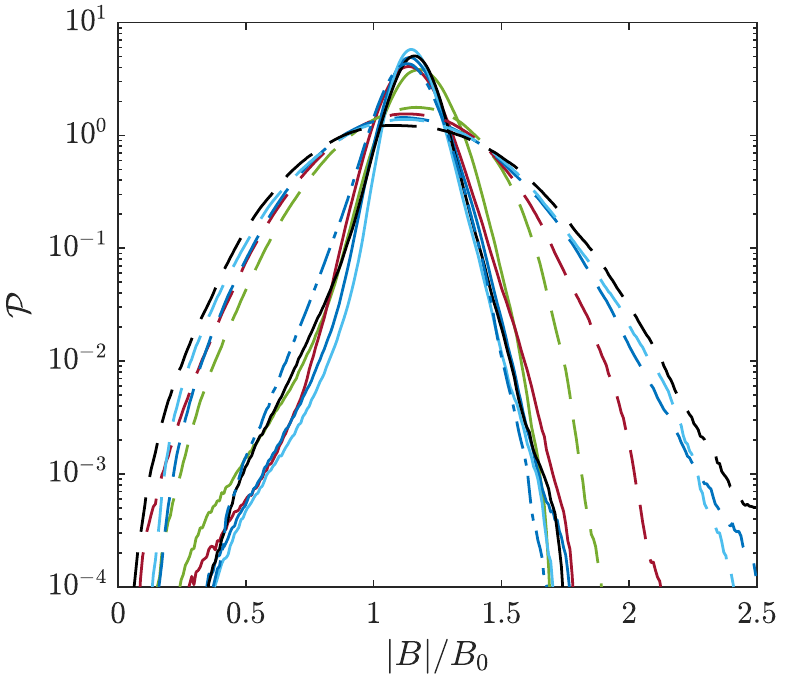}
\caption{Same as in~\cref{fig: Dp PDFs brag} but for the \emph{lrCL} series of collisionless simulations with varying interruption number (see~\cref{sub:study design}). For very small interruption numbers (e.g., in the simulation with $\beta_{0}=100$, $\nu_{\rm c}=0$,  $\It\simeq0.04$), the proportion of the box at the mirror/firehose thresholds is larger (${\simeq}5\%$) because
it is naturally driven to much larger values by the dynamics (cf.~the passive simulation shown with the black dashed line). We also include a simulation with $\Lambda_{\rm FH}=1.4$ for comparison, 
to explore the possible effect of  kinetic oblique-firehose modes limiting $\Dp$ before the fluid firehose threshold (see~\cref{subsub: microinst computational}).}
\label{fig: Dp PDFs cl}
\end{center}
\end{figure}

A central result -- that changes to the pressure anisotropy are suppressed by its feedback on the flow (magneto-immutability) -- 
is illustrated in~\crefrange{fig: Brazil plot}{fig: Dp PDFs cl}.  \Cref{fig: Brazil plot} shows the joint probability distribution function (PDF)  of  
$\beta$ and $p_{\perp}/p_{\|}$ (the `Brazil plot') for all simulations from the `\emph{lrCL}' and `\emph{lrB}' sets (18 in all; see~\cref{tab:sims}). We compare 
the set of  CGL-LF simulations (left-hand panel) to the identical passive-$\Dp$ simulations where the pressure-anisotropy feedback has been 
suppressed (right-hand panel). This diagnostic, although only  qualitative (the total probability is normalised separately at each $\beta$ value for illustrative purposes, but done so identically for both simulation sets), 
  clearly demonstrating the  difference between the two. In particular,
we see that the force  on the flow from the pressure anisotropy  causes most of the plasma to  remain well within the microinstability limits. In contrast, the passive case, without  $\Dp$ feedback, has most of the volume stuck at the  microinstability limits (the edges of the PDFs at high and low $p_{\perp}/p_{\|}$) because 
the turbulent fluctuations continuously push the plasma to positive and negative $\Dp$.  Similar results are demonstrated with a 
snapshot of the \emph{CL10} simulation in~\cref{fig: visualization}, where we again compare the active- and passive-$\Dp$ simulations with otherwise identical 
parameters. While the perpendicular flows are of similar magnitudes, indicating similar turbulent fluctuation levels (top panels), 
the pressure-anisotropy variation (middle panels) is much smaller in the presence of pressure-anisotropy feedback. As a consequence, the variation in $B=|\bm{B}|$
is also suppressed: both $\Dp$ and $B$ are driven by $\bbgu$, so suppressing one suppresses the other (more fundamentally, $\Dp$ is driven 
by changes in $B$ and $\rho$).

%%%%%%%%%%%%%%%%%%%%%%%%%%%%%%%%%%
%%%%%%%%%%%%%%%%%%%%%%%%%%%%%%%%%%
%%%%%%%%%%%%%%%%%%%%%%%%%%%%%%%%%%
%%%%%%%%%%%%%%%%%%%%%%%%%%%%%%%%%%

\Cref{fig: Dp PDFs brag,fig: Dp PDFs cl} demonstrate the same ideas quantitatively,  showing PDFs of $4\pi\Dp/B^{2}$ and $B$ from the
\emph{lrB} simulation set in~\cref{fig: Dp PDFs brag} (each with similar $\It\approx 0.4$) and from the collisionless \emph{lrCL} simulation set in~\cref{fig: Dp PDFs cl} (covering $\It\approx 0.04$, at $\beta_{0}=100$, to $\It>1$, at $\beta_{0}=1$). For each 
of the pressure-anisotropy PDFs (left-hand panels), we indicate the volume fraction of the box that sits at the mirror or firehose thresholds as a percentage (label
colours match those of the curves, and the passive-$\Dp$ cases are the larger numbers listed in the upper part of each panel). The PDFs have  qualitatively 
different  shapes between the active- and passive-$\Dp$ simulations, with pressure-anisotropy feedback causing $\Dp/B^{2}$ to peak near zero in the active-$\Dp$ runs, as opposed to being
nearly flat through the stable regions in the passive runs. As a consequence, the fraction of the plasma that sits at the microinstability thresholds 
is strongly reduced due to the pressure-anisotropy feedback on the flow. This fraction does increase modestly with decreasing 
interruption number  (increasing $\beta_{0}$ in~\cref{fig: Dp PDFs cl}), but remains
very small given that $\Dp/B^{2}$ is effectively forced across a wider range with decreasing $\It$. 
We do not see any significant change with $\nu_{\rm c}$
across the \emph{lrB} simulations  (\cref{fig: Dp PDFs brag}),  all of which have $\It\approx0.4$, showing that the collisionality regime (collisionless, weakly collisional, or Braginskii MHD) is of subsidiary importance. 
One exception to this is the mean negative $\Dp$ in the collisionless cases, which
becomes modestly more negative with increasing $\beta$ in collisionless simulations (\cref{fig: Dp PDFs cl}), but is not seen in  the weakly collisional or Braginskii regimes because the bulk $\nu_{\rm c}$ \revchng{also depletes any mean $\Dp$ that might otherwise develop}. As discussed in \aksqs, this feature is likely 
related to  
turbulent heating occurring predominantly in the parallel direction via Landau damping (as approximated by the Landau-fluid closure).
In~\cref{fig: Dp PDFs cl}, we also test the effect of a reduced firehose limit $\Lambda_{\rm FH}=1.4$, which is likely a better approximation  for scattering from kinetic oblique firehose modes at very large scale separations $L_{\perp}/\rho_{i}$  (see~\cref{subsub: microinst computational}; \citealp{Bott2021}). This increases
the proportion of the domain at the limiter thresholds (as expected, since the range between them is smaller) and decreases the mean $|\Dp|$. 
The latter effect  suggests that  the asymmetry of the microinstability limits (i.e., the fact that $\Lambda_{\rm FH}>\Lambda_{\rm M}$) 
also contributes to the mean pressure anisotropy, but modest changes to $\Lambda_{\rm FH}$ do not seem to lead to any other qualitatively important 
changes, so we will not consider this further.

%%%%%%%%%%%%%%%%%%%%%%%%%%%%%%%%%%
\begin{figure}
\begin{center}
\includegraphics[width=0.5\columnwidth]{\ffold/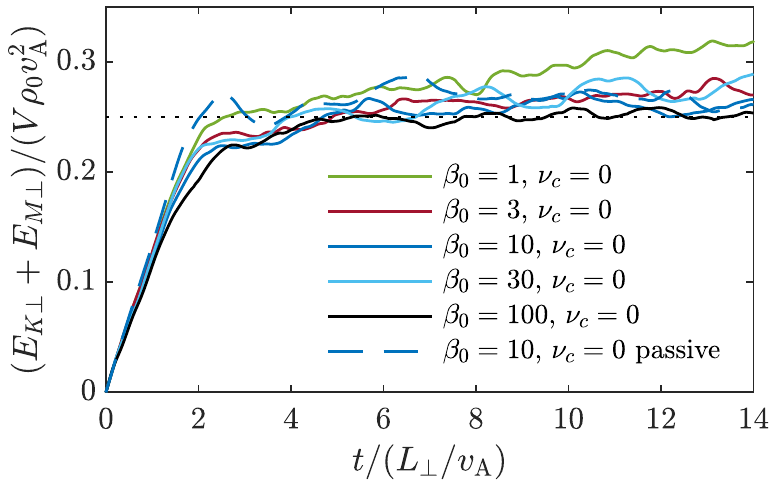}\includegraphics[width=0.5\columnwidth]{\ffold/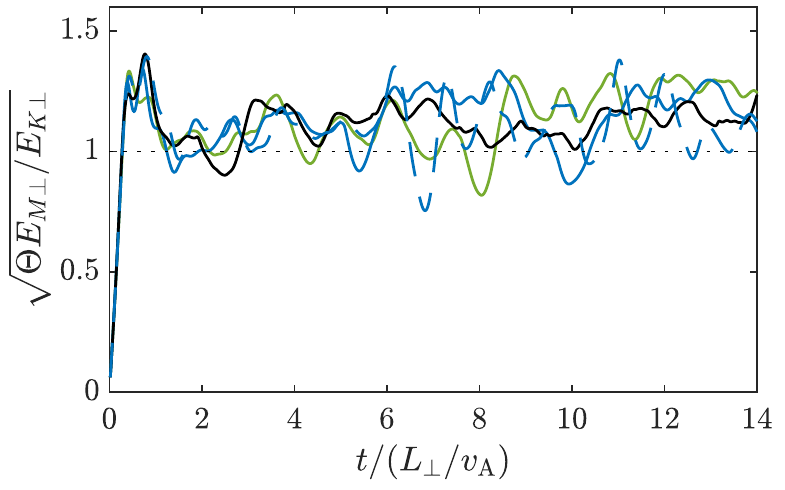}
%\psfragfig[width=0.5\columnwidth]{\ffold/time-en}\psfragfig[width=0.5\columnwidth]{\ffold/time-er}
\caption{Time evolution of the total perpendicular energy (left panel) and the Alfv\'en ratio $r_{\rm A}=\sqrt{\fh E_{M\perp }/E_{K\perp }}$ (right panel) for the \emph{lrCL} set of collisionless simulations
with varying interruption number. Note that $r_{\rm A}$ defined in this way (with $\fh$ computed as a volume average) accounts for the effect of the mean pressure 
anisotropy on Alfv\'enic fluctuations (see \cref{eq: dp alfven speed}). The dotted lines in each panel show the `expected' values assuming $\delta u_\perp/\va=\delta B_\perp/B_0=1/2$: $E_{K\perp} + E_{M\perp}=V\rho_0 \va^2/4$ and $r_{\rm A}=1$. We see that the bulk properties of collisionless turbulence are rather similar to standard MHD, even for $\It\ll1$. }
\label{fig: time evolution}
\end{center}
\end{figure}
%%%%%%%%%%%%%%%%%%%%%%%%%%%%%%%%%%

%
%
%
\subsubsection{Turbulent energy}

While the results shown in~\crefrange{fig: Brazil plot}{fig: Dp PDFs cl} clearly demonstrate the significant impact of
the pressure-anisotropy feedback on the statistics of $B$ and $\Dp$, it is important
to confirm that this does not occur purely  as a result of the viscous damping of those motions that would  otherwise drive significant $\Dp$.
Because all simulations are driven with the same energy-injection rate~$\varepsilon$, the 
simplest diagnostic of this is the  turbulent fluctuation energy,  which is shown in~\cref{fig: time evolution} for the \emph{lrCL} runs (we plot $E_{K\perp} + E_{M\perp} \doteq \int \rmd \bm{x}\, (\rho |\bm{u}_\perp|^2 + |\bm{B}_\perp|^2/4\pi)/2$, 
which includes only the fluctuating $y$ and $z$ components of $\bm{u}$ and $\bm{B}$). A 
fluctuation energy that decreased with decreasing $\It$ would imply a  cascade efficiency that decreased with $\It$, as would be na\"{i}vely expected if fluctuations were increasingly strongly damped above the interruption limit (see~\cref{sub: magneto-immutability}). 
Instead, we see in the left panel of~\cref{fig: time evolution}  that such a decrease is quite small (compare, e.g.,  the active and passive cases at 
$\beta_{0}=10$), with a similar steady-state energy reached  even at  $\beta_{0}=100$ (the lowest $\It$ that we explore). 
Thus, we see that there is only modest viscous damping of pressure-anisotropy-generating fluctuations, a feature that we explore in greater depth below. Note that, in contrast, a  single linearly polarised shear-Alfv\'en wave with similar amplitude would be strongly damped under these conditions \revchng{because its fluctuations are confined to the plane set by the initial conditions and so it cannot rearrange itself to
avoid generating large pressure anisotropies} (\citealp{Squire2016}; \skqs).

The right panel of~\cref{fig: time evolution} shows the Alfv\'en ratio, $r_{\rm A} = \sqrt{\fh E_{M\perp} /E_{K\perp} }$, for some of the simulations. This should be approximately unity for Alfv\'enic turbulence. The time-dependent box-averaged mean anisotropy parameter $\fh$ is included in $r_{\rm A}$ because an Alfv\'en wave satisfies $\delta B_{\perp}=\delta u_{\perp}\sqrt{4\upi\rho/\fh}$
in the presence of a mean pressure anisotropy $\Dp$ (see~\cref{subsub: lf waves}; we set $\fh=1$ for the passive simulation). \Cref{fig: time evolution} shows that, after accounting for the 
change in the Alfv\'en ratio due to variation of the mean $\Dp$ with $\It$ (see~\cref{fig: Dp PDFs cl}), the bulk properties of the turbulence are rather similar to MHD, with just a slight excess 
of magnetic energy (also observed in MHD; \citealp[see, e.g.,][]{Mueller2005,Chen2013a}).

\subsection{Alfv\'enic turbulence structure}\label{sub: results 2}

We now diagnose the structure of the Alfv\'enic fluctuations in more detail, with the goal of understanding the similarities and differences between turbulence in high-$\beta$, weakly 
collisional plasmas and turbulence in standard MHD \citep{Schekochihin2020}. 
We will see that the basic statistics of the flow and magnetic field are surprisingly similar to MHD. More detailed diagnostics, which focus on compressive fields 
and components of the rate-of-strain tensor (see~\cref{sub: flow rearrangement results}), are needed to reveal
the changes to the turbulent flow structure that enable the suppression of $\Dp$ discussed in \cref{sub: results 1}.

\subsubsection{Turbulent energy spectra at $\It\lesssim1$}

%%%%%%%%%%%%%%%%%%%%%%%%%%%%%%%%%%
\begin{figure}
\begin{center}
\includegraphics[width=0.7\columnwidth]{\ffold/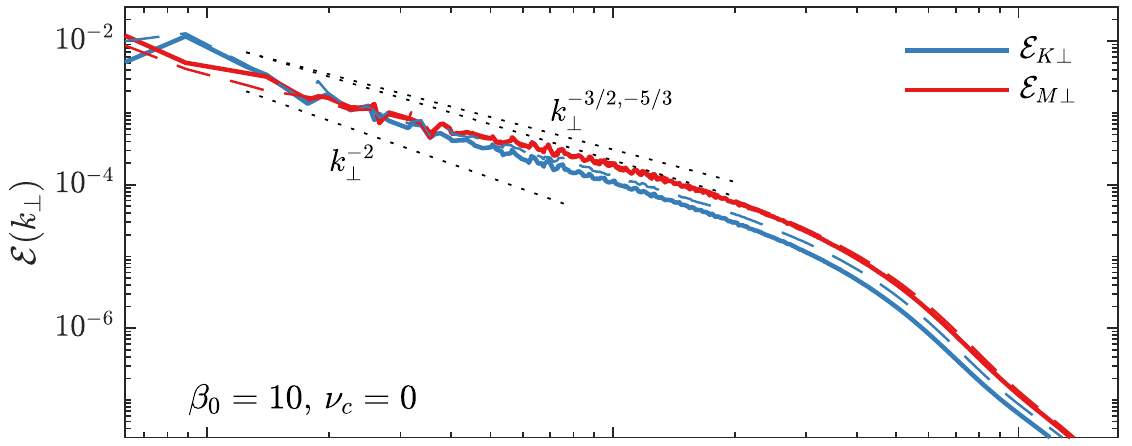}\\
\includegraphics[width=0.7\columnwidth]{\ffold/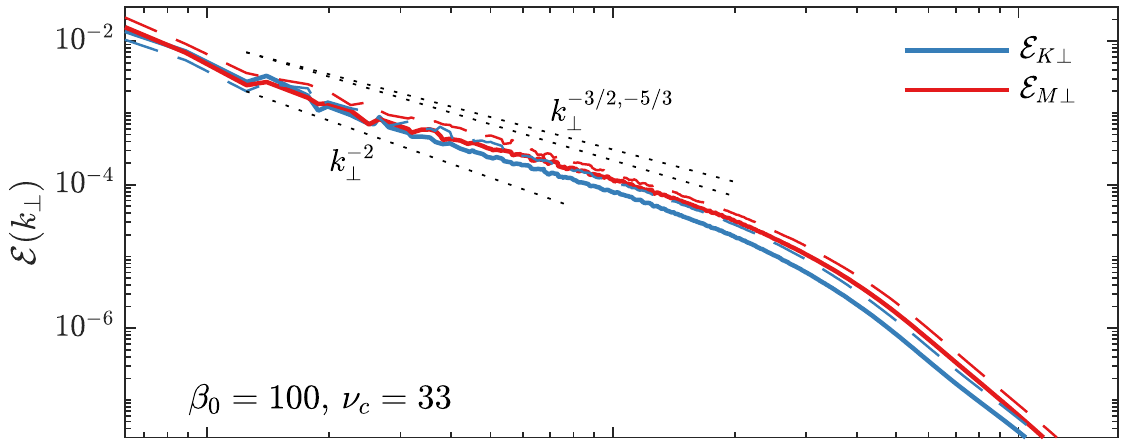}\\
\includegraphics[width=0.7\columnwidth]{\ffold/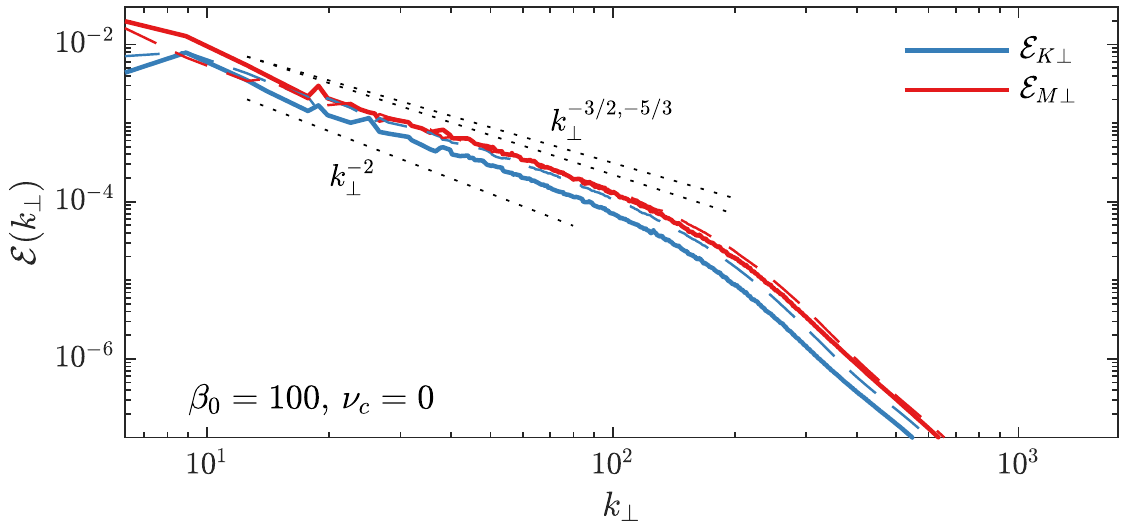}
\caption{Kinetic (blue) and magnetic (red) energy spectra for simulations \emph{CL10}, \emph{B100}, and \emph{CL100}, as labelled. Solid lines show CGL-LF (active-$\Dp$) simulations, and dashed lines show isothermal MHD (passive-$\Dp$) simulations. The dotted black lines indicate slopes of $k_{\perp}^{-3/2}$, $k_{\perp}^{-5/3}$, and $k_{\perp}^{-2}$ for comparison.  We see spectra broadly consistent with a  $k_{\perp}^{-3/2}$ scaling, although  velocity spectra are  steepened modestly by the effect of pressure anisotropy in the active-$\Dp$ runs
(however, they are clearly flatter than the ${\propto}k_{\perp}^{-2}$ spectrum observed in the hybrid-kinetic simulations of \aksqs). This demonstrates that vigorous turbulence, broadly 
resembling MHD, is maintained even when the feedback from the pressure anisotropy is strong ($\It<1$ in all cases). Note that the cutoff of the spectra at relatively larger scales seen in \emph{CL100} is simply a result of that run's lower numerical resolution.}
\label{fig: basic spectra}
\end{center}
\end{figure}
%%%%%%%%%%%%%%%%%%%%%%%%%%%%%%%%%%
%%%%%%%%%%%%%%%%%%%%%%%%%%%%%%%%%%
\begin{figure}
\begin{center}
\includegraphics[height=0.3\columnwidth]{\ffold/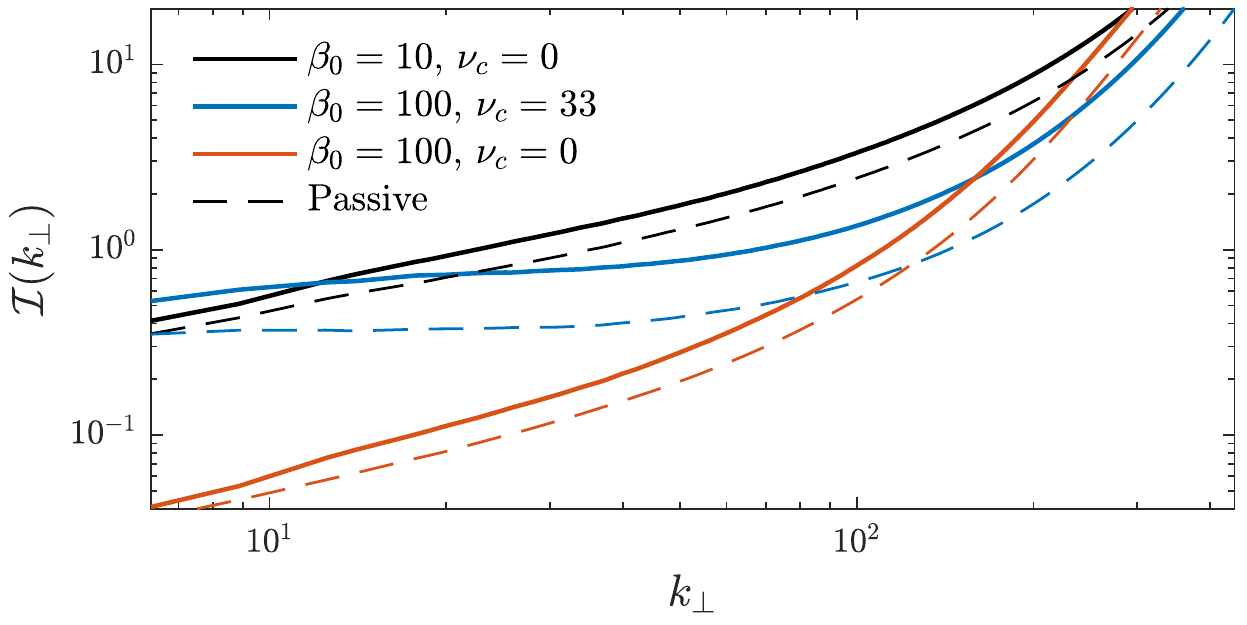}
\caption{`Local' interruption number defined by~\cref{eq: local interruption number} (see also~\cref{subsub: interruption number}). We show the three active-$\Dp$ simulations whose spectra are given in~\cref{fig: basic spectra}, comparing to the passive (MHD) runs with dashed lines. In the \emph{B100} simulation ($\beta_{0}=100$, $\nu_{\rm c}=33\va/L_{\perp}$), $\omegaa$ depends on scale as per critical balance, $\omegaa=k \sqrt{\delta u_{\perp}\delta B_{\perp}/\rho_{0}^{1/2}}$, which causes a slower increase in $\It$ with scale. 
Clearly, $\It(k_{\perp})\lesssim 1$ across a reasonable range in each simulation, indicating that much larger pressure anisotropies would be created without
dynamical feedback from $\Dp$ (as is the case for the passive simulations).
}
\label{fig: It}
\end{center}
\end{figure}
%%%%%%%%%%%%%%%%%%%%%%%%%%%%%%%%%%

\Cref{fig: basic spectra} shows the kinetic (blue lines) and magnetic (red lines) spectra  obtained in the \emph{CL10}, \emph{B100}, and \emph{CL100} simulations, 
again comparing active-$\Dp$ simulations (solid lines) to the passive-$\Dp$ simulations (dashed lines; note that this case is just isothermal MHD for velocity and magnetic 
fluctuations). 
The purpose of this comparison is to exhibit the extremely similar spectra in each case, demonstrating that 
vigorous turbulence survives even when the expected damping from pressure anisotropy  is strong across a wide range of scales starting at the outer scale (as
is the case for all three simulations shown here since $\It<1$). A careful examination reveals minor differences caused by 
the pressure-anisotropy feedback, in particular 
a slight steepening of the kinetic-energy spectrum compared to MHD. 
\revchng{In all cases the magnetic spectra exhibit a scaling that is flatter than ${\sim}k_{\perp}^{-5/3}$  and closer to ${\sim}k_{\perp}^{-3/2}$  (shown with the dashed lines), although
the exact slopes vary somewhat across the inertial range. We shall study the eddies' 3-D structure and dynamic alignment  in \cref{sec: alignment}.}

Similar information in a different form is shown in~\cref{fig: It}. We integrate the spectra to obtain the scale-dependent amplitudes
\begin{equation}
\delta B_{\perp}(k_{\perp})^{2} = \frac{2}{V} \int_{k_{\perp}}^{\infty} \rmd k_{\perp}'\,  \mathcal{E}_{M\perp}(k_{\perp}'),\quad \delta u_{\perp}(k_{\perp})^{2} = \frac{2}{V{\rho}_{0}} \int_{k_{\perp}}^{\infty} \rmd k_{\perp}'\,  \mathcal{E}_{K\perp}(k_{\perp}'),
\end{equation}
and then use these to compute the scale-dependent interruption number  (see~\cref{eq:interruption number}),
\begin{equation}
\It(k_{\perp}) = \left( \beta \frac{\delta B_{\perp}}{B_{0}}\frac{\delta u_{\perp}}{\va}\right)^{-1} \max\left(1, \frac{\nu_{\rm c}}{\omegaa}\right)\label{eq: local interruption number}
\end{equation}
(we ignore the effect of the mean $\fh$ on $\va$ and $\omegaa$, as it does not make a significant difference for these qualitative estimates). Here $\omegaa$ should be considered a function of scale as per critical balance, so we use $\omegaa(k_{\perp}) = k_{\perp} \sqrt{\delta u_{\perp}\delta B_{\perp}/{\rho}_{0}^{1/2}}$.
We see that, while the estimated local interruption number is slightly larger for the CGL-LF simulations, because 
the turbulence amplitudes are modestly smaller, $\It(k_{\perp})$ nonetheless remains ${\lesssim}1$ throughout
a wide portion of the turbulent cascade in each case. Since the  damping rate of a linearly polarised Alfv\'enic perturbation 
is comparable to its frequency for for $\It\lesssim 1$ (i.e., \revchng{for amplitude $\delta B_{\perp}/B_{0}\gtrsim \beta^{-1/2} \max[1,(\nu_{c}/\omegaa)^{1/2}]$), but
the turbulent fluctuations exceed this amplitude throughout most scales of the simulations,} the implication is that the fluctuations must be avoiding 
those motions that would  cause strong damping in order for the cascade to proceed as observed. Also of note is the
difference between the collisionless and weakly collisional regimes, as shown by the effectively flat $\It(k_{\perp})$ for the \emph{B100} simulation up to $k_{\perp}\approx 100$.
This occurs because when $\omegaa\ll\nu_{\rm c}$ (the weakly collisional and Braginskii MHD regimes; see~\cref{sub: different regimes}), the increasing frequency of the motions towards smaller scales  cancels 
their decreasing amplitudes, conspiring  to keep $\omegaa\delta u_{\perp} \delta B_{\perp}$ approximately constant through the inertial range (see~\cref{subsub: interruption number}).
This breaks down once $\omegaa\gtrsim \nu_{\rm c}$, which happens here for $k_{\perp}L_{\perp}\gtrsim 100$, around the expected scale based on the
parameters (see~\cref{tab:sims}). Below this, $\It$ increases at smaller scales since the turbulent eddies are fast enough to be effectively collisionless.

%%%%%%%%%%%%%%%%%%%%%%%%%%%%%%%%%%
\begin{figure}
\begin{center}
\includegraphics[height=0.37\columnwidth]{\ffold/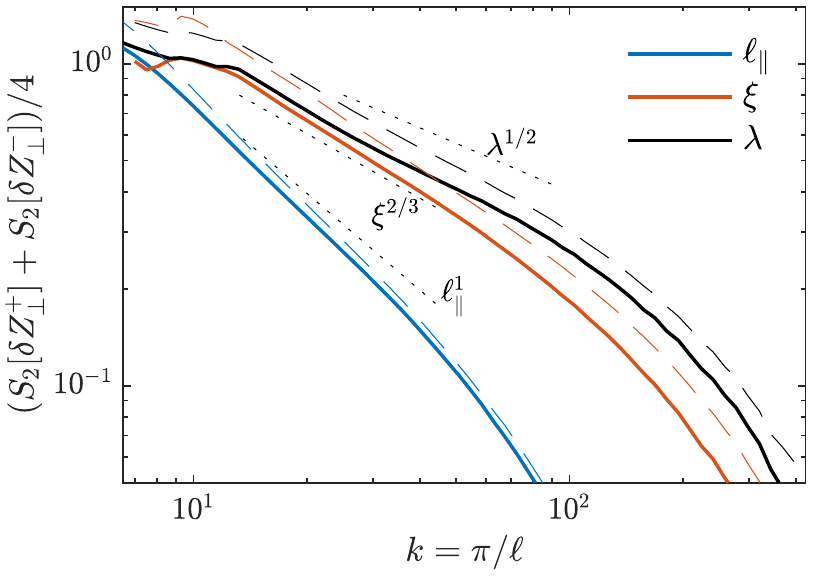}~\includegraphics[height=0.37\columnwidth]{\ffold/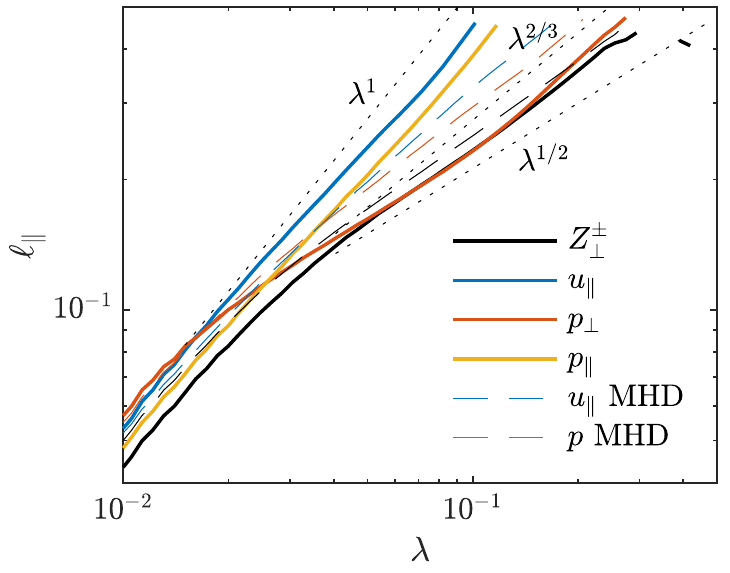}
\caption{Structure functions in the $\beta_{0}=10$, $\nu_{\rm c}=0$ simulations. \revchng{The left panel shows (3-point) second-order structure functions of $\bm{Z}_{\perp}^{\pm}$ (see~\cref{subsub: structure functions}) with increments taken along the field (blue; $\ell_{\|}$), parallel to the local fluctuation in the opposite Els\"asser variable $\bm{Z}_{\perp}^{\mp}$  ($\xi$; red), and  perpendicular to the local fluctuation in $\bm{Z}_{\perp}^{\mp}$ ($\lambda$; black). Solid and dashed lines show the active-$\Dp$ and  passive-$\Dp$ cases, respectively, with all length scales normalised to $L_\perp$. The scalings are very similar in both active and passive runs, and are broadly consistent with the expected $\lambda^{1/2}$, $\xi^{2/3}$, and $\ell_{\|}^{1}$ scalings of dynamic alignment} (shown with dotted black lines).  The right-hand panel shows the measured $\ell_{\|}(\ell_{\perp})$ for the same simulation, which is computed from the second-order structure functions of a variety of different quantities as labeled (this provides a measure of the average shape of an eddy for different variables). While $\bm{Z}_{\perp}^{\pm}$ and $p_{\perp}$ show the expected scale-dependent anisotropy $\ell_{\|}\sim \ell_{\perp}^{1/2}$ of a dynamically aligned cascade, $p_{\|}$ and $u_{\|}$ have nearly constant anisotropy $\ell_{\|}\sim 5\ell_{\perp}$ throughout the inertial range. This differs from MHD (dashed lines), where $u_{\|}$ and $p$ seem to follow $\ell_{\|}\sim \ell_{\perp}^{2/3}$, as expected for non-aligned eddies in a critically balanced cascade.}
\label{fig: alignment}
\end{center}
\end{figure}
%%%%%%%%%%%%%%%%%%%%%%%%%%%%%%%%%%

\subsubsection{Three-dimensional anisotropy and alignment}\label{sec: alignment}

Many previous works have studied the 3-D statistical structure of eddies in MHD turbulence, 
which has important implications for the cascade rate and intermittency \citep[see][and references therein]{Schekochihin2020}. Starting with \citet{Boldyrev2006}, these have argued that eddies become increasingly `dynamically aligned' towards smaller scales, evolving into elongated sheet-like structures satisfying $\ell_\|\gg\xi \gg \lambda$ (where $\ell_\|$, $\xi$, and $\lambda$ are the eddy's correlation lengths in the field-parallel direction, in the direction of the turbulent fluctuation, and in the mutually perpendicular direction, respectively). Motivated by its utility as a sensitive diagnostic of
the Alfv\'enic fluctuations' structure, in
\Cref{fig: alignment} we illustrate the 3-D anisotropy of the  \emph{CL10} simulations, computed using three-point  structure functions (\cref{subsub: structure functions}).
In the left panel, we show the directionally conditioned structure functions of $\bm{Z}^{\pm}_{\perp} = \bm{Z}^{\pm} - \hat{\bm{b}}(\hat{\bm{b}}\bcdot\bm{Z}^{\pm})$. Here  $\bm{Z}^{\pm} = \bm{u}\pm \bm{\va}$ and  $\bm{v}_{\rm A}$ is the local Alfv\'en speed $\bm{v}_{\rm A,eff}=\bm{B}\fh^{1/2}(4\upi\rho)^{-1/2}$, with $\fh$ and $\rho$  evaluated using their local three-point average ($\fh=1$ for the passive simulation). 
The colours show different directions of the separation vector $\bm{\ell}=(\lambda,\xi,\ell_{\|})$, with $\ell_{\|}$ applying to separations within $15^{\circ}$ of the local $\hat{\bm{b}}$, $\xi$ to separations that are perpendicular ($>45^{\circ}$) to $\hat{\bm{b}}$ and are within $15^{\circ}$ \revchng{of the direction of the local increment of $\bm{Z}^{\mp}_{\perp}$},\footnote{The choice of $15^{\circ}$ to define a `parallel' fluctuation
is arbitrary. Any value below ${\simeq}20^\circ$ gives similar results \citep{Chen2012}, though  
smaller values give fewer point-separation pairs and thus noisier results.}
and $\lambda $ to separations that are perpendicular to both $\hat{\bm{b}}$ and  \revchng{the $\bm{Z}^{\mp}_{\perp}$ increment} (note that 
we use $\bm{Z}^{\mp}$ to define the directions of $\bm{Z}^{\pm}$ because nonlinear interactions are controlled by the opposite Els\"asser variable). We see results 
that are quite similar to standard MHD turbulence (dashed lines), with $S_{2}$ flatter in the $\lambda$ direction than in the $\xi$ direction, and
broadly consistent with  (though modestly steeper than)  the dynamically-aligned-MHD-turbulence scalings $S_{2}\sim \lambda^{1/2}$, $S_{2}\sim \xi^{2/3}$, $S_{2}\sim \ell_{\|}^{1}$
indicated by the dotted lines \citep{Boldyrev2006,Mallet2015}. The right panel compares the parallel-perpendicular anisotropy of eddies measured from different quantities in the turbulence. 
As seen in the left panel, the Alfv\'enic eddies are relatively similar to isothermal MHD, with $\ell_{\|}\sim \ell_{\perp}^{1/2}$ as expected for an aligned cascade. 
However, while in MHD the compressive quantities $u_{\|}$ and $p = T_{i}\rho$ seem to scale as $\ell_{\|}\sim \ell_{\perp}^{2/3}$, suggesting  such fluctuations  are unaligned and passively mixed by the critically balanced Alfvénic cascade \citep{Lithwick2001,Schekochihin2009,Chen2012}, the compressive fields in CGL-LF turbulence scale quite differently, with  $p_{\|}$ and $u_{\|}$ having nearly constant anisotropy $\ell_{\|}\simeq5\ell_{\perp}$ throughout the entire box. This provides the first hint of the differences caused by magneto-immutability, which will be discussed in more detail in the next section (\cref{sub: flow rearrangement results}).

\subsection{The rearrangement of flows and fields in magneto-immutable turbulence}\label{sub: flow rearrangement results}

%%%%%%%%%%%%%%%%%%%%%%%%%%%%%%%%%%
\begin{figure}
\begin{center}
\includegraphics[width=0.7\columnwidth]{\ffold/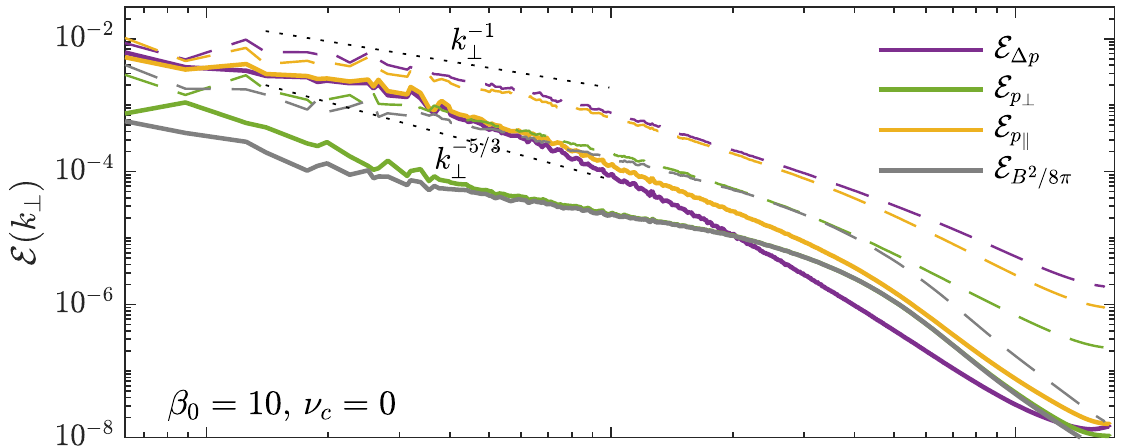}\\\includegraphics[width=0.7\columnwidth]{\ffold/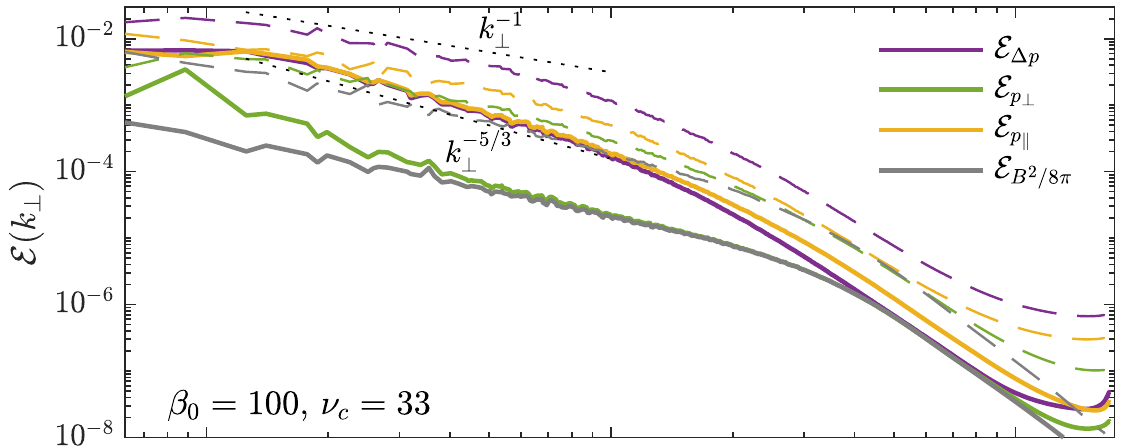}\\\includegraphics[width=0.7\columnwidth]{\ffold/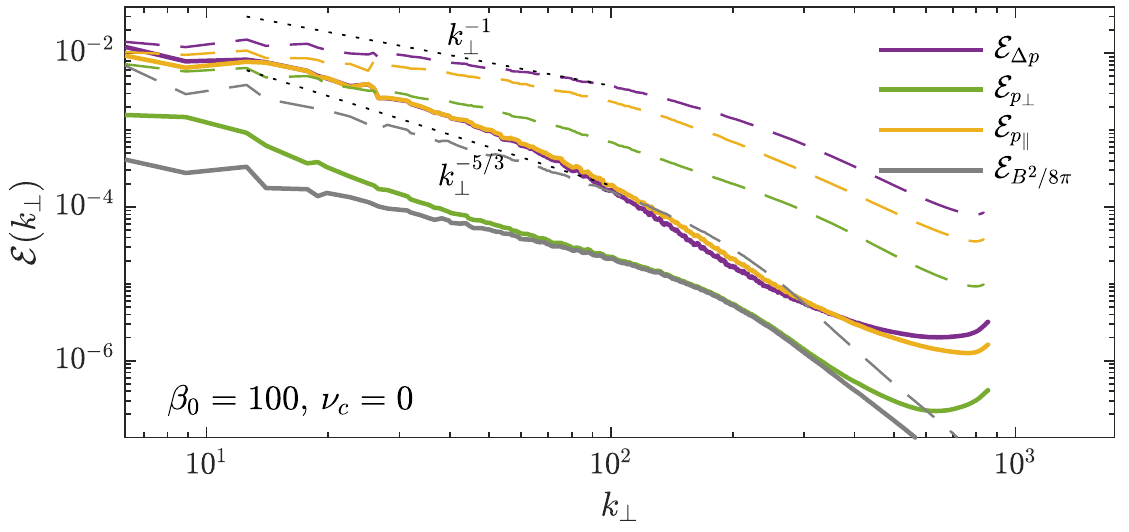}
\caption{Spectra of $\Dp$ (purple lines), $p_{\perp}$ (green lines),  $p_{\|}$ (yellow lines), and the magnetic pressure $B^{2}/8\upi$ (grey lines) for the same simulations as in~\cref{fig: basic spectra}, and again with the passive-$\Dp$ simulations shown with dashed lines. The dotted lines indicate the spectral scalings $\mathcal{E}\propto k_{\perp}^{-1}$ and $\mathcal{E}\propto k_{\perp}^{-5/3}$ (see~\cref{subsub: compressive theory} for discussion). We 
see a significant difference between active- and passive-$\Dp$ cases, with $\mathcal{E}_{\Dp}\sim \mathcal{E}_{p_{\|}}\gg \mathcal{E}_{p_{\perp}}$ observed most clearly  in the active-$\Dp$ simulations (and, to a lesser degree, collisionless cases).  The active-$\Dp$ spectra bear  encouraging resemblance to those seen in the hybrid-kinetic simulations of \aksqs. }
\label{fig: dp spectra}
\end{center}
\end{figure}
%%%%%%%%%%%%%%%%%%%%%%%%%%%%%%%%%%

\Crefrange{fig: basic spectra}{fig: alignment} show that the CGL-LF system can sustain random, turbulent-like motions across a wide range of 
scales, even when those motions might be expected to be strongly  damped by the effects of pressure anisotropy.
The properties of the kinetic and magnetic fluctuations appear  broadly similar to standard MHD, including relatively detailed measures such as the 
scale-dependent anisotropy and eddy-alignment  intermittency. 
In this section, we focus on
the differences compared to MHD, understanding how the turbulent cascade rearranges itself due to the pressure-anisotropy stress.
We first present the numerical results, then speculate on theoretical explanations in~\cref{subsub: compressive theory}.

\subsubsection{Pressure statistics}

Let us first examine the spectra of pressure fluctuations, which are shown in~\cref{fig: dp spectra}, again comparing active-
and passive-$\Dp$  for the \emph{CL10}, \emph{B100}, and \emph{CL100} simulations. Here, unlike for the kinetic- and magnetic-energy spectra, we see
substantial differences due to the feedback of $\Dp$ on the flow. The most obvious feature is  the significant reduction and steepening of 
the pressure spectra in all active-$\Dp$ runs. While the passive-$\Dp$ simulations have quite flat spectra -- approximately ${\propto}k_{\perp}^{-1}$ in \emph{B100} and yet flatter in the collisionless simulations --
all three active cases exhibit pressure spectra ${\propto}k_{\perp}^{-5/3}$ across a reasonable range near the outer scales. Notably, the latter is in agreement with the hybrid-kinetic 
simulations of \aksqs, providing evidence that similar physics operates in their more realistic simulations.
 The steepening of pressure spectra due to the pressure-anisotropy feedback is clear evidence of
the modified flow structure in the active-$\Dp$ cases, despite their similar velocity spectra.

Another interesting feature of the active-$\Dp$ pressure spectra is that $\mathcal{E}_{\Dp}\sim \mathcal{E}_{p_{\|}}\gg\mathcal{E}_{p_{\perp}}$, while in the passive runs the difference between $\mathcal{E}_{p_{\|}}$ and $\mathcal{E}_{p_{\perp}}$ is far less pronounced (particularly for  \emph{B100}, where $\mathcal{E}_{p_{\|}}\sim\mathcal{E}_{p_{\perp}}$, as might be na\"{i}vely expected).\footnote{In the passive-$\Dp$ collisionless simulations $\delta p_{\|}\approx 2\delta p_{\perp}$ ($\mathcal{E}_{p_{\|}}\approx 4 \mathcal{E}_{p_{\perp}}$). This ratio can be explained simply by the fact that $p_{\|}$ is driven by $-2\bbgu$, while  $p_{\perp}$ is driven by $\bbgu$ (see~\crefrange{eq:KMHD pp}{eq:KMHD pl}).}
This feature was also clearly observed in the hybrid-kinetic simulations of \aksqs.
As discussed in more detail in \cref{subsub: compressive theory} and in \aksqs, it likely results from perpendicular pressure balance between $\grad p_{\perp}$ and the magnetic pressure $\grad B^{2}/8\upi$, while $ p_{\|}$ is driven more strongly by $\bbgu$, thus dominating $\Dp$. \revchng{The basic feature seems generic to anisotropic  
collisionless (CGL) dynamics, occurring because $\delta p_{\perp}$ remains highly constrained by perpendicular pressure balance, while $\delta p_{\|}$ is not, affecting the plasma only through the parallel force (see \cref{appsub: dispersion relation} for a simple linear calculation to illustrate the effect).}
We provide evidence for the scenario in~\cref{fig: dp spectra} by plotting the spectra of the magnetic pressure $\mathcal{E}_{B^{2}/8\upi}$ (grey lines), which matches $\mathcal{E}_{p_{\perp}}$ very well below the forcing scales. 
In the \emph{CL10} simulation, where the feedback of the pressure anisotropy is only important at 
larger scales because $\It(k_{\perp})$ increases to ${\gg}1$ at small scales (see~\cref{fig: It}), the  feature reverses to $\mathcal{E}_{p_{\|}}\simeq \mathcal{E}_{p_{\perp}}> \mathcal{E}_{\Dp}$ for $k_{\perp}\gtrsim 200$ (there is no similar reversal for the passive-$\Dp$ simulation). This is
presumably the signature pressure-anisotropy feedback becoming subdominant (i.e., magneto-immutability  no longer being important) once
the driving of $\Dp$ via $\bbgu$ becomes subdominant to standard MHD effects (the perpendicular pressure balance). We see hints that similar behaviour would occur in \emph{B100} and \emph{CL100}, 
 but at smaller scales than in \emph{CL10} (as expected), which unfortunately are unresolved in these simulations.
Overall, the general similarity of
the active-$\Dp$ spectra, but not the passive-$\Dp$ spectra, to the large-scale hybrid-kinetic results of \aksqs\ is encouraging for the
applicability of the CGL-LF model.

%%%%%%%%%%%%%%%%%%%%%%%%%%%%%%%%%%
\begin{figure}
\begin{center}
\includegraphics[height=0.43\columnwidth]{\ffold/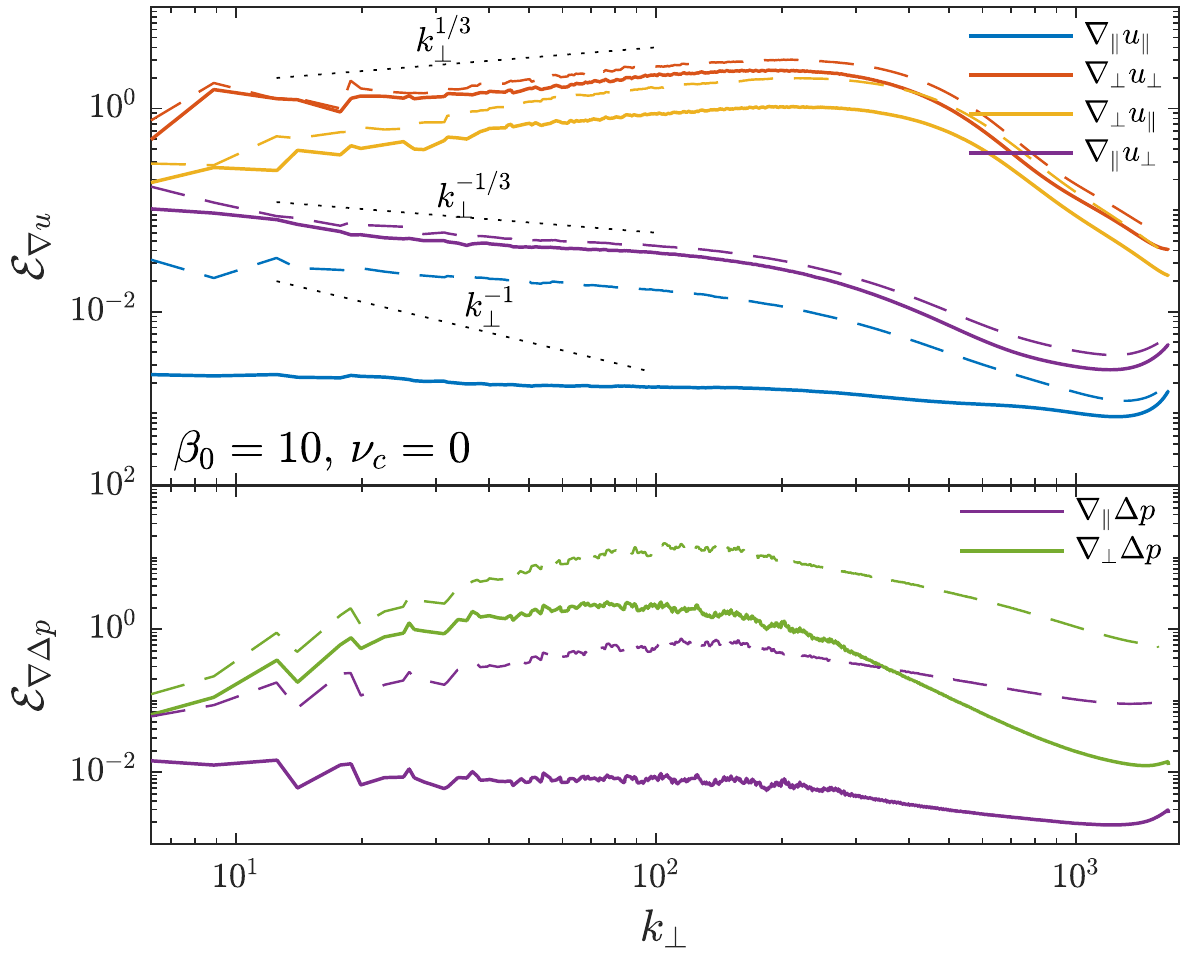}\includegraphics[height=0.43\columnwidth]{\ffold/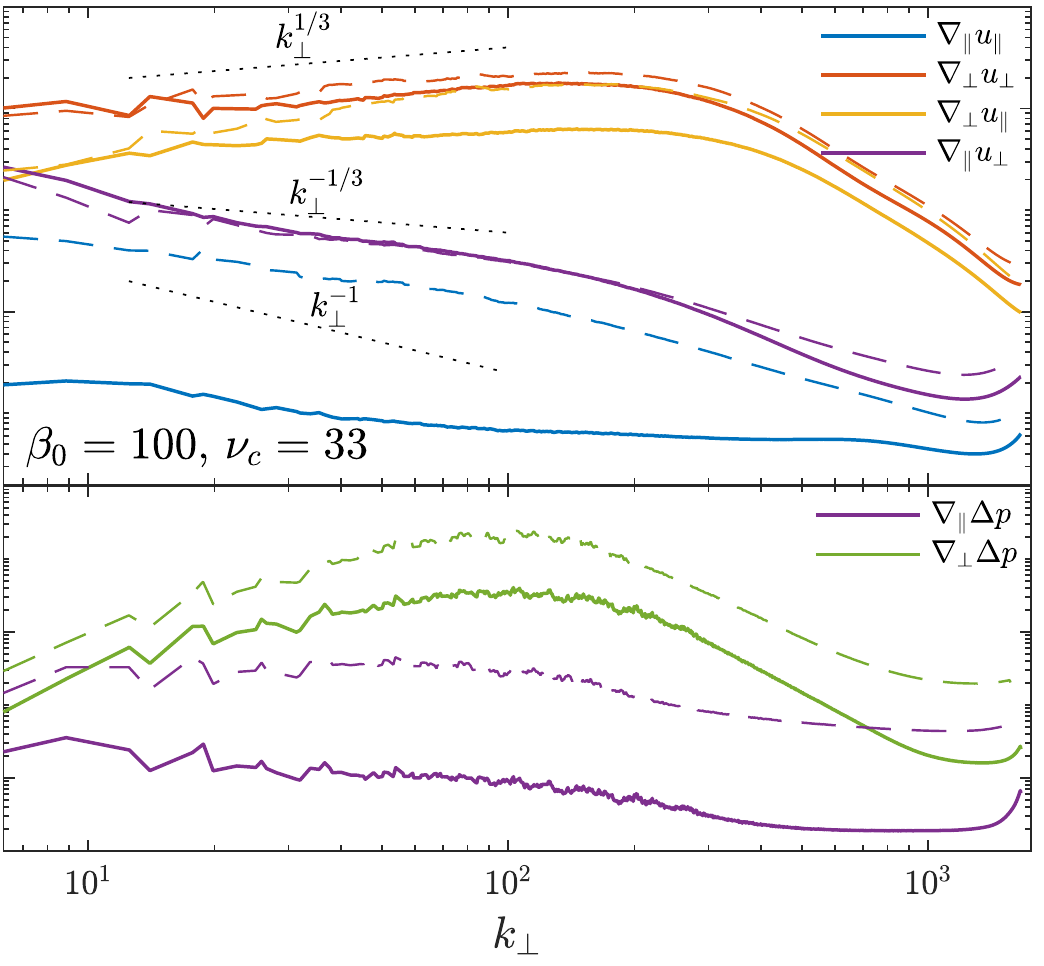}
\caption{Rate-of-strain and pressure-gradient spectra in the \emph{CL10}  and \emph{B100} simulations (left and right panels, respectively). Dashed lines show the equivalent 
passive-$\Dp$ runs, while the colours indicate the specific strain direction considered. Details and definitions are described in~\cref{subsub: ros spectra defs} (see~\cref{eq: grad prp prl defs,eq: grad prp prl Dp}); the dotted black lines indicate the ${\sim}k_{\perp}^{1/3}$, ${\sim}k_{\perp}^{-1/3}$, and $k_{\perp}^{-1}$ scalings (see text). While $\nabla_{\perp}u_{\perp}$, $\nabla_{\perp}u_{\|}$, and $\nabla_{\|}u_{\perp}$ are each relatively similar in active- and passive-$\Dp$ runs, $\nabla_{\|}u_{\|}=\bbgu$, which is responsible for the creation of $\Dp$, is markedly reduced when $\Dp$ is active. Note that neither $\nabla_{\|}u_{\perp}$ nor $\nabla_{\perp}u_{\|}$ is similarly reduced, showing that the effect is not just a reduction in $u_{\|}$ or in $\nabla_{\|}$. The bottom panels show that $\nabla_{\|}\Dp$ is suppressed more than $\nabla_{\perp}\Dp$ in active-$\Dp$ turbulence, which is a signal that the 
pressure-anisotropy stress is reduced beyond just the reduction in the variance of $\Dp$, {\em viz.}, $\grad\bcdot(\bh\bh\Dp)/\Dp_{\rm rms}$ is smaller in active-$\Dp$ than in passive-$\Dp$ turbulence.  }
\label{fig: rates of strain}
\end{center}
\end{figure}
%%%%%%%%%%%%%%%%%%%%%%%%%%%%%%%%%%

\subsubsection{Rate-of-strain spectra}

Given that the pressure anisotropy is driven by the plasma motions, the pressure-anisotropy 
feedback must be driving important changes in the flow and magnetic-field structure, despite having only a small influence on the kinetic-energy spectrum (\cref{fig: basic spectra}). 
We diagnose this in~\cref{fig: rates of strain} via the spectra of the local rates of strain, as defined in~\cref{subsub: ros spectra defs}. 
These reveal the expected substantial suppression of $\nabla_{\|}u_{\|}=\bbgu$
by the pressure anisotropy feedback, as needed to suppress the turbulently-driven fluctuations in $\Dp$ and $B$. 
Of some interest is the comparison of $\nabla_{\|}u_{\|}$ to $\grad_{\perp}u_{\|}$ and $\nabla_{\|}\bm{u}_{\perp}$,
with neither of the latter two significantly suppressed. This demonstrates that the system does not reduce $\bbgu$ purely via a reduction in field-parallel 
flows, or via the reduction of field-parallel gradients, but  through the combination of the two. 
We see broadly similar features in the collisionless (\emph{CL10}) and weakly collisional (\emph{B100}) regimes, with the approximate spectral scalings
 $\mathcal{E}_{\nabla_{\perp}u_{\perp}}\propto k_{\perp}^{1/3}$,  $\mathcal{E}_{\nabla_{\perp}u_{\|}}\propto k_{\perp}^{1/3}$,  $\mathcal{E}_{\nabla_{\|}u_{\perp}}\propto k_{\perp}^{-1/3}$, and $\mathcal{E}_{\nabla_{\|}u_{\|}}\propto k_{\perp}^{-1/3}$
 for both the active- and passive-$\Dp$ simulations.
 As discussed below (\cref{subsub: compressive theory}), the relatively flat $\mathcal{E}_{\nabla_{\|}u_{\|}}$ scaling may be a signature of some form of cascade of compressive fluctuations,
 since a $\bbgu$ spectrum that is dominated by linearly polarised Alfv\'enic fluctuations  would  be na\"{i}vely expected to scale as ${\sim}k_{\perp}^{-1}$. 
 The most interesting difference between the collisionless and weakly collisional spectra
is the closer-to-constant offset between the active and passive $\nabla_{\|}u_{\|}$ spectra in the weakly collisional case (i.e., the modestly steeper $\nabla_{\|}u_{\|}$ spectrum in \emph{B100}), 
which may be a result of the pressure-anisotropy feedback being approximately constant across 
scales when   $\omegaa<\nu_{\rm c}$, rather than being dominated by the outer scale (see~\cref{fig: It}).

The lower panels in~\cref{fig: rates of strain} show spectra of  gradients of the pressure anisotropy. As is clear from the $\bm{u}$ evolution equation~\eqref{eq:KMHD u}, 
it is not the pressure anisotropy itself that feeds back on the flow, but only its divergence
\begin{equation}
\grad\bcdot(\bh\bh\Dp) = \bh \nabla_{\|}\Dp - \bh \Dp \nabla_{\|}\ln B + \Dp \bh\bcdot\grad\bh.\label{eq: dp gradient expansion}
\end{equation}
Computing each of these terms separately, one finds that $\nabla_{\|}\Dp$ is  larger than the other terms, and there are no significant correlations between the different terms. Thus, if an active-$\Dp$ run exhibits a  lower  ratio of $\nabla_{\|}\Dp$  to $\grad_{\perp}\Dp$ than an equivalent passive-$\Dp$ run, this signifies that the effect of 
pressure anisotropy on the flow, $\grad\bcdot(\bh\bh\Dp)$, is reduced beyond what would be assumed by just looking at the variance of $\Dp$ or $\nabla_{\|}u_{\|}$. 
We see from~\cref{fig: rates of strain} that this is indeed the case, \emph{viz.,} there is a secondary mechanism -- the turbulence reduces  $\grad\bcdot(\bh\bh\Dp)$ beyond just
$\Dp$ (or $\grad\Dp$) -- which will further suppress the damping of plasma motions by pressure anisotropy. 

We have examined a wide variety of  correlations between the different terms 
that drive the  pressure-anisotropy evolution~\eqref{eq: Dp equation}. For example, one might imagine that a positive correlation between $\bbgu$ and $\grad\bcdot\bm{u}$ would
aid in reducing $\Dp$ generation, particularly at lower $\beta$ (see~\cref{eq: B and bbgu}); or similarly for correlations  between the 
heat fluxes and other terms in the $\Dp$ evolution~\eqref{eq: Dp equation}.
However, directly comparing such measures between the active- and passive-$\Dp$ simulations has not 
yielded any further effects that are worthy of note, so we do not discuss these here.

\subsubsection{Possible theoretical interpretation}\label{subsub: compressive theory}

In this section we speculate on possible phenomenologies that  could explain the observed pressure and rate-of-strain 
spectra in the inertial range below the forcing scales. 
We examine two qualitatively different possibilities: the first is that 
the pressures and $\bbgu$ are dominated by the residual magnetic-field variation that arises from a collection of linearly polarised Alfv\'enic fluctuations; 
the second is that they are dominated by a cascade of compressive fluctuations, which is
set up around the forcing scales as the system rearranges itself. In the first scenario, 
the $\Dp$ and $B^{2}$ spectra are the residual local `left-overs' that cannot quite be eliminated by magneto-immutability; 
in the second, they are diagnosing another type of cascade or mixing of larger-scale fluctuations, similar to the passive slow-mode cascade in reduced MHD or gyrokinetics (\citealp{Schekochihin2009,Kunz2015,Meyrand2019}; however, unlike these gyrokinetic cases, the compressive and Alfv\'enic cascades could lack individually conserved invariants, meaning the distinction between them may be less precise than implied above).\footnote{Note that the theory of \citet{Kunz2015} and \citet{Kunz2018} applies in the presence of 
a mean pressure anisotropy, but not when the fluctuations themselves contribute to a  dynamically important $\Dp$, as is the case for turbulence with $\It\lesssim1$.} 
Motivated by the observation in \cref{fig: dp spectra} that $\mathcal{E}_{p_\|}\gg \mathcal{E}_{p_\perp}$,
we will 
imagine the compressive cascade to consist of oblique slow-magnetosonic-like
modes,\footnote{Another possibility, a cascade of non-propagating modes, seems to be ruled out immediately because these modes satisfy  $\delta p_{\perp}> \delta p_{\|}$ at high $\beta$ \citep{Schekochihin2009,Majeski2023}.} which, because they maintain perpendiclar pressure balance, satisfy $\delta p_{\perp}\ll \delta p_{\|}$. This property  
causes the mode's frequency to scale as $k_{\|}v_{\rm th}$, rather than $k_{\|}\va$ (as for the MHD slow mode), as well as leading to significant damping by heat fluxes and/or collisions. Further details are given in~\cref{appsub: dispersion relation} and \citet{Majeski2023}.

Let us examine  predictions for the rate-of-strain and pressure  spectra for each possibility in turn, starting with the idea that finite-amplitude Alfv\'enic fluctuations dominate.
In this case, a simple estimate for the spectrum of $\bbgu$ comes from~\cref{eq: B and bbgu}, using $\bbgu\approx (1/2) B_{0}^{-2} \rmd(\delta B_{\perp}^{2})/\rmd t \sim \omegaa \delta B_{\perp}^{2}/B_{0}^{2}$. For a critically balanced Goldreich--Sridhar cascade ($\delta B_{\perp}\propto k_{\perp}^{-1/3}$, $\omegaa\propto k_{\perp}^{2/3}$), this  suggests that $\bbgu\sim k_{\perp}^{0}$, or a $\bbgu$ spectrum $\mathcal{E}_{\nabla_{\|}u_{\|}}\propto k_{\perp}^{-1}$,  which is clearly significantly steeper than observed (cf.~the steepest dotted line in~\cref{fig: rates of strain}). However, this prediction is less rigorous than it seems because ignores the influence of  magneto-immutability:
larger $\delta B_{\perp}$, with larger local $\It(k_{\perp})$,
 will be more affected by $\Dp$ forces, thus reducing $\bbgu$ below  $\omegaa \delta B_{\perp}^{2}/B_{0}^{2}$ and flattening its spectrum, perhaps to what is observed. 
 However, the most  straightforward estimate for this effect -- 
that magneto-immutability would lead to $\Dp$ fluctuations of approximately  constant dynamical importance  across scale (see~\cref{subsub: interruption number}) -- would suggest $\delta \Dp\propto B_{0}^{2}$ or $\mathcal{E}_{\Dp}\sim \mathcal{E}_{p_\|}\sim k_{\perp}^{-1}$, which also does not agree with what is observed (\cref{fig: dp spectra}). So, there is  no obvious phenomenological explanation for the measured spectral slopes within this framework.
As described above, the observation that $\delta p_{\perp}\ll \delta p_{\|}$ can be naturally explained by postulating that the finite-amplitude Alfv\'enic fluctuations are in perpendicular pressure balance with 
the  magnetic-field-strength fluctuations, $\delta p_{\perp}\sim \delta (B^{2})$ (as observed in \cref{fig: dp spectra}). Because 
 $\delta p_{\|}$ is instead driven by $\bbgu$ to $\delta p_{\|} \sim \beta \max(1,\omegaa/\nu_{\rm c})\delta (B^{2})\gg \delta (B^{2})$, 
this suggests  that $\delta \Dp\sim \delta p_{\|}\gg \delta p_{\perp}$.  

The other possibility -- a compressive magnetosonic cascade -- 
more naturally explains some spectral features, but appears to disagree with  other key properties. Its most problematic aspect is that oblique CGL  magnetosonic slow waves satisfy  $\delta p_{\|}/\delta p_{\perp}\approx (5\beta+6)/2$ (see~\cref{appsub: dispersion relation}), a far larger ratio than what is observed
in all three simulations shown in~\cref{fig: dp spectra}.\footnote{Heat fluxes and collisions modify $\delta p_{\|}/\delta p_{\perp}$ modestly, but
the predicted ratio remains far too large to match \cref{fig: dp spectra} in all relevant regimes.} 
\revchng{Furthermore, although the observed ratio $\delta p_{\|}/\delta p_{\perp} $ increases modestly with $\beta$, the 
increase is far slower than the predicted linear dependence, which, for example, would yield $\mathcal{E}_{p_{\|}}/\mathcal{E}_{p_{\perp}}\sim 1000$ in the \emph{CL100} simulation.}
This suggests that at least some other component or complication is necessary to decrease $\delta p_{\|}/\delta p_{\perp}$ well below the linear prediction to match the simulations.
On the other hand, the observed spectral scalings of the rates of strain and
pressure seem to be most naturally explained via a compressive cascade: 
if one postulates $u_{\|}\propto u_{\perp}\sim k_{\perp}^{-1/3}$ (assuming the compressive and Alfv\'enic cascades scale similarly),
$\bbgu$  should scale as ${\sim}k_{\|}u_{\|}\propto k_{\perp}^{1/3}$, giving $\mathcal{E}_{\nabla_{\|}u_{\|}}\propto k_{\perp}^{-1/3}$, which is
close to what is seen in~\cref{fig: rates of strain}.
Predicted scalings of the other rates of strain ($\nabla_{\|}u_{\perp}\propto k_\perp^{1/3}$,
$\nabla_{\perp}u_{\|}\propto k_\perp^{2/3}$, and $\nabla_{\perp}u_{\perp}\propto k_\perp^{2/3}$)
are also consistent with what is observed, being 
the same as what is expected for passive slow-mode cascade  in isothermal MHD (and indeed, the passive-$\Dp$ rates of strain exhibit similar scalings in \cref{fig: rates of strain}).
In this scenario, pressure spectra would result predominantly from the passive mixing of larger-scale fluctuations, as opposed to local driving by $\bbgu$, which also naturally yields the observed scaling $\mathcal{E}_{p_\|}\propto \mathcal{E}_{p_\perp}\sim k_\perp^{-5/3}$ through the inertial range where $u_{\perp}\sim k_{\perp}^{-1/3}$, as well as explaining the similarity of the collisionless and  weakly collisional simulations.
Another interesting observation is the long parallel 
scale of $ p_{\|}$ and $u_{\|}$ fluctuations (see  $\ell_{\|}(\lambda)$  in~\cref{fig: alignment}b),  
which would be expected if for some reason   the frequencies of Alfv\'enic and compressive fluctuations were matched around the outer scale.

Overall, we
see a plausible agreement of the Alfv\'enic hypothesis with most diagnostics, 
although its predictions remain too qualitative to say much of substance. Its biggest flaw is the similarity of the collisionless and weakly collisional rate-of-strain and pressure spectra, which  argues against pressure fluctuations being driven locally in scale by $\bbgu$. A compressive cascade alone cannot explain 
the observed spectra because the $\delta p_\|/\delta p_\perp$ ratio characteristic of linear slow-mode fluctuations is too large to fit the data (particularly the 
dependence on $\beta$); however, we cannot 
rule out their importance in some form (e.g., if $\delta p_\perp$ was dominated by residual Alfv\'enic fluctuations),
and various other properties, particularly spectral scalings, seem to be more naturally explained via the compressive-cascade hypothesis.
Clearly, more work  is needed to make further progress.

\subsection{Heating and energy fluxes}\label{sub: heating}

Perhaps the most interesting and macroscopically relevant impact of the pressure-anisotropy reduction through magneto-immutability is the
suppression of viscous (pressure-anisotropy) heating that it entails. As discussed above, the effect implies that 
 viscous  damping  is suppressed compared to na\"{i}ve estimates for individual
Alfv\'enic perturbations of similar amplitudes to the observed turbulent fluctuations, leading to a larger turbulent-cascade efficiency, with more energy transfered
 to small scales. Consequently, a larger fraction of the heating 
 will occur through processes mediated by the small-scale turbulent cascade (e.g., Landau damping and nonlinear phase mixing around the ion gyroscale; 
 \citealp{Schekochihin2009,Kunz2018}), as opposed to viscous damping of  outer-scale eddies. These  different heating mechanisms could in turn could have important thermodynamic consequences, modifying, for instance, the perpendicular-to-parallel ion heating fraction, or the ion-to-electron heating fraction.
 
In this section, we quantify these ideas by  measuring directly the turbulent transfers and fluxes between scales. This allows us 
to measure the local damping due to pressure anisotropy across each scale in the cascade, and to determine the cascade efficiency by measuring the fraction of input energy that is lost into heat near the outer scale. This will again highlight a large difference between active- and passive-$\Dp$ simulations, demonstrating that magneto-immutable turbulence is effectively conservative (no damping)
below the outer scale. 
%The total viscous damping rate, while interesting for its dependence on different parameters, is unfortunately always dependent
%on the forcing mechanism, since it is controlled by the statistics of the outer-scale (forced) eddies.

\subsubsection{Energy transfers and fluxes}
As discussed in~\cref{subsub: energy transfers}, the turbulence energetics can be usefully  quantified using  
`transfer functions' $\mathcal{T}^{\rm AB}_{q\rightarrow k}$, which measure the energy transfer between shells in Fourier space due to different types of nonlinear interactions. 
Each of these transfers can involve several terms in the compressible system,  so, as detailed in \citet{Grete2017}, we combine relevant compressive terms into the non-compressive terms in order to reduce the number of quantities under consideration. 
 The compressive terms are all individually very small and do not show interesting features, although they are necessary to include in order to respect energy conservation.
Their full forms, as we compute them, are as defined in \citet{Grete2017}, but we simply use the placeholder  $\mathcal{C}_{\rm AB}$ to refer to them below. 
We  follow \citet{Grete2017} in our nomenclature; ${\rm AB}={\rm UU}$ refers to the transfer between shells of the kinetic energy through $\bm{u}\bcdot\grad\bm{u} + \mathcal{C}_{\rm UU}$ (see~\cref{eq: UU  transfer}),  ${\rm AB}={\rm BU}$ refers to the transfer from magnetic to kinetic energy through $\bm{B}\bcdot\grad\bm{B} + \mathcal{C}_{\rm BU}$, ${\rm AB}={\rm UB}$  refers to the transfer from kinetic to magnetic energy through $\bm{B}\bcdot\grad\bm{u}+\mathcal{C}_{\rm UB}$, ${\rm AB}={\rm BB}$   refers to the transfer between shells of the magnetic energy through $\bm{B}\bcdot\grad\bm{B}+\mathcal{C}_{\rm BB}$,  ${\rm AB}={ \Dp {\rm U}}$ refers to the transfer from the kinetic to thermal energy via the pressure anisotropy through $\grad\bcdot(\bh\bh\Dp)$ (see below), and ${\rm AB}={\rm F U}$ refers to the transfer to kinetic energy from the external forcing.\footnote{We do not plot the pressure-stress transfer from $-\grad p_{\perp}$, ${\rm AB}={\rm PU}$, because it is small and uninteresting, although it is included when computing sums of all terms. } 
We compute transfers as a function of the isotropic $k$ as opposed to $k_{\perp}$, but results versus $k_{\perp}$ are  similar.

As introduced in \aksqs, the only important new term compared to  \citet{Grete2017} is that due to the pressure-anisotropy stress $\mathcal{T}_{q\rightarrow k}^{ \Dp {\rm U}}$.  There
is some freedom in the definition of this;  \aksqs\ use 
\begin{equation}
\mathcal{T}_{q\rightarrow k}^{\Dp\mathrm{U}} = \int\mathrm{d} \boldsymbol{x}  \left\langle\sqrt{\rho}\bm{u}\right\rangle_{k} \bcdot {\rm sign}({\Delta p}) \frac{\sqrt{|\Dp|} }{B}\frac{\bm{B}}{\sqrt{4\upi \rho}}  \bcdot \grad\langle\sqrt{|\Dp|}\hat{\bm{b}}\rangle_{q},\label{eq: T dp lev}
\end{equation}
 which has the advantage that the square of  $\langle \sqrt{|\Dp|}\hat{\bm{b}}\rangle_{k}$ can be interpreted as the pressure anisotropy's contribution to the thermal energy \citep{Kunz2015}, but
 the disadvantage of sign discontinuities that arise from $\sqrt{|\Dp|}$. If we instead interpret the pressure-anisotropy stress as a damping of kinetic energy (as opposed to the transfer between energy reservoirs) a more natural definition is 
\begin{equation}
\mathcal{T}_{q\rightarrow k}^{\Dp\mathrm{U}} = \int\mathrm{d} \boldsymbol{x}  \left\langle\sqrt{\rho}\bm{u}\right\rangle_{k} \bcdot \frac{\bm{B}}{\sqrt{4\upi \rho}} \bcdot \grad\left\langle \frac{\Dp}{B^{2}}{\bm{B}}\right\rangle_{q}.\label{eq: T dp mine}
\end{equation}
We have computed both versions, finding qualitatively similar results, but will focus on~\eqref{eq: T dp mine}, because it fits more naturally with our focus on pressure-anisotropy damping. 
Note also that  the influence of the sign-discontinuity issue of~\cref{eq: T dp lev} was mitigated in \aksqs\ because the $\Dp$ distribution was mostly confined to negative values, but the spread of $\Dp$ is somewhat broader in our simulations (see~\cref{fig: Dp PDFs brag,fig: Dp PDFs cl}).

%%%%%%%%%%%%%%%%%%%%%%%%%%%%%%%%%%
\begin{figure}
\begin{center}
\includegraphics[width=1.0\columnwidth]{\ffold/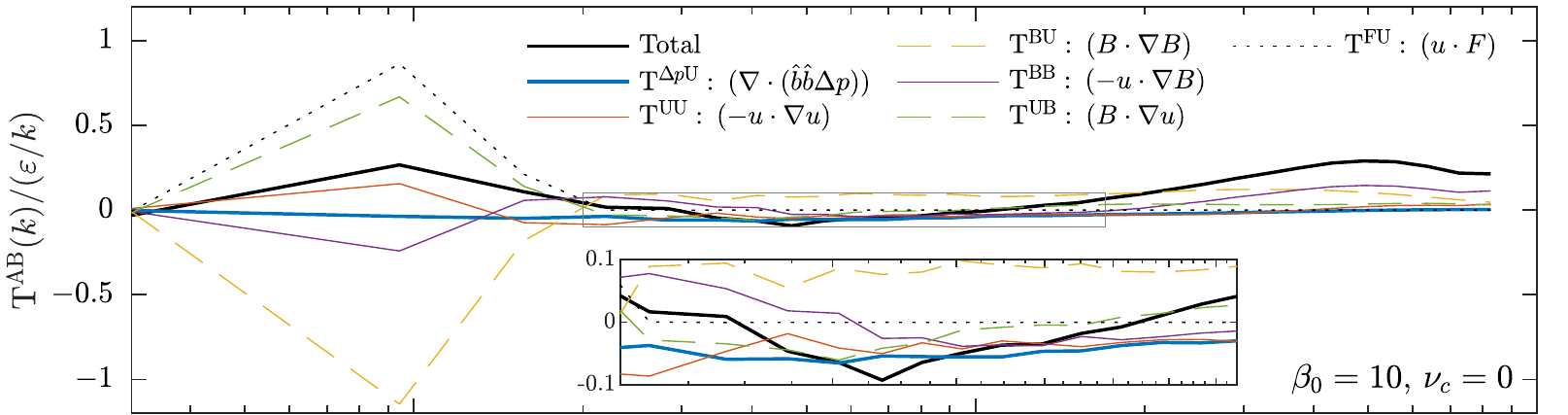}\\
\includegraphics[width=1.0\columnwidth]{\ffold/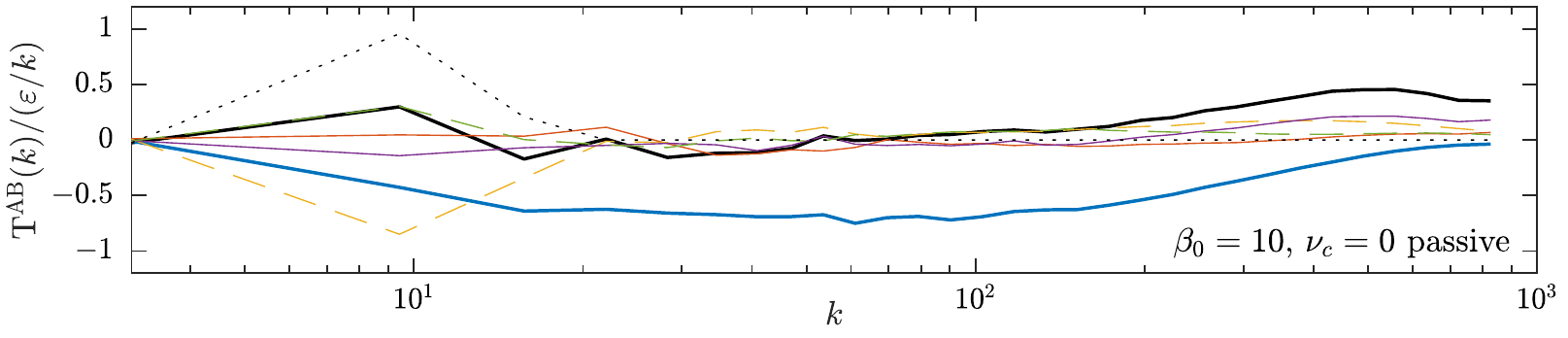}\\
\includegraphics[width=1.0\columnwidth]{\ffold/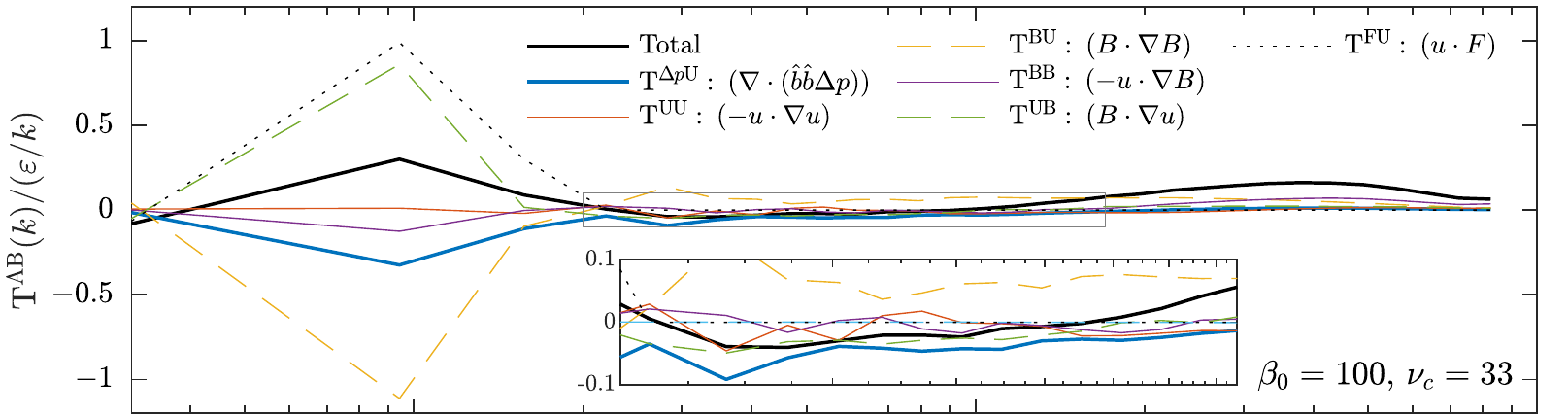}\\
\includegraphics[width=1.0\columnwidth]{\ffold/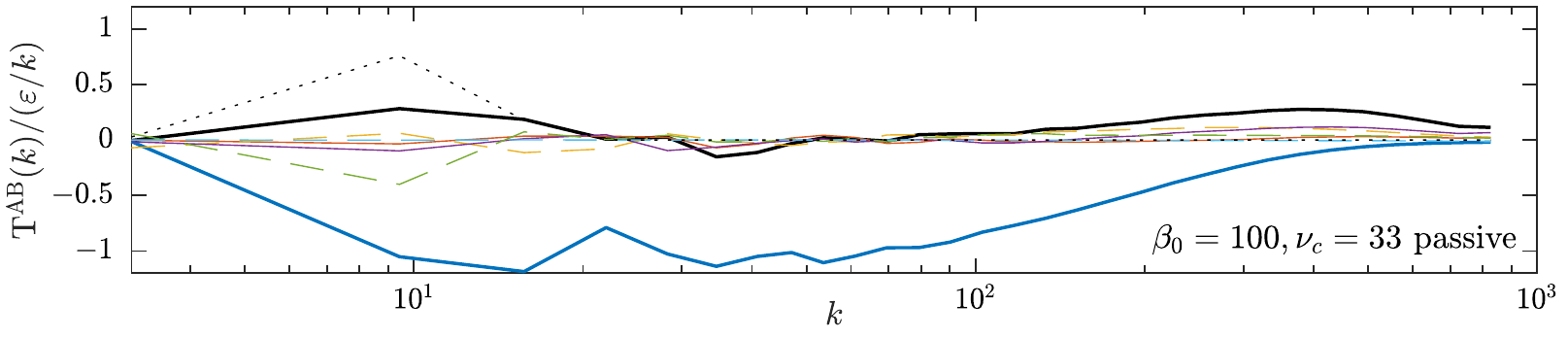}
\caption{Important terms in the net energy-transfer spectra, as defined in~\cref{subsub: energy transfers} and~\cref{eq: total energy transfer}. The top panels show the simulations with  $\beta_{0}=10$, $\nu_{\rm c}=0$ and the bottom panels show the simulations with  $\beta_{0}=100$, $\nu_{\rm c}=33$, with the upper and lower subpanels for each showing the active- and passive-$\Dp$ cases, respectively. Insets in the plots of the two active-$\Dp$ cases show a zoom of the grey-box region. All transfers
 are normalised to $\varepsilon/(k/k_{f})$, which approximately represents the local cascade rate (see text). The different colours show different energy-transfer terms as labeled in the top panel: $\mathrm{T}^{\Dp {\rm U}}$ is the transfer from $\bm{u}$ to thermal energy due to $\dpstress$; $\mathrm{T}^{\rm UU}$ is momentum transfer via $\bm{u}\bcdot\grad\bm{u}$; $\mathrm{T}^{\rm BU}$ is the contribution to kinetic energy from magnetic tension and pressure; $\mathrm{T}^{\rm BB}$ is the advection of $\bm{B}$ in the induction equation; $\mathrm{T}^{\rm UB}$ is the contribution to magnetic energy from field stretching; and $\mathrm{T}^{\rm FU}$ is the forcing contribution. In the active-$\Dp$ simulations below the outer scale of the turbulence, we see little contribution of $\Dp$ to the energy transfer, which shows directly that the effect of the pressure-anisotropy stress is minimised below the outer scale, even though  the local interruption number is still smaller than unity in this range (see~\cref{fig: It}). In contrast, in the
passive-$\Dp$ simulations (bottom panel), the pressure anisotropy that develops is such that, if it did feed back on the flow, it would cause $\mathcal{O}(1)$ dissipation at all scales in the cascade (blue line), which would completely damp the turbulence.}
\label{fig: transfer spectra}
\end{center}
\end{figure}
%%%%%%%%%%%%%%%%%%%%%%%%%%%%%%%%%%
%%%%%%%%%%%%%%%%%%%%%%%%%%%%%%%%%%
\begin{figure}
\begin{center}
\includegraphics[width=1.0\columnwidth]{\ffold/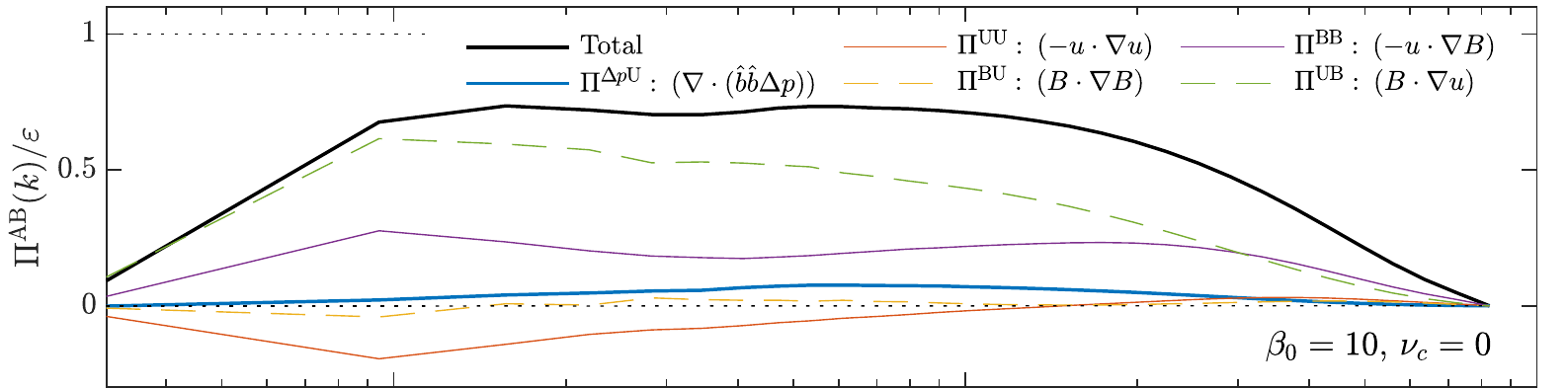}\\
\includegraphics[width=1.0\columnwidth]{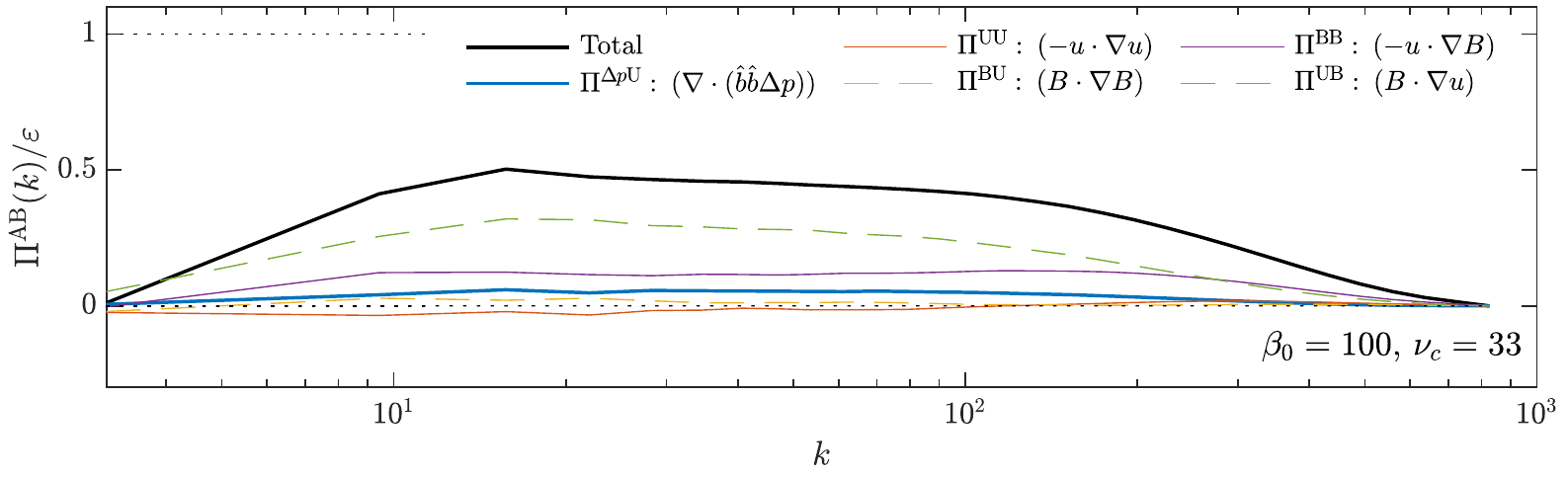}%\\
\caption{The energy fluxes (see~\cref{eq: energy fluxes}), which measure the  transfer of energy across a particular $k$, for the two active-$\Dp$ simulations with $\beta_{0}=10$, $\nu_{\rm c}=0$ (top panel) and $\beta_{0}=100$, $\nu_{\rm c}=33\va/L_{\perp}$ (bottom panel). We normalise to the forcing input $\varepsilon$, which
implies that the total flux would be $\Pi/\varepsilon=1$ across the full inertial range for a conservative cascade with no damping. As also seen in~\cref{fig: transfer spectra}, although there is some energy loss due to $\Dp$ at the outer (forcing) scale, at smaller scales the flux remains constant, showing that there is little energy damping through the inertial range.}
\label{fig: energy fluxes}
\end{center}
\end{figure}
%%%%%%%%%%%%%%%%%%%%%%%%%%%%%%%%%%

In~\cref{fig: transfer spectra}, we plot the net energy transfers ${\rm T}^{\rm AB}(k)$ (see~\cref{eq: total energy transfer}), which measure the net contribution from each term to the  rate of change of the kinetic or magnetic energy spectrum
at each $k$. For a conservative cascade, these are zero, because the transfer into a given shell from larger scales is balanced by the transfer out of the shell to smaller scales. Thus, ${\rm T}^{\rm AB}(k)$ provides a direct measure of the damping of kinetic or magnetic energy at each scale due to pressure 
anisotropy or other terms. To  plot the results clearly, we normalise ${\rm T}^{\rm AB}(k)$ by the dimensional (Kolmogorov) estimate  $\partial_{t}\mathcal{E}\sim k u_{k}\mathcal{E}\sim \varepsilon (k/k_{0})^{-1}$ (where $k_{0}=2\upi/L_{\perp}$); with this normalisation, a line that is constant and nonzero with $k$ symbolises a contribution 
that is of approximately constant importance compared a conservative cascade towards smaller scales. As above, 
we show the \emph{CL10} and \emph{B100} simulations, comparing to the results from the corresponding  passive-$\Dp$ simulations (lower subpanels). The thick black lines, which show the sum of all terms (excluding ${\rm T}^{ \Dp {\rm U}}$ for the passive runs), are approximately zero throughout the inertial range, as expected, with no clear difference 
between the active and passive runs. While the contribution from ${\rm T}^{\Dp {\rm U}}$ is slightly negative in both \emph{CL10} and \emph{B100}, indicating a slight damping of inertial-range motions,  
it is  small compared to the cascade rate. Contrast this to the equivalent ${\rm T}^{\Dp {\rm U}}$ from the passive simulations, which gives an indication 
of what the damping would be in the absence of feedback from $\Dp$. Although this measurement is  counterfactual -- the cascade could not have proceeded in the
presence of such strong damping -- it concisely illustrates how
motions driven by normal turbulence would be strongly damped, at a rate comparable to the cascade rate, across  all scales. 
This  is not entirely obvious for the collisionless case particularly, because 
 the contribution to the kinetic energy from $\grad\bcdot(\bh\bh\Dp)$ can be both positive and negative (it is negative definite in the Braginskii-MHD regime, but not otherwise; see~\cref{eq: energy conservation}).  The effect of magneto-immutability all but eliminates such damping  below the outer scale, leaving only a modest damping of outer-scale motions for the weakly collisional case. 
 
 \Cref{fig: energy fluxes} shows similar information in a different form by plotting the contributions to the turbulent energy flux~\eqref{eq: energy fluxes}. We show only the active-$\Dp$ runs because the
 comparison with passive simulations is less interesting in this case. Unlike the net transfers (\cref{fig: transfer spectra}), individual terms in the fluxes are 
 not  easily interpreted  because they do not
include the diagonal $\mathcal{T}^{\rm AB}_{k\rightarrow k}$ transfer. While this diagonal contribution necessarily   vanishes for the total energy flux,
 it  dominates the transfer between different energy reservoirs, including the damping
 of the flow by the pressure anisotropy (blue line). Thus the fact that  $\Pi^{\Dp {\rm U}}$ is small and positive is not particularly relevant,
  while the individual lines are of interest only 
 insofar as they indicate a contribution to the total flux (black lines). The most important feature that we observe is
 that the total energy fluxes are nearly constant for $10\lesssim k\lesssim200$, consistent with the result
 of~\cref{fig: transfer spectra} that there is little pressure-anisotropy damping through the inertial range (a slight decrease in $\Pi(k)$ is observed for \emph{B100}, indicating a
 slight damping). The value of the $\Pi/\varepsilon$ in the inertial range is a direct measure of the cascade efficiency, which, as 
 expected, is less than unity because there is
 some pressure-anisotropy damping near the forcing scales (the reduction of ${\simeq} 25\%$ for C100 and ${\simeq}50\%$
 for \emph{B100} is consistent with the total inferred pressure anisotropy heating, which is discussed below)
 
% \footnote{The lower flux in the passive \emph{B100}
% simulation arises because the total energy is slowly growing in time over the period that we ran this simulation. This 
% is a consequence of the initial conditions used for the refinement to get to high resolution. Its high computational
% expense compared  }

%%%%%%%%%%%%%%%%%%%%%%%%%%%%%%%%%%

\subsubsection{Pressure-anisotropy heating fraction and cascade efficiency}

The net energy transfer ${\rm T}^{\Dp {\rm U}} (k)$ also gives a convenient way 
to evaluate the total heating rate due to pressure anisotropy. Because  the pressure-anisotropy (viscous) heating 
is confined to the outer scales (\cref{fig: transfer spectra}),  this is also a measure of the cascade efficiency -- the energy that is available to cascade in the usual way to heat via small-scale collisionless mechanisms. Its
precise value  depends on the forcing scheme, since it mostly occurs at the outer scales where the 
forcing drives the flow, \revchng{and could thus differ more significantly in systems with more realistic turbulence generation mechanisms (e.g., 
magneto-rotational turbulence; \citealp{Kunz2016,Kempski2019}). Nonetheless,
its dependence on parameters ($\beta$ and the collisionality) is interesting to consider, as is a comparison to \aksqs.}

We define
\begin{equation}
\mathcal{H}_{\Dp} = \int_{k_{0}}^{k_{\rm max}} \rmd k \,{\rm T}^{\Dp {\rm U}}(k),\label{eq: Hdp definition}
\end{equation}
so $\mathcal{H}_{\Dp} /\varepsilon$ measures the fraction of the energy input that is  converted into heat via pressure anisotropy up 
 to wavenumber $k=k_{\rm max}$. Because the grid-scale dissipation in our simulations is non-physical -- it is supposed
 to represent energy absorbed by the smaller scales, which would eventually cascade to mostly dissipate around the ion gyro-radius scale \citep{Schekochihin2009} -- it is most appropriate to choose $k_{\rm max}$ to lie near the bottom of the inertial range. 
 We set $k_{\rm max}=100$ as appropriate for the  resolution $400\times 200^{2}$, but the choice  hardly affects the results for the active-$\Dp$ simulations anyway, because almost all of the viscous heating is
 confined near the outer scales.\footnote{For  passive-$\Dp$ cases, $\mathcal{H}_{\Dp}$ increases with $k_{\rm max}$ through the inertial range because there is more and more damping (see~\cref{fig: transfer spectra}).}

%%%%%%%%%%%%%%%%%%%%%%%%%%%%%%%%%%
\begin{figure}
\begin{center}
\includegraphics[width=0.7\columnwidth]{\ffold/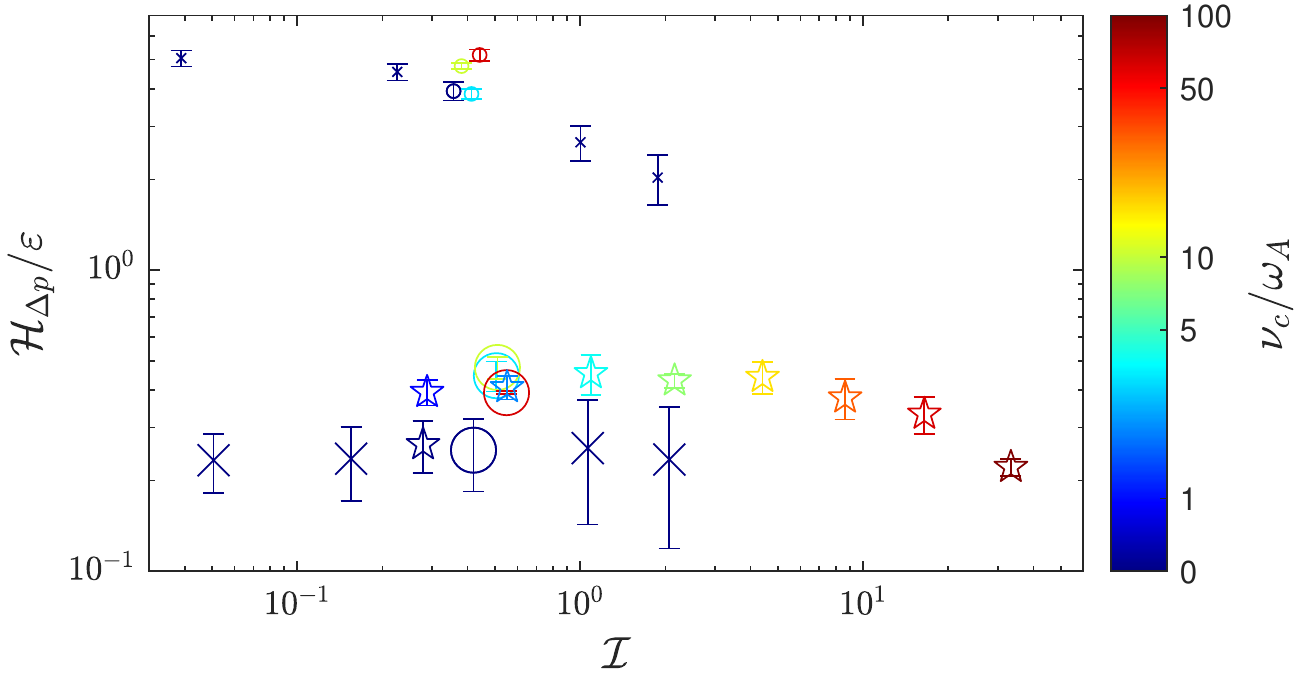}
\caption{Net pressure-anisotropy heating rate $\mathcal{H}_{\Dp}$ normalised to the cascade rate $\varepsilon$ from all simulations. The cascade efficiency, defined as the fraction of energy that participates in the cascade to small scales,  is $1 - \mathcal{H}_{\Dp}/\varepsilon$. This is computed using the net transfer functions (\cref{fig: transfer spectra}) by summing up the contributions from all $k\leq 100$, \emph{viz.,} $\mathcal{H}_{\Dp}=\sum_{k\leq100} \mathrm{T}^{{\rm U}\Dp}(k)$,  so as to exclude the grid scales. The marker style denotes the simulation set (see~\cref{sub:study design} and~\cref{tab:sims}): circles denote the \emph{lrB} set at constant $\It$, crosses denote the \emph{lrCL} set with $\nu_{\rm c}=0$, and  stars denote the $\beta16$ set. We see that viscous (pressure-anisotropy) heating damps up to ${\simeq} 45\%$ of the cascade in collisional regimes ($\nu_{\rm c}/\omegaa\gtrsim1$), and less in  collisionless cases. This value is approximately constant for $\It\lesssim 10$ and  independent of $\beta$ and  $\nu_{\rm c}/\omegaa$ (so long as $\nu_{\rm c}/\omegaa\gtrsim 1$). The small symbols show the same computation from the passive-$\Dp$ simulations; in this case the heating rate is larger than unity, because $\Dp$ does not actually feed back on the flow.  Such large values imply that if $\Dp$ had not been reduced by magneto-immutability, it would have nearly completely damped the cascade.}
\label{fig: heating in all sims}
\end{center}
\end{figure}

We evaluate $\mathcal{H}_{\Dp} $ for all of the \emph{lrCL}, \emph{lrB}, and $\beta16$ simulations 
 (see~\cref{tab:sims}), time averaging across the steady state of each.  Results are shown in~\cref{fig: heating in all sims}, plotted against the measured interruption number, with marker colour
 indicating the collisionality and marker style indicating the simulation set (\emph{lrCL}, \emph{lrB}, or $\beta16$). 
 The CGL-LF (active-$\Dp$)
simulations, shown with large symbols, have $ \mathcal{H}_{\Dp}$ between $\mathcal{H}_{\Dp}\simeq0.2\varepsilon$ for the collisionless simulations and 
 $\mathcal{H}_{\Dp}\simeq0.45\varepsilon$ for the weakly collisional and Braginskii-MHD simulations at $\It\sim1$, \revchng{with no significant
 dependence on $\beta$.} 
 In other words, for this choice of forcing, the cascade efficiency is always above  ${\simeq}50\%$, meaning most of the input energy 
can participate in an MHD-like turbulent cascade. 
 Contrast this with the small symbols, which show the passive-$\Dp$ simulations for \emph{lrCL} and \emph{lrB}, indicating 
 what the pressure-anisotropy heating rate would be  in the absence of magneto-immutability. The numerical values are of course meaningless in this case -- 
 the situation is counterfactual and having $\mathcal{H}_{\Dp}>\varepsilon$ is clearly not possible -- but, as above, it demonstrates that 
 the motions involved in standard MHD Alfv\'enic turbulence would be strongly damped 
 by pressure anisotropy were it not for the pressure-anisotropy feedback modifying their structure. 
From the $\beta16$ simulations, we see also that $\mathcal{H}_{\Dp}(\It)$ remains relatively large up to 
large $\It$, despite the effective driving of pressure anisotropy fluctuations decreasing towards large $\It$  due 
to their larger $\nu_{\rm c}$. This shows that as $\It$ is decreased below ${\simeq}10$, the suppression 
of viscous heating is sufficiently strong that it balances the stronger driving of $\Dp$ fluctuations that results from  lower $\nu_{\rm c}$ (otherwise $\mathcal{H}_{\Dp} $ would continue increasing below $\It\simeq10$).

\subsection{Effect of electron pressure}

\begin{figure}
\begin{center}
\includegraphics[height=0.5\columnwidth]{\ffold/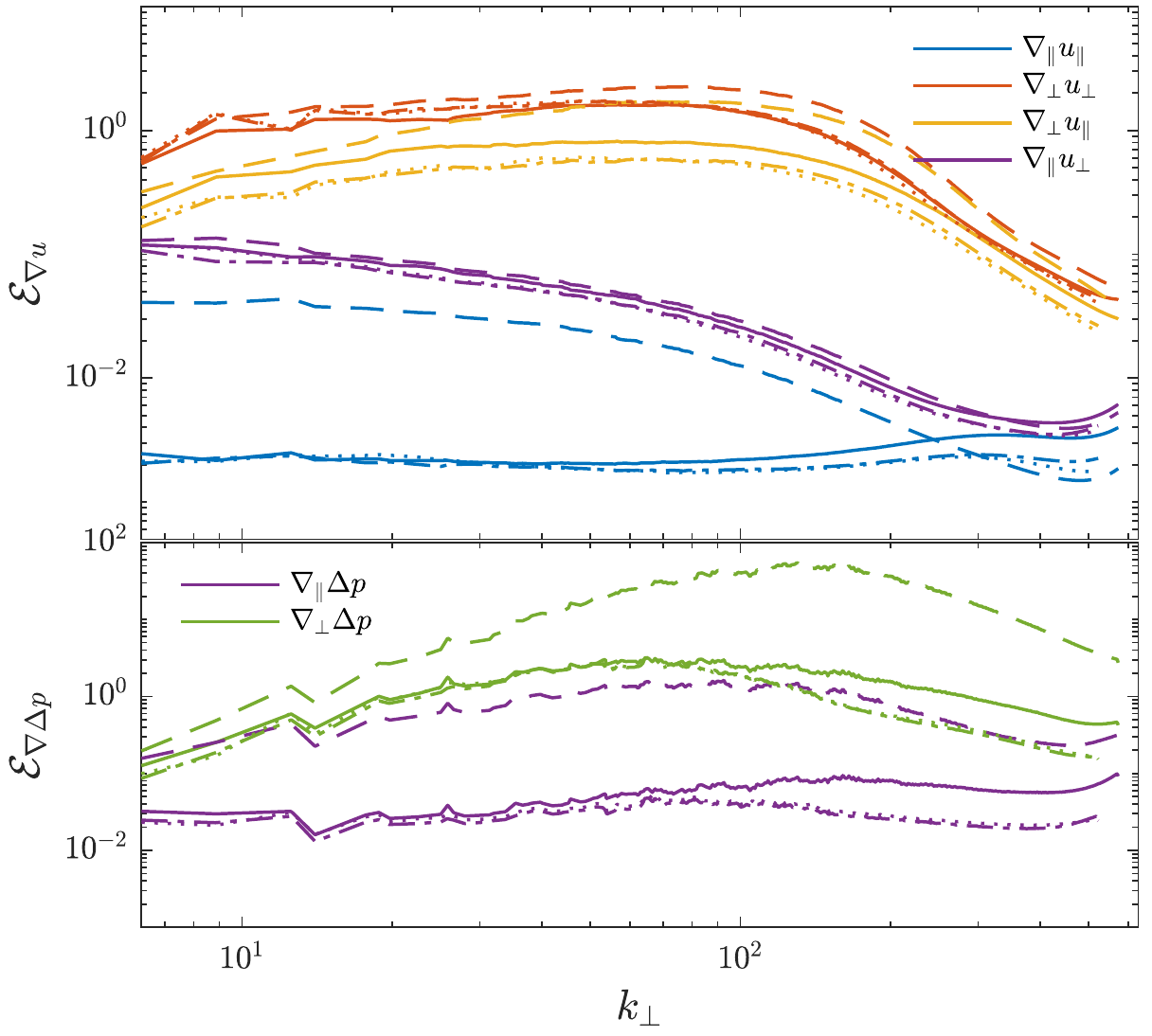}
\caption{Rate-of-strain spectra at $\beta_{0}=10$, $\nu_{\rm c}=0$, comparing the effect of including an (isothermal) electron pressure with different $T_{e}$. We show $T_{e}=0$ (solid line), $T_{e}/T_{i0}=1$ (dot-dashed lines), $T_{e}/T_{i0}=5$ (dotted lines),  and the passive-$\Dp$ simulation (dashed lines).   The plasma's behaviour is almost identical to the $T_{e}=0$ case, because the isotropic electron pressure has little effect once motions are already nearly incompressible.}
\label{fig: rates of strain Te}
\end{center}
\end{figure}

In all of the simulations presented so far, we have set $p_{e}=\rho T_{e}=0$, thus ignoring the influence of electrons. 
These add an additional isotropic pressure response 
that is worthy of exploration in case it interferes
with magneto-immutability once  it starts to dominate over the anisotropic ion pressure response. 
We explore this in~\cref{fig: rates of strain Te} via the rate-of-strain spectra, motivated by the finding above that these 
showed a clear difference between CGL-LF and standard MHD (passive) simulations. 
We show the low-resolution versions of the spectra 
for the \emph{CL10} case already shown in~\cref{fig: rates of strain}, comparing them with two additional simulations 
that include electrons with  $T_{e}/T_{i0}=1$ and $T_{e}/T_{i0}=5$ (here $T_{i0}$ is the initial ion temperature).
We see  no important differences, even for $T_{e}\gg T_{i}$, \revchng{which further implies that the cascade efficiency is independent of $T_{e}/T_{i}$}. This  
is expected since the plasma is already almost incompressible anyway, and an additional isotropic  
pressure should simply help it be even more so. The simulations of \skqs\ and \citet{Kempski2019} used a fully incompressible Braginskii-MHD
model and found similar results.

%%%%%%%%%%%%%%%%%%%%%%%%%%%%%%%%%%
%%%%%%%%%%%%%%%%%%%%%%%%%%%%%%%%%%

The isotropic  fluid electron model is formally valid either when the plasma is semi-collisional (e.g., the ICM) or when  electrons are cold. For 
the case of hot collisionless electrons, a separate electron pressure-anisotropy equation must  be solved, which will  add  
an electron contribution to the pressure anisotropy. Presumably this will generically act to enhance the influence 
of pressure anisotropy, even at large scales, while also bringing in a plethora of electron-scale kinetic  instabilities 
driven by pressure-anisotropies, heat fluxes, and other non-Maxwellian features \citep[e.g.,][]{Gary2003a,Verscharen2022}.
\revchng{However, because the Coulomb interspecies temperature equilibration 
 is extremely slow and (as far as we are aware) there  exist no plasma instabilities that act to decrease it \citep[e.g.,][]{Zhdankin2021},
there is not an obvious connection between the ion and electron thermodynamics.
Thus, even though electron  instabilities could strongly influence the electron temperatures (e.g., by limiting pressure anisotropies and heat fluxes; \citealt{Verscharen2022}),  insofar as they remain very small-scale and unable to interact directly with ions, 
we do not expect them to have a strong influence on the large-scale ion dynamics other than through the pressure-anisotropy stress.\footnote{Perhaps, by enhancing the electron scattering
rate, electron instabilities could improve the validity of a collisional electron closure.} 
That said, there is clearly more work  needed to  understand this complex physics better.}

\section{Discussion: uncertainties and observable effects}\label{sec: discussion}

The numerical results presented herein have  demonstrated how pressure anisotropy 
feeds back on Alfv\'enic turbulent flows at high $\beta$, minimising its own influence to reduce the
variance of $B^{2}$ and the associated parallel viscous heating. The basic effect 
is neither surprising nor escapable, arising  from the large pressure-anisotropy forces that are rapidly generated with 
any change of $B$ and act to oppose this change, in the same way that the large isotropic pressure forces that result from changes to density render fluids incompressible.
However, our more detailed conclusions and quantitative features (e.g., the differences between collisional and collisionless results)
 do depend on the assumptions of the CGL-LF with hard-wall pressure-anisotropy limiters.
In this section, we consider these in more detail by reference and comparison to the hybrid-kinetic simulations 
of \aksqs, particularly focusing on uncertainties relating to the interplay of magneto-immutability 
and microinstability scattering. We argue that most results are robust and compare well to \aksqs, 
which is promising more generally for the use of CGL-LF models in the study of weakly collisional plasmas. 
We also discuss various diagnostics that can be used in observations or kinetic simulations, where the comparison to passive-$\Dp$ simulations -- so useful in our analysis above -- is not available. 

\subsection{Interplay between magneto-immutability and microinstabilities}\label{subsub: mi and microinstabilities}

Perhaps the largest uncertainty concerning the CGL-LF model relates to the 
influence of particle scattering through microinstabilities. As discussed in~\cref{sub: theory microinstabilities}, 
the `hard-wall-limiter' method used here  and in previous works  is based on the idea
that large-scale turbulent motions are extremely slow compared to microinstability-saturation time scales. This suggests that
 microinstabilities  should instantaneously halt the growth of $|\Dp|$ when it is driven beyond the stability thresholds, then instantaneously disappear again 
when the driving reverses and the plasma returns towards stability. In its practical implementation, the hard-wall-limiter model achieves this
by taking $\nu_{\rm c}^{\rm lim}$ to be very large in unstable regions; however,
 kinetic theory and simulations show that the saturated scattering rate should actually be $\nu_{\rm c}\sim S\beta$, where $S\sim\bbgu$ is the shear driven by the turbulence.
The effect on $\Dp$ is similar -- both hard-wall limiters and  scattering with $\nu_{\rm c}\sim S\beta$ act to pin $\Dp$ at the stability thresholds -- but the differences could lead to inaccuracies.\footnote{Most
importantly, the physics  of particles  moving spatially between stable and unstable regions of the plasma 
can, at least in principle, be captured through the heat fluxes in the CGL-LF model; but, with 
large  $\nu_{\rm c}^{\rm lim}$, the heat fluxes are completely suppressed in unstable regions meaning such physics is missed. 
Nevertheless, while kinetic simulations find
that heat fluxes are well approximated by the collisional estimate in firehose regions, in mirror regions, they seem
to be nearly completely suppressed \citep{Kunz2020}. Thus, it is likely that a more accurate model  ought to treat $\Dp>0$ and $\Dp<0$ regions differently, meaning that the complete
suppression of heat fluxes  via a large $\nu_{\rm c}^{\rm lim}$  in limited regions may be no worse than using $\nu_{\rm c}\sim S\beta$ anyway.} Furthermore, kinetic 
simulations have shown that the assumption that microinstabilities saturate and decay instantaneously compared
to the turbulent motions is questionable. This is most acute for the mirror instability, which can limit $\Dp$ growth via particle
trapping through much of a macroscopic shearing time \citep{Schekochihin2008b,Kunz2014,Rincon2015}. Decaying
firehose and mirror fluctuations can also be extremely long lived, continuing to scatter particles even when not actively driven by the large-scale shear \citep{Squire2017a,Kunz2020}. Indeed, their decay rate, as it 
involves changing $B^{2}$, may itself be limited by the need to avoid creating unstable pressure anisotropies \citep{Melville2016}. 
Thus, it may not be appropriate to consider scattering only in unstable regions -- rather, 
the continual driving will create a `soup' of magnetic fluctuations that drive 
scattering across the entire plasma, putting it in the weakly collisional or Braginskii regime \citep{Ley2022}. This situation applies to almost any foreseeable kinetic numerical simulation, including
those of \aksqs, but is at odds with the assumptions of our collisionless simulations, where we set $\nu_{\rm c}=0$ in stable regions.

We are thus left with some uncertainty between two limiting scenarios: in the first, scattering sites  pervade the plasma; in the second,  they  exist only 
in spatiotemporal locations where the plasma is unstable. Reality, for astrophysically relevant scale 
separations, likely lies somewhere between these extremes.\footnote{The scale separation in the solar wind, which is 
${\simeq}10^{4}$ between the outer scales and $\rho_{i}$, is definitely not large enough to lie in the second regime, being foreseeably approachable 
with kinetic simulations \citep[e.g.,][]{Bott2021}.} Comparing the microinstability-scattering estimate above, $\nu_{\rm c}\sim \beta S \sim \beta \omegaa \ma^{2}$, with the Alfv\'enic interruption estimates discussed in~\cref{sub: magneto-immutability}, we see that the $\nu_{\rm c}$ expected in the first scenario is exactly the collisionality required to maintain $\It\simeq 1$. Indeed, the correspondence is inherent in the estimate, since this $\nu_{\rm c}$ is simply that which is necessary to suppress the production of $\Dp$ to a level where its
influence becomes comparable to the magnetic forces. This shows that if the first scenario applies,
all collisionless plasmas effectively become weakly collisional with $\nu_{\rm c}/\omegaa\sim \ma^{2}$ and $\It\gtrsim1$, meaning our 
$\It<1$ collisional and collisionless simulations are never formally applicable (see further discussion in \aksqs). 
More generally, in both scenarios, the scattering from microinstabilities and the pressure anisotropy stress (magneto-immutability)
must  be of similar importance to the plasma's dynamics.
 Both 
effects  limit $\Dp$ to $|\Dp|\lesssim B^{2}/4\upi$, and scale in the same way with $\beta$ and turbulence amplitude. 
As a corollary, there is no limit in which one or the other can be ignored, except in the presence of mean pressure anisotropy (resulting
in microinstabilities but with no corresponding pressure-anisotropy stress; \citealp{Bott2021}). 

%However, in order to provide more detailed estimates, there are  subtleties that make the differing effects of magneto-immutability and microinstability scattering  non-obvious, particularly in the limit of large scale separations. In effect, we 
%are left with estimating the relative importance of two effects of that scale in the same way. Various effects could tip the balance 
%one way or the other; for example, 
%For example, if magnetic fluctuations
%are very slow to decay they could build 
% If this effect remains strong and important at very large scale separations (as suggested by  \citealt{Kunz2020}), it may be that the scattering is so strong across the full plasma that the turbulence could reach a steady state with $\It\gtrsim 1$, 
%implying a dominance of scattering over magneto-immutability. On the 
%other hand, if the microinstability scattering is spatio-temporally localised, magneto-immutability effects will dominate in stable regions where the collisionality is small. Second, there are various order-unity factors in both $\nu_{\rm c}$ and the definition of $\It$, and which effect wins out could depend on these. For example, we see a significant reduction in $\Dp$ production even for $\It\gtrsim 1$ (see e.g., $\beta=1$ case in~\cref{fig: Dp PDFs cl}), 
%while on the other hand scattering may saturate modestly above the level $\nu_{\rm c}\approx S\beta$ if it decays   slowly and is continuously driven by different regions 
%going unstable. 

In order to provide a more detailed theory, we must estimate the relative importance of two effects  that scale in the same way with 
key parameters.
This is  difficult to do  without kinetic simulations at asymptotically large scale separation, currently  not feasible. 
The kinetic simulations of \aksqs\ exhibited a box-averaged scattering rate consistent with
 $\nu_{\rm c}\simeq \beta S$ at $\beta=16$, or $\It\simeq1$ (see their figure~12).\footnote{Their $\beta=4$ simulation exhibited 
 a larger scattering rate, which does not fit with this scaling, but that simulation also had several other peculiarities that may explain this
 discrepancy.} While consistent with the first scenario (`scattering everywhere'), given the modest separation (a factor ${\simeq}20$ between the outer scales and $\rho_{i}$), this does not constitute strong evidence against the second scenario in general, as there is no reasonable way to separate stable and unstable regions in the later stages of their simulations. 
This measured scattering rate, $\nu_{\rm c}\simeq \beta S$, also shows that   either our $\beta16\nu6$ or  $\beta16\nu12$ simulation should 
 be directly comparable to the results of \aksqs\ (see~\cref{tab:sims}). Indeed, for scales above $\rho_{i}$,  most diagnostics appear relatively similar (e.g., pressure anisotropy spectra, the standard deviation of $B$ and $\Dp$, and the viscous heating rate), with perhaps the most obviously significant large-scale
 difference being that the kinetic simulation drives itself to   negative mean pressure anisotropy $\langle \Dp\rangle$, while our $\beta16$ simulations 
 have $\langle \Dp\rangle\approx0$. A likely explanation for this lies in the crudeness of the scattering process from $\nu_{\rm c}$ in the CGL-LF model. In the kinetic simulation, negative pressure anisotropy builds  over time as a result of
the preferentially parallel heating due to Landau damping, seemingly reaching steady state once it triggers firehose scattering across the box (see their figures~3 and 5). 
However, once particle scattering from firehose fluctuations is triggered, rather than driving the plasma towards isotropy, it should be expected to maintain the plasma near the firehose threshold, allowing the system to maintain $\langle \Dp\rangle<0$ even with a relatively high scattering rate. In contrast, the scattering included in our weakly collisional simulations with the  CGL-LF model obviously drives the system towards $\langle \Dp\rangle=0$, making it much harder for the system to maintain 
$\langle \Dp\rangle<0$ in the weakly collisional case.

Given all of these uncertainties,   our approach in this paper has been to explore a range of options within the confines of the fluid model. 
We have found that the pressure-anisotropy stress has a strong influence on the plasma's dynamics across all collisionality regimes, for $\It\lesssim10$ (see, e.g.,~\cref{fig: heating in all sims,fig: beta16 scan}). Thus even if the first scenario applies at astrophysically relevant scale separations, 
magneto-immutability should play an important role in the plasma's dynamics, allowing a nearly conservative 
turbulent cascade to be set up below the forcing scales.  Promisingly, \aksqs\ measured a cascade efficiency
of $55$--$60\%$ (meaning that $55$--$60\%$ of the input energy is processed via the turbulent cascade to kinetic scales), which  compares very well to all of our weakly collisional and Braginskii-MHD simulations with $\It\lesssim10$ (all of which have similar $\mathcal{H}_{\Dp}$; see~\cref{fig: heating in all sims}). Given 
the significant differences in physics and forcing, the correspondence  could hint at ${\simeq}50\%$ being a quasi-universal (or at least minimum) cascade efficiency for weakly collisional plasmas, although an  exploration of more realistic forcing
mechanisms would be  needed to confirm this. 
Overall, the use of the semi-phenomenological fluid model has been both a blessing and a curse: while 
we clearly miss very important kinetic effects, leading to the uncertainties discussed above, the influence of different physics can be 
explored more thoroughly by selectively studying different options, spanning wider parameter ranges,  and comparing with counterfactual scenarios.
%In addition -- perhaps more importantly --
%we have been able to directly compare situations with and without the pressure-anisotropy feedback by artificially 
%removing it from the equations (the passive-$\Dp$ simulations), which is clearly not possible in a kinetic simulation.

% Our collisionless simulations, with $\nu_{\rm c}=0$ in stable regions, are likely not truly applicable, since they would apply
% to a real kinetic system only if the effect of the scattering was strictly confined to regions with large $|\Dp|$.
%  While this may be possible at extremely large scale separations, it is likely an oversimplification.
% For the opposite case, where scattering pervades the entirety of the plasma, one should  
% include the collisionality  
%   $\nu_{\rm c}\sim \beta \omegaa \delta b_{\rm rms}^{2} $ in~\cref{eq: Dp equation} throughout the plasma volume (in addition to the collisionality of the limiters in unstable regions). 

%\textcolor{red}{Indication of how to see this in the solar wind. Compare to Lev's sims with $\beta=16$ scan in $\nu_{\rm c}$. Maybe show some results from $\beta=0.2$ run that heats in time.}

%
%
%
\subsection{Observable consequences and diagnostics}\label{sub: observations}

Much of the evidence for magneto-immutability that we presented in this paper has relied on the direct comparison to the equivalent passive-$\Dp$ or MHD simulations, a diagnostic luxury that is clearly 
not available in observations or even in kinetic simulations. In this section, we use the $\beta16$ simulation set, which scans between the 
 collisionless and MHD-like regimes by changing~$\nu_{\rm c}$, to suggest various methods for diagnosing magneto-immutability in simulations
 or spacecraft data.  
There are several difficulties in this endeavour. First, the general
effect of magneto-immutability is to make the weakly collisional turbulent  plasma look more similar to MHD than it otherwise would; the corollary  
is that those differences that do persist are rather subtle. Secondly, there are other effects in MHD that tend to reduce the variance of $B$, which are
unrelated to pressure anisotropy but very important in the near-Sun (low-$\beta$) solar wind especially. Thirdly, in either the 
solar wind or kinetic simulations, we will often not know \emph{a priori} the effective particle scattering rate \citep[see, e.g.,][]{Hellinger2011,Coburn2022}, making it hard to disentangle the
relative contributions of scattering and magneto-immutability to the reduction in the variance of $|\Dp|$ (see~\cref{subsub: mi and microinstabilities} above).

\subsubsection{Constant-$B$ states in the solar wind -- imbalance and spherical polarisation}

\emph{In-situ} spacecraft measurements in the solar wind and magnetosphere provide us with an unparalleled laboratory for studying fundamental 
plasma physics, and especially collisionless plasma turbulence \citep{Chen2016,Verscharen2019}. Regions with $\beta\gtrsim1$  are regularly observed in the
bulk solar wind, with values as high as $\beta\sim 10^{2}$ in specific regions \citep[e.g.,][]{Cohen2017,Chen2021}, making
it a promising arena for studying the physics of pressure anisotropy and magneto-immutability.
However, when working in this laboratory, we must recognise and consider other possible physics at play. 
In this context, a key point to note is that constant-$B$ `spherically polarised' fluctuations, which 
are observed ubiquitously near  the Sun \citep[e.g.,][]{Belcher1971,Kasper2019}, are most likely \revchng{not related} to pressure anisotropy 
and magneto-immutability. Here we discuss these and other possible mechanisms for reducing the variance of $|\bm{B}|$, which 
should be taken into account when interpreting observational data.

The robustness of spherically polarised  states is related to 
the fact that any perturbation that satisfies
\begin{equation}
  \frac{\delta\bm{B}}{\sqrt{4\upi \rho}} =  \pm\bm{u},\quad B={\rm const},\quad p_{\perp},p_{\|},\rho={\rm const}\label{eq: nl alfven wave}
\end{equation}
is an exact nonlinear solution of the Kinetic MHD system~\eqref{eq:KMHD rho}--\eqref{eq:KMHD pl} (or even the standard MHD equations). Any such solution propagates at velocity $\mp\bm{B}_{0}\sqrt{\fh/4\upi\rho}$, where $\bm{B}_{0}$
is the mean of $\bm{B}$ and $\delta \bm{B}$ is the remainder \citep{Barnes1974}. Since the solution~\eqref{eq: nl alfven wave} holds even if $|\delta \bm{B}|\gg |\bm{B}_{0}|$, these states are the natural nonlinear generalisation of 
 linear Alfv\'en waves to large amplitudes. 
 Their ubiquitous presence in the near-Sun solar wind is not unexpected given that Alfv\'enic perturbations
 grow in amplitude as they propagate outwards from the low corona due to the decreasing Alfv\'en speed \citep[e.g.,][]{Voelk1973a,Hollweg1974}. 
 Given that they are observed consistently when $\beta\ll 1$, which necessarily implies $\It \gg1$ for trans-Alfv\'enic motions, 
 it seems unlikely  that pressure anisotropy is playing an important role in their formation and sustenance (but see \citealt{Tenerani2020a}).
 Indeed, 
most aspects of their evolution can be well described by isothermal MHD \citep{Hollweg1974,Vasquez1998,Squire2020,Mallet2021}.
 
Clearly, as exact nonlinear solutions, such states are not meaningfully turbulent, although there do exist turbulent-like states with very small $B$ variance that
are similar to~\eqref{eq: nl alfven wave} \citep{Dunn2023}.
But in order for such states to occur, the turbulence must be strongly \emph{imbalanced}, with $|\bm{z}_{+}|=|\bm{u}+\bm{B}/\sqrt{4\upi\rho} |\gg |\bm{z}_{-}|=|\bm{u}-\bm{B}/\sqrt{4\upi\rho}|$ (or vice versa). \revchng{This is the norm in the fast solar wind, but is less prevalent 
in the slow wind -- presumably turbulence is also generically less imbalanced in other astrophysical plasmas with less ordered global structure.}
In this paper, we have not considered imbalanced turbulence, taking  $|\bm{z}_{+}| \sim|\bm{z}_{-}|$ in all simulations, a limitation of our study that should be relaxed in future work. At any rate, the basic message is that there exist both MHD-related and pressure-anisotropy-related  effects that 
reduce $B$ fluctuations in turbulence. In order to diagnose the influence of pressure anisotropy by considering the variance of $B$ between different regions, the
imbalance must be taken into account, lest one inadvertently study the prevalence of nearly spherically polarised solutions instead.
\revchng{Similarly, in highly imbalanced high-$\beta$  turbulence, both pressure anisotropy and other reasons (e.g., wave growth in a plasma with a decreasing 
Alfv\'en speed) could cause the plasma to develop constant-$B$  spherically polarised fluctuations. Once formed, 
such fluctuations would not generate significant turbulent pressure anisotropy,
in which case  magneto-immutability might not play such an important role even for $\It\lesssim1$.}

%%%%%%%%%%%%%%%%%%%%%%%%%%%%%%%%%%
\begin{figure}
\begin{center}
\includegraphics[width=0.6\columnwidth]{\ffold/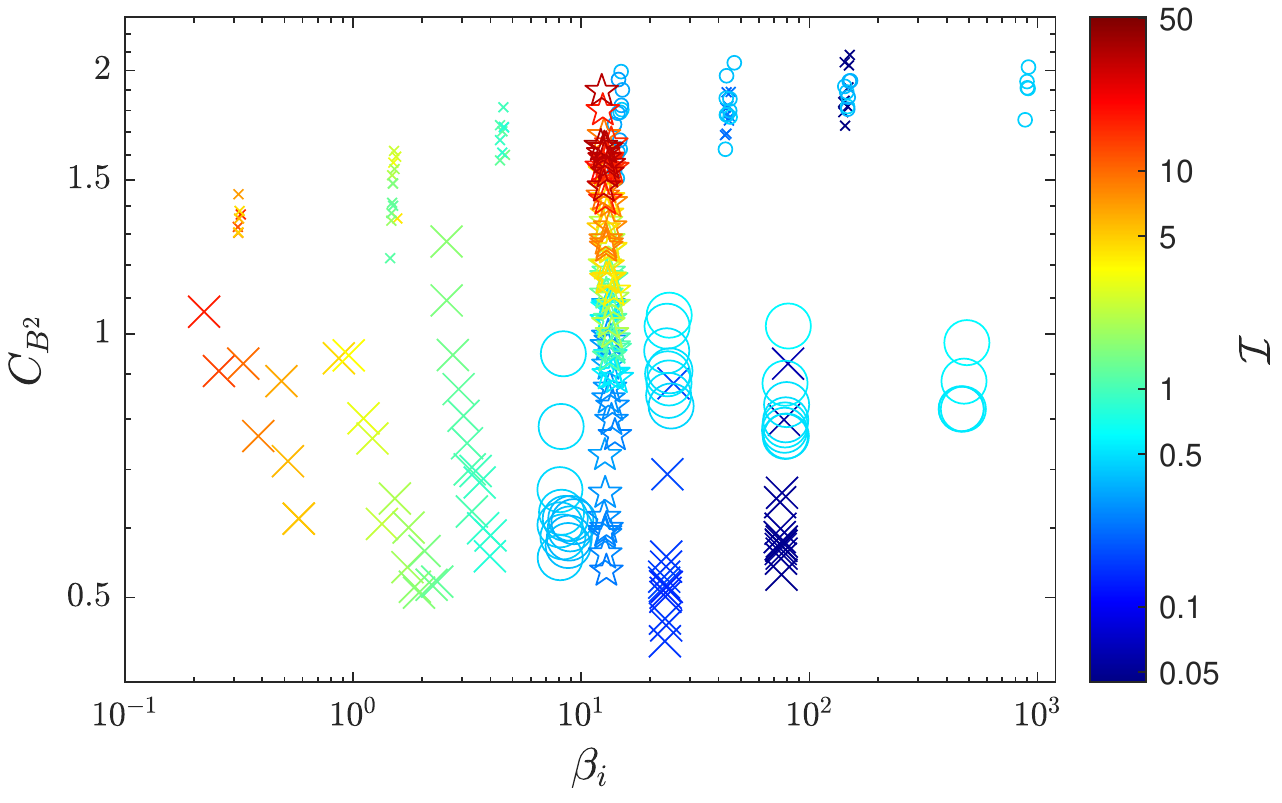}
\caption{The variation in magnetic-field strength, or spherical polarisation, for all simulations, quantified in terms of $C_{B^{2}}$ given by~\cref{eq: cb2}. This measures the relative tendency for the components of $\bm{B}$ to be correlated in order to reduce the variation in $|\bm{B}|$. We plot multiple snapshots for each simulation, with the marker colour showing $\It$, which changes modestly during simulations due to heating and random fluctuations. Marker styles denote the simulation set and the small markers show the equivalent passive-$\Dp$ simulations, with the same styles as in~\cref{fig: heating in all sims}. We see a clear correlation between the magneto-immutability and the interruption number, even within individual simulations, and without any clear dependence on the collisionality regime.  Further, even for $\beta\sim1$, $C_{B^{2}}$ is reduced significantly by pressure-anisotropy effects. However, even the lowest values of $C_{B^2}$ observed here are larger than those regularly observed in the near-Sun solar wind due to 
the predominance of  highly imbalanced, spherically polarised states~\cref{eq: nl alfven wave}.}
\label{fig: CB2}
\end{center}
\end{figure}
%%%%%%%%%%%%%%%%%%%%%%%%%%%%%%%%%%

To demonstrate more directly the reduction in the variance of $B$ across our simulations, 
in~\cref{fig: CB2} we plot
\begin{equation}
C_{B^{2}}=\frac{(\delta B^{2})_{\rm rms}}{ (\delta \bm{B}_{\rm rms})^{2}}=\frac{\langle(|\bm{B}|^{2}-\langle|\bm{B}|^{2}\rangle)^{2}\rangle^{1/2}}{\langle|\bm{B}-\langle\bm{B}\rangle|^{2}\rangle},\label{eq: cb2}
\end{equation}
which was introduced by \citet{Squire2020} as a simple measure of the relative tendency of the components of $\bm{B}$ to become correlated in order to reduce
the variance of $B$. This measure is mostly insensitive to the amplitude of the turbulence (small-amplitude, linearly polarised Alfv\'en waves have $C_{B^{2}}\approx1/2$).\footnote{A number of other reasonable choices exist to quantify features of the spherical polarisation. The similar statistic $C_{B} = (\delta B_{\rm rms})^{2}/(\delta \bm{B}_{\rm rms})^{2} $ decreases with turbulence amplitude, meaning it is more appropriately interpreted as a total magnetic compressibility  \citep{Chen2020}. Another possibility, which may be preferable because it is likely to depend less strongly on the fluctuations' intermittency,  is $\widetilde{C}_{B^{2}} = (\delta B^{2})_{\rm rms}/\langle |\bm{B}-\langle \bm{B}\rangle|^{4}\rangle^{1/2}$. This produces very similar  results to~\cref{eq: cb2}, aside from all points being reduced by a factor ${\approx}2$. } For reference, {\it Parker Solar Probe} (PSP)  observations and highly imbalanced expanding-box MHD simulations have $C_{B^{2}}$ in the range of $0.1$ to $0.3 $ \citep{Squire2020,Johnston2022}. We plot $C_{B^{2}}$ from a number of snapshots of the $\beta16$, \emph{lrCL}, and \emph{lrB} simulation sets, with each point coloured by the interruption 
number $\It$ (this can vary modestly between snapshots in a given simulation). As in~\cref{fig: heating in all sims}, the MHD (passive-$\Dp$) simulations are plotted with small markers, showing the expected higher variance of $B$ for otherwise similar conditions (see~\cref{fig: Dp PDFs brag,fig: Dp PDFs cl}).
The dependence on $\It$ is obvious, with lower-$\It$ simulations generally showing lower $C_{B^{2}}$ at otherwise identical parameters. 
We also see a tendency for modestly lower $C_{B^{2}}$ at lower $\beta$, which is likely related to the 
larger influence of magnetic pressure forces compared to thermal pressure forces at low $\beta$ \citep{Vasquez1998}.
However, also of note is that all simulations, including those where pressure-anisotropy forces are very strong (low $\It$), show values of
$C_{B^{2}}$ that are rather large compared to what is routinely observed in the solar wind or imbalanced turbulence \citep{Kasper2019,Squire2020}.
It seems that magneto-immutability, while an important influence on the evolution of $B$, can never drive balanced turbulence arbitrarily 
close to keeping  $B$ truly constant across the domain -- imbalanced spherically polarised solutions are much better at this task.

\subsubsection{Spectral diagnostics}

%%%%%%%%%%%%%%%%%%%%%%%%%%%%%%%%%%
\begin{figure}
\begin{center}
\includegraphics[width=0.46\columnwidth]{\ffold/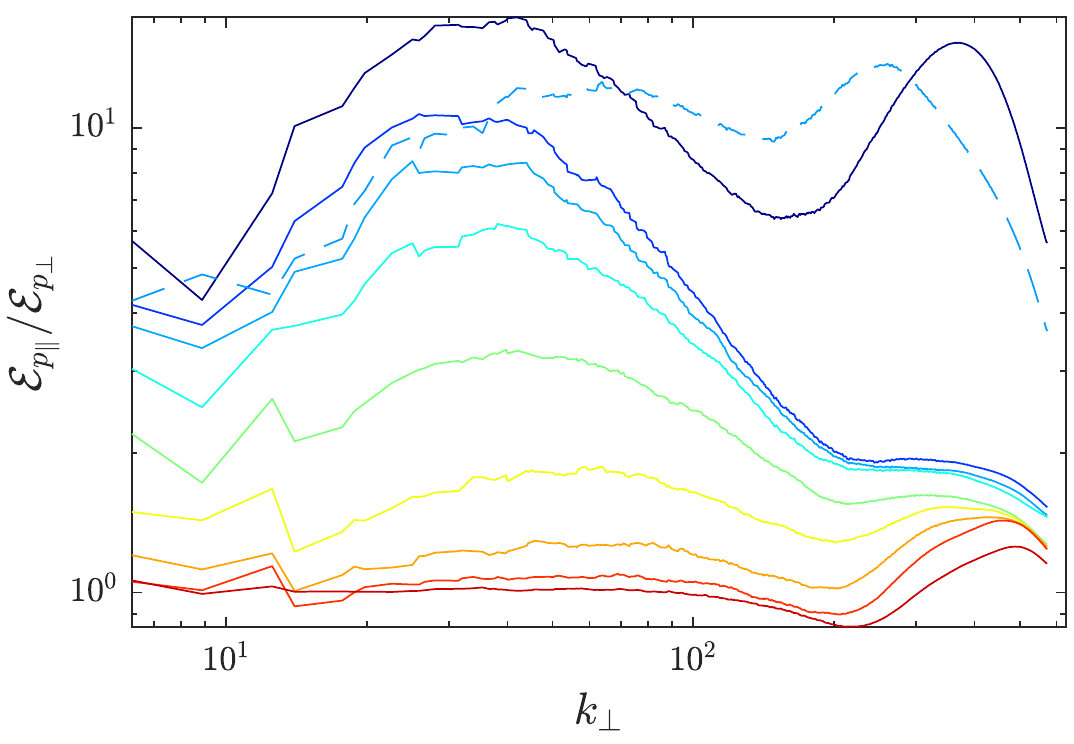}~~\includegraphics[width=0.542\columnwidth]{\ffold/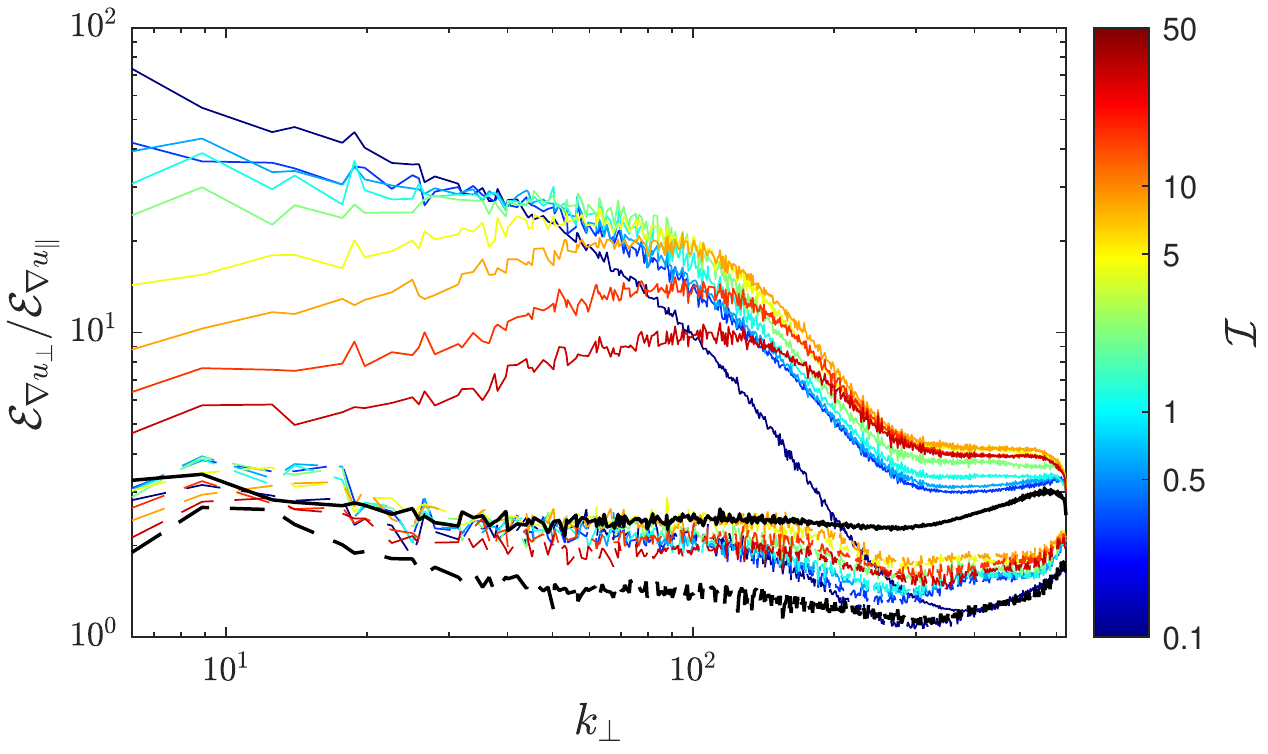}
\caption{
\revchng{The 
left panel shows the ratio of pressure spectra $\mathcal{E}_{p_{\|}}/\mathcal{E}_{p_{\perp}}$, a simple diagnostic of magneto-immutability that should 
be readily observable in the solar wind.} We show the simulations from the $\beta16$ simulation set  (see~\cref{sub:study design} and~\cref{tab:sims}), with the line colour showing $\It$, and the black lines showing the corresponding results for isothermal MHD. 
The dashed line shows the \emph{lrB}600, $\beta_{0}=600$ Braginskii-MHD simulation, to demonstrate that the
observed decrease in $\delta p_{\|}/\delta p_{\perp}$ is not due to the change from the weakly collisional to the Braginskii-MHD regime at high $\It$ and $\beta_{0}=16$.
The right panel shows the ratio of the spectra of $u_{\|}$ and $u_{\perp}$ in the parallel and perpendicular direction; $\mathcal{E}_{\nabla_{\|}u_{\perp}}/\mathcal{E}_{\nabla_{\|}u_{\|}}$ is shown with solid lines, and $\mathcal{E}_{\nabla_{\perp}u_{\perp}}/\mathcal{E}_{\nabla_{\perp|}u_{\|}}$ is shown with dashed lines (see~\cref{fig: rates of strain} and~\cref{subsub: ros spectra defs}).  The increase in $\nabla_{\|}u_{\perp}/\nabla_{\|}u_{\|}$ with decreasing 
$\It$ shows the change in the structure of the flow that results from the turbulence  becoming more magneto-immutable at smaller collisionality.
The lack of the same change in  $\nabla_{\perp}u_{\perp}/\nabla_{\perp|}u_{\|}$ allows the effect 
to be considered separately from $u_{\|}$ itself, which may depend on the forcing and other parameters. } \label{fig: beta16 scan}
\end{center}
\end{figure}
%%%%%%%%%%%%%%%%%%%%%%%%%%%%%%%%%%

In~\cref{sec: results}, we saw that, while kinetic- and magnetic-energy spectra remained rather 
similar in magneto-immutable turbulence compared to MHD, the  rate-of-strain spectra and pressure anisotropy spectra
did not. The $\beta16$ simulation set provides a useful scan from low-$\It$ (small $\nu_{\rm c}$) to large-$\It$ collisional MHD, 
while other parameters ($\ma$ and $\beta$) remain  similar. In~\cref{fig: beta16 scan} we plot two interesting ratios of spectra 
that  clearly show the change in the structure of the turbulence at low $\It$.
\revchng{In the left panel,} we show  the ratio of parallel- to perpendicular-pressure spectra, \revchng{which should be readily 
observable with \emph{in-situ} solar-wind measurements}. As discussed in~\cref{sub: flow rearrangement results}, it
becomes  larger than unity in magneto-immutable turbulence because of perpendicular pressure balance.
We see the clear trend with~$\It$: the higher-collisionality simulations ($\nu_{\rm c}/\omegaa\gtrsim50$, $\It\gtrsim 8$) all collapse down 
to $\delta p_{\|}\sim \delta p_{\perp}$, because the pressure anisotropy ceases to have a  strong dynamical 
influence. In addition, the dashed line shows the \emph{lrB}600 simulation, which is in the Braginskii regime but with $\It\approx0.4$.
This verifies that the observed change is not related to a transition between collisionality regimes (collisionless, weakly collisional, or Braginskii MHD).
A similar diagnostic could be constructed from the slopes of the pressure spectra: in low-$\It$ turbulence, 
$\mathcal{E}_{p_{\perp}}$ flattens with scale while $\mathcal{E}_{p_{\|}}$ steepens, with the two spectra converging at small scales (see~\cref{fig: dp spectra}); in contrast, when unaffected by the pressure-anisotropy feedback, the $p_{\perp}$ and $p_{\|}$ spectra 
have similar shapes.

 Since we saw that the reduction 
in $\nabla_{\|}u_{\|}$ was one of the more obvious consequences of the pressure anisotropy feedback, the right panel compares   
the ratios of parallel-flow gradients $\mathcal{E}_{\nabla_{\|}u_{\perp}}/\mathcal{E}_{\nabla_{\|}u_{\|}}$ (solid lines) to 
those of perpendicular-flow gradients $\mathcal{E}_{\nabla_{\perp}u_{\perp}}/\mathcal{E}_{\nabla_{\perp}u_{\|}}$ (dashed lines).
While the latter ratio is almost constant as a function of $\It$ and looks very similar to MHD (black lines), the former varies by a factor of at least 
${\sim}10$ as the plasma transitions to collisional MHD. We also see that 
even at the highest collisionality ($\nu_{\rm c}/\omegaa=200$, $\It\approx50$), there remain clear differences in the flow structure compared to MHD.

In summary, these diagnostics could provide a useful way to understand and measure the influence of pressure-anisotropy feedback
in more realistic simulations or observations.\footnote{Rate-of-strain spectra are not possible to measure with
a single spacecraft, but could be computed from data taken by multi-spacecraft constellations such as the upcoming NASA MidEx mission HelioSwarm \citep{Klein2021a}.} While the exact values of these metrics will depend on details (e.g., the forcing, $\ma$, and the imbalance), we expect that the qualitative features should be robust, especially the fact that $\mathcal{E}_{\nabla_{\|}u_{\perp}}/\mathcal{E}_{\nabla_{\|}u_{\|}}\gg \mathcal{E}_{\nabla_{\perp}u_{\perp}}/\mathcal{E}_{\nabla_{\perp}u_{\|}}$ and $\delta p_{\|}\gg \delta p_{\perp}$.
This may allow one, for example, to diagnose and understand a transition to $\It\gg1$ turbulence at small scales (as seen, e.g.,  for \emph{CL10} in~\cref{fig: dp spectra}).

\section{Conclusions}\label{sec:conclusions}

In this paper, we have studied the influence of fluctuation-generated pressure anisotropy ($\Dp$) on magnetised plasma turbulence. The simulations and theory are based on 
a CGL-Landau-fluid model, which  is effectively drift kinetics \citep{Kulsrud1983} with  heat fluxes approximated by a fluid closure  that captures linear parallel Landau damping \citep{Snyder1997}. This
makes  it a reasonable model for collisionless plasma dynamics on scales far above the proton gyroradius. Indeed, our results compare well to the recent hybrid-kinetic simulations  of \aksqs, which capture a much 
more comprehensive array of kinetic processes at the price of having a limited inertial range. 
Our results focus on the $\beta>1$ regime, where small changes of the magnetic-field 
strength lead to  large pressure-anisotropy stresses on the plasma. This regime is relevant to many
hot and diffuse astrophysical plasmas, including the intracluster medium and regions of the solar wind. 

We show numerically, and argue theoretically, that the primary effect of the pressure-anisotropy stress on the fluid 
is  to reduce its own influence via a dynamical feedback. This makes high-$\beta$ weakly collisional turbulence behave more like standard collisional MHD than would be na\"{i}vely 
expected, aside from a few important differences. The effect is straightforwardly understood by analogy with 
the origin of incompressibility in fluids, which results from the large $-\grad p$ force   that rapidly opposes any flow with $\grad\bcdot\bm{u}\neq0$ attempting to change the fluid's pressure. With pressure and density tied together, this feedback eliminates density fluctuations. Analogously, 
there is a large force from $\grad\bcdot(\bh\bh\Dp)$ that rapidly opposes any flow with $\bbgu\neq 0$ attempting to change the pressure anisotropy; 
 since $\Dp$ and $B$ are tied together, the feedback minimises  magnetic-field-strength fluctuations in the turbulence. 
In the same way that the Mach number of a hydrodynamic flow quantifies the importance of pressure compared to inertial 
forces,  the `interruption number' $\It$ (introduced in \skqs) quantifies the importance of pressure-anisotropic forces  compared to magnetic tension 
(see~\cref{subsub: interruption number}). 
The most important consequence of this physics is to reduce viscous heating from pressure  anisotropy,
confining it to a small range of scales near the outer scale  so that most of the cascade remains nearly conservative.
This increase in `cascade efficiency' will in turn increase the fraction of heating that occurs via kinetic processes near and below the
ion gyroscale \citep{Schekochihin2009}, thus influencing the thermodynamics of the plasma.

Compared to the previous work on Braginskii-MHD Alfv\'enic turbulence (\skqs; \citealp{Kempski2019}), the results here explore plasmas in more realistic regimes at more modest $\beta$ and lower collisionality. \revchng{Our results highlight the resilience of the pressure-anisotropy feedback mechanism:}
despite the much more complex evolution of the pressure 
anisotropy in the regimes studied here, particularly the effect of heat fluxes that act quasi-diffusively to smooth out $\Dp$, we find
that the basic features of magneto-immutable turbulence are robust. Our detailed analysis is aided  by  direct comparisons to `passive-$\Dp$' simulations, which are identical in setup but artificially remove the
feedback of the pressure anisotropy on the flow. This allows us to compare, for example, the pressure-anisotropy distribution and viscous heating to the counterfactual 
situation in which $\Dp$ evolved in standard MHD turbulence. 
A familiar illustration of this comparison is given in~\cref{fig: Brazil plot}, where we plot the classic `Brazil plot' of the joint PDF of $\beta$ and pressure anisotropy \citep[e.g.,][]{Kasper2002,Hellinger2006}. 
In the CGL-LF simulations, the pressure anisotropy is naturally constrained close to zero by magneto-immutability; in the passive-$\Dp$ simulations, 
most of the volume sits instead at the microinstability boundaries, enforced here by artificial limiters.

A primary result of this study is that magneto-immutable turbulence
behaves surprisingly similarly to standard MHD, \revchng{illustrating the fundamental robustness of the MHD turbulent cascade. Specifically, 
despite the system having to rearrange itself to avoid dissipating most of the injected energy, the simulations show that}
saturated fluctuation amplitudes are similar to MHD for identical forcing, and the magnetic- and kinetic-energy spectra have slopes 
close to the expected  $k_{\perp}^{-3/2}$ or $k_{\perp}^{-5/3}$.
This has the effect of making the influence of pressure anisotropy rather subtle: rather than it
causing some dramatic, obvious modifications to the system, magneto-immutability 
interferes and causes such turbulence to resemble MHD rather closely.
That said, there are important differences that enable this behaviour, particularly that the system 
significantly suppresses the spectrum of the parallel strain $\bbgu$ (see~\cref{fig: rates of strain}), 
which reduces the magnitude of the $\Dp$ and $B$ fluctuations (e.g.,~\cref{fig: Dp PDFs cl,fig: CB2}) 
and steepens the pressure spectra, leaving a characteristic signature with $\delta \Dp\sim \delta p_{\|}\gg \delta p_{\perp}$ (\cref{fig: dp spectra}).
These changes are summarised in~\cref{fig: beta16 scan}, which shows how turbulence transitions from being magneto-immutable (low $\It$, blue)
to collisional MHD (high $\It$, red and black), \revchng{demonstrating how such effects could be diagnosed and studied with \emph{in-situ} solar-wind observations (see \cref{sub: observations}).}

Macroscopically, the most important effect of magneto-immutability is to  
suppress strongly the viscous heating that would otherwise almost completely damp the turbulent cascade. This leaves an effectively conservative    cascade below the forcing scales that  processes ${\approx}50\%$ to  ${\approx}80\%$ of the input energy (depending on the regime; see~\cref{fig: transfer spectra,fig: heating in all sims}). 
\revchng{While the particular values for the cascade efficiency will depend on the forcing scheme, so will certainly differ
for more realistic systems (e.g., those driven by large-scale instabilities), the general point is that
the magneto-immutability modifies}  the conversion between  mechanical and thermal energy in the plasma. This
 could have interesting consequences for its macroscopic thermodynamics. For example, important quantities like the fraction of turbulent energy that heats ions versus electrons, or 
in the perpendicular versus parallel directions, can depend on the cascade efficiency and viscous heating. 
A specific application where such physics could have direct consequence is the intracluster medium, in which turbulent heating is thought to offset a large fraction of radiative losses to mitigate cluster cooling flows \citep[e.g.,][]{Churazov2004,Zhuravleva2014}. \citet{Kunz2010} suggested that, by forcing the plasma to adjust its parallel rate of strain,  pressure anisotropy could regulate turbulent heating to maintain naturally the plasma's thermal stability against Bremsstrahlung cooling (with viscous heating ${\propto}B^4 T^{-5/2}$ at fixed gas pressure increasing faster than cooling ${\propto}T^{-3/2}$ as the temperature drops). While the details of their scheme differ from the turbulent heating studied here (specifically, \citealt{Kunz2010} assumed that kinetic microinstabilities provide the primary regulation mechanism by contributing to, and thereby regulating, the total parallel rate of strain), the 
basic idea -- that suppression of the parallel rate of strain provides a strong constraint on the heating rate -- is analogous to magneto-immutability. Thus, similar thermal-stability considerations may apply to magneto-immutable turbulence, with the modification that it is the large-scale (as opposed to the small-scale) adjustment to the parallel rate of strain that regulates the viscous heating. At the 
same time, the mechanism would allow some of the injected energy (the cascade efficiency) to pass 
into a turbulent cascade, as  needed 
to explain observations of small-scale fluctuations on scales below the Coulomb mean-free path \citep{Zhuravleva2019,Li2020a}. The effectiveness 
of such a mechanism likely depends on how fluctuations are driven (the forcing), as well as complications such as anomalous particle scattering (see \cref{subsub: mi and microinstabilities}), so further study in more realistic settings is clearly needed.

Due to the {\it ad-hoc} nature of some approximations that were needed for the CGL-LF model, 
our study is beset with some serious (but, we think, ultimately not game-changing) uncertainties. The most important one is the influence of kinetic microinstabilities, e.g., the firehose and
mirror instabilities, which grow extremely rapidly compared to  large-scale motions in regions of large $|\Dp|$.
We use simplified hard-wall limiters \citep{Sharma2006}, but it is clear from kinetic simulations that these could miss important effects
such as long-lived fluctuations that continue to scatter particles even in stable regions.
In this context, our study complements the recent hybrid-kinetic simulations of \aksqs, which 
capture all such complexities (except the electron physics), but suffer from  
unavoidable limitations  related to low scale separation. This complicates the exploration 
of the inertial-range $k_{\perp}\rho_{i}\ll1$ turbulence in the system, both because the microinstabilities' growth and
saturation are not particularly fast compared to the integral-scale motions, and because $\rho_{i}$-scale 
physics (e.g., the damping of perturbations; \citealp{Foote1979}) can start to interfere directly with scales comparable to the
size of the box. Nonetheless, 
our comparable simulations that have  an imposed collisionality to approximate the effects of microinstabilites (see~\cref{subsub: mi and microinstabilities}), provide a good match to most features observed  in \aksqs. This  
includes most spectra of Alfv\'enic and compressive quantities, $\Dp$ and $B$ distributions (aside from a mean $\langle \Dp\rangle<0$ in \aksqs), and the cascade efficiency, which 
is  ${\simeq}55\%$ in \aksqs\ (matching effectively all of the weakly collisional and Braginskii-MHD simulations; see~\cref{fig: heating in all sims}). 
Thus, a secondary result of our work is that the CGL-LF model provides an accurate and relatively simple way 
to model weakly collisional plasmas in high-$\beta$ regimes. That said, there remain significant caveats, and
 working to build a better  bridge between the regimes accessible via the 
CGL-LF and hybrid-kinetic approaches should be a priority for future work.

\acknowledgements  
We thank  Chris Chen, Ryan Davis,  Sergey Komarov, Stephen Majeski, and Romain Meyrand   for useful and illuminating discussions, 
\revchng{as well as the anonymous referees who provided helpful comments that improved the presentation of this article.}
Support for  J.S. was
provided by Rutherford Discovery Fellowship RDF-U001804, which is managed through the Royal Society Te Ap\=arangi. M.W.K. was supported in part by NSF CAREER Award No.~1944972. Support for L.A. was provided by the Institute for Advanced Study. The work of A.A.S. was supported in part by grants from STFC (ST/W000903/1) and EPSRC (SP/R034737/1), as well as by the Simons Foundation via a Simons Investigator award.  EQ was also supported in part by a Simons Investigator award from the Simons Foundation. We wish to acknowledge the use of New Zealand eScience Infrastructure (NeSI) high performance computing facilities as part of this research. New Zealand's national facilities are provided by NeSI and funded jointly by NeSI's collaborator institutions and through the Ministry of Business, Innovation \& Employment's Research Infrastructure programme. We also wish to acknowledge the generous hospitality of the Wolfgang Pauli Institute, Vienna, where these ideas were discussed during several `Plasma Kinetics' workshops.

\appendix

%\label{----------------- APPENDIX STARTS HERE}

\section{Numerical method and validation}\label{app: numerical tests }

In this appendix we discuss various features of the numerical implementation of our equations. 
We first  give the relevant equations and other results that are needed
for the conservative part of the equations (i.e., for the CGL equations;~\cref{app:sub: cgl equations}), followed by those 
needed for the heat fluxes and collisions (\cref{app: heat fluxes and collisions}). 
We then present some simple wave tests  and discuss a collection of numerical problems  
that arose for this system, speculating on their possible causes. As part of this discussion,
we present a more accurate Riemann solver that works well for some problems, but has turned out to be too 
unreliable for large-amplitude (large-$\delta B_{\perp}$) turbulence simulations.

\subsection{Conservative form}\label{app:sub: cgl equations}

As is standard for finite-volume implementations, we solve~\cref{eq:KMHD rho}--\cref{eq:KMHD pl} in the conservative
form 
\begin{equation}
\pD{t}{\bm{U}} + \pD{x}{\bm{F}_{x}}  + \pD{y}{\bm{F}_{y}}  + \pD{z}{\bm{F}_{z}} + \grad\bcdot\bm{Q}= 0,\label{eq:conserv.form}
\end{equation}
where
\begin{equation}
\bm{U}=\begin{pmatrix} \rho \\ m_{x} \\ m_{y} \\ m_{z} \\ E \\ \an \\ B_{x} \\ B_{y}\\ B_{z}\end{pmatrix} 
= 
\begin{pmatrix} \rho \\ \rho u_{x} \\ \rho u_{y} \\ \rho u_{z}\\ p_{\perp} + \frac{1}{2}p_{\|} + \frac{1}{2}B^{2} + \frac{1}{2} \rho |\bm{u}|^{2} +p_{e}\ln p_{e}\\ \rho \ln (p_{\perp} \rho^{2}/p_{\|}B^{3}) \\ B_{x} \\ B_{y}\\ B_{z}\end{pmatrix};\label{eq:u.def} 
\end{equation} 
%
% Note mu/\rho, or p_{\|}B^{2}/rho^{3} are both passively advected by CGL, so their ratio is also, which is A. A passively advected quantity becomes a conserved quantity when multiplied by $\rho$. So, ln(A) is also passively advected, and rho*lnA is conserved. Could also use rho*A, which is p_{\perp}/p_{\|}*rho^{3}/B^{3} -- but I never really tried this... Also worth trying might be rho/A, which wouldn't have a singularity at B->0?
 $\bm{F}_{x}$, $\bm{F}_{y}$, and $\bm{F}_{z}$ are the CGL fluxes; and $\bm{Q}$ are the heat fluxes, which are defined below (for simplicity of notation, we normalise $\bm{B}$ by $\sqrt{4\upi}$ through this section). In the limit of $T_e=0$, the $p_e\ln p_e$ contribution to $E$ disappears. There is a fundamental degeneracy in 
the choice of the second conserved thermodynamic variable (in addition to energy), since both $\mu = p_{\perp}/\rho B$ and $\mathcal{J}=p_{\|}B^{2}/\rho^{3}$ are passively advected by CGL dynamics. Our choice of the logarithm of their ratio  $\an$ was inspired by \cite{Santos-Lima2014} and also by extensive numerical testing of solvers that use $\rho\mu$ instead, which
can become highly unstable for certain types of problems. We suspect the superiority of $\an$ compared to $\mu$ is related to the fact that $\an$ treats $p_{\perp}$ and $p_{\|}$ on approximately equal footing, rather than assigning artificial importance to one over the other (see further discussion below). Note that, in principle, $\rho$ multiplied by any function of $\mu$ and/or $\mathcal{J}$
could be used as the conserved variable, and it is possible that there exist other options with better numerical properties.

The fluxes are 
\begin{equation}
\bm{F} = \begin{pmatrix} \rho u_{x}      \\      \rho u_{x}^{2} + P^{\star} - B_{x}^{2}\fh \\
\rho u_{x} u_{y} - B_{x} B_{y} \fh  \\     \rho u_{x} u_{z} - B_{x} B_{z} \fh  \\ 
(E+P^{\star} )u_{x} - B_{x}(\bm{u}\bcdot\bm{B}) \fh   \\ u_{x}\an \\
0 \\ B_{y} u_{x}-B_{x}u_{y}    \\    B_{z} u_{x}-B_{x}u_{z} \end{pmatrix}, \label{eq:fluxes}
\end{equation}
with the obvious permutation of $x$, $y$, and $z$ used to get $\bm{G}$ and $\bm{H}$.
Here $P^{\star}\doteq p_{\perp} + B^{2}/2 + p_{e}$
%See derivation in the `CGL conservation laws and electron pressure' note for discussion of ln(pe) term, derivation of the conservative form, and derivation of the matrix.
 and 
$\fh \doteq 1+ {\Dp}/{B^{2}}$ is the anisotropy parameter.

\subsubsection{Heat fluxes and collisions}\label{app: heat fluxes and collisions}

A downside of using $\an$, as opposed to $\mu$, is that the heat fluxes take a rather complex form that does not seem possible to write as a total divergence $\grad\bcdot\bm{Q}$. 
% Specifically, I don't think you can write the parallel invariant heat fluxes as a total divergence, starting from equation 7 of Sharma 2006. Is this telling us something fundamental about why magnetic moment is more important? perhaps... Just that magnetic moment is globally conserved by the actual (with HFs) collisionless dynamics, unlike p_{\|}
In order to maintain this property, thus simplifying the numerical implementation of the heat fluxes, we  transform $\an$ to $\mu$, add the heat fluxes 
to $e$ and $\mu$, then transform back to $\an$. For this purpose, the $e$-component of the heat flux is $Q_{e}=\hat{\bm{b}}(q_{\perp}+q_{\|}/2)$, while the $\mu$-component is $Q_{\mu}=\hat{\bm{b}}q_{\perp}/B$, where $q_{\perp}$ and $q_{\|}$ are given by~\cref{eq:GL heat fluxes p,eq:GL heat fluxes l}, respectively (all other components of $\bm{Q}$ are zero). Further information is provided in~\cref{sub: methods cgl lf}.

Collisions, including microinstabilities, are evaluated implicitly on the primitive variables  $p_{\perp}$ and $p_{\|}$ at the end of each global time step. This is done via the exact solution of $\partial_{t} p_{\perp} = -(\nu_{\rm c}/3)\, \Dp$, $\partial_{t} p_{\|} = (2\nu_{\rm c}/3)\, \Dp$ across time $\delta t$:
\begin{gather}
p_{\perp}(t+\delta t) = \frac{1}{3}(2+\rme^{-\nu_{\rm c}\delta t})p_{\perp}(t) +  \frac{1}{3}(1-\rme^{-\nu_{\rm c}\delta t})p_{\|}(t),\nonumber\\
p_{\|}(t+\delta t) = \frac{2}{3}(1-\rme^{-\nu_{\rm c}\delta t})p_{\perp}(t) +  \frac{1}{3}(1+2\rme^{-\nu_{\rm c}\delta t})p_{\|}(t).\label{eq: pprp pprl coll soln}
\end{gather}
The microinstability limiters are also implemented implicitly, via the first-order in $\delta t$ solution of $\partial_{t} p_{\perp} = -(\nu_{\rm c}^{\rm lim}/3)\, (\Dp-\Lambda_{\rm MI} B^{2}/8\upi)$, $\partial_{t} p_{\|} = (2\nu_{\rm c}^{\rm lim}/3)\, (\Dp-\Lambda_{\rm MI} B^{2}/8\upi)$, applied only in regions
where $\Dp$ lies beyond the relevant instability threshold (here $\Lambda_{\rm MI}=-2$ and $\Lambda_{\rm MI}=1$ for the firehose and mirror thresholds, respectively). The methods relax $\Dp$ back to isotropy or to the relevant instability threshold accurately and with no 
constraint on the time step, meaning that adiabatic MHD can be recovered by setting $\nu_{\rm c}\rightarrow\infty$, and true `hard wall' limiters are recovered by setting 
$\nu_{\rm c}^{\rm lim}\rightarrow\infty$.

\subsubsection{Dispersion relation}\label{appsub: dispersion relation}

The HLL Riemann solver and its relatives \citep{Toro2009}, which are used to approximate the solutions of~\cref{eq:conserv.form} across cell boundaries, require the wave speeds for the hyperbolic double-adiabatic system ($\bm{Q}=0,\,\nu_{\rm c}=0$). 
Assuming a mode of wavenumber $\bm{k} = k \bm{\hat{x}}$ and linearizing 
about the  equilibrium, $\rho=\rho_{0}$, $\bm{u}=u_{x}\bm{\hat{x}}$, $p_{\perp}=p_{\perp0}$, $p_{\|}=p_{\|0}$, and $\bm{B} = (B_{x}, B_{y},0)$, one
finds  the set of eight eigenvalues \citep{Baranov1970,Meng2012},
\begin{gather}
k^{-1}\omega_{C} = u_{x}, \qquad k^{-1}\omega_{\Dp} = u_{x},\label{eq:entropy.omega}   \\
k^{-1}\omegaa = u_{x} \pm \left(\frac{\fh_{0}}{\rho_{0}}\right)^{1/2}B_{x},\label{eq:alfven.omega} 
\end{gather}
\begin{align}
k^{-1}\omega_{{\rm MS}\pm} =  u_{x} & \pm \left(\frac{1}{2\rho_{0}}\right)^{1/2 } \Bigg\{ 2p_{\perp0} + B^{2}+\hat{b}_x^2 (2p_{\|0}-p_{\perp0}) + T_{e}\rho_{0}  \nonumber \\
&\pm \bigg[ \left(2p_{\perp0} + B^{2}+\hat{b}_x^2 (2p_{\|0}-p_{\perp0} ) + T_{e}\rho_{0}\right)^{2} + 4 p_{\perp0}^{2}\hat{b}_x^2(1-\hat{b}_x^2)  \nonumber \\
&\qquad \:\:-12p_{\perp0 }p_{\|0} \hat{b}_x^2(2-\hat{b}_x^2) + 12p_{\|0}^{2}\hat{b}_x^{4}-12p_{\|0}B^{2}\hat{b}_x^2\bigg]^{1/2}\Bigg\}^{1/2},\label{eq:fast.and.slow.om}
\end{align}
where $\fh_{0} = 1 + (p_{\perp0} - p_{\|0})/B^{2}$. Here $\omegaa$ is like the MHD 
shear-Alfv\'en wave, modified by the pressure anisotropy, while $\omega_{C}$ and $\omega_{\Dp}$ are two entropy-like waves. One of these entropy waves ($\omega_{C}$) involves only density perturbations and no $\Dp$, like the MHD entropy mode, but is strongly damped by heat fluxes.  The other mode ($\omega_{\Dp}$) involves balanced perturbations to $p_{\perp}$, $p_{\parallel}$, $\rho$, and $B$ in general and becomes the gyrokinetic non-propagating mode with the inclusion of heat fluxes \citep{Howes2006}. 

The CGL magnetosonic waves $\omega_{{\rm MS}+}$ and $\omega_{{\rm MS}-}$, which correspond to the $+$ and $-$ on the second line of~\eqref{eq:fast.and.slow.om}, respectively, are compressive waves that also involve perturbations to the magnetic field. In general, for $u_{x}=0$, $|\omega_{{\rm MS}+}|\geq |\omega_{{\rm MS}-}|$ and $|\omega_{{\rm MS}+}|\geq |\omegaa|$, but for $\beta\gtrsim 1$,  $|\omega_{{\rm MS}-}|>|\omegaa|$, meaning that CGL `slow' waves propagate faster than shear-Alfv\'en waves, unlike  in standard MHD. 
As discussed in~\cref{subsub: compressive theory}, the properties of obliquely propagating CGL slow waves are potentially of interest to explain
the  compressive features of the cascade. A key characteristic is their perpendicular pressure balance, $\delta p_{\perp}+\delta (B^{2})/8\upi\approx 0$  or $ \delta p_{\perp} \approx -  (B_{0}^{2}/4\upi) \delta B_{\|}/B_{0} $ (just like the MHD slow
mode), which is derived by applying the standard RMHD ordering \citep{Schekochihin2009} to the perpendicular part of the momentum  equation~\cref{eq:KMHD u} (with $T_{e}=0$). Then, using the linearised form of the CGL invariants ($d\mu/dt=d\mathcal{J}/dt=0$) yields $\delta p_{\perp}/p_{0} - \delta \rho/\rho_{0} - \delta B_{\|}/B_{0} = 0 $ and $\delta p_{\|}/p_{0} - 3\delta \rho/\rho_{0} +2 \delta B_{\|}/B_{0} = 0 $, which 
can be combined with the $\delta p_{\perp}$ constraint to yield $\delta p_{\|} = -(5+6/\beta)\delta B_{\|}/B_{0}$.
Thus we see that $\delta p_{\|}/\delta p_{\perp} =  (5\beta+6)/2 \gg1$, a strong dominance of parallel pressure, as observed in~\cref{fig: dp spectra} (although this prediction is much stronger than that observed). Then, because $\delta p_{\|}$ provides the restoring force for the wave and it is not constrained by $\delta p_{\perp}$ (unlike
MHD, where $p_{\perp}=p_{\|}$), the frequency scales with the thermal speed, becoming $\omega_{{\rm MS}-}^{2} = (5/2)k_{\|}^{2}p_{0}/\rho_{0} $ for $\beta\gg1$. In the presence of heat fluxes, this general structure is maintained, although the wave becomes strongly damped at a rate that
is approximately half that of its propagation frequency. Weak collisions with $\nu_{\rm c}\lesssim k_{\|}v_{\rm th}$ cause even 
stronger damping by weakly coupling together $\delta p_{\perp}$ and $\delta p_{\|}$;  the mode becomes non-propagating around $\nu_{\rm c}\sim k_{\|}v_{\rm th}$ before reverting towards the MHD slow mode for $\nu_{\rm c}\gg k_{\|}v_{\rm th}$ \citep[see][]{Majeski2023}.

Full expressions for the eigenvectors for all angles are inconveniently complex and not important for our purposes, so we do not list them here.

\subsection{Linear-wave convergence }
%%%%%%%%%%%%%%%%%%%%%%%%%%%%%%%%%%
\begin{figure}
\centering
\includegraphics[width=0.52\columnwidth]{\ffold/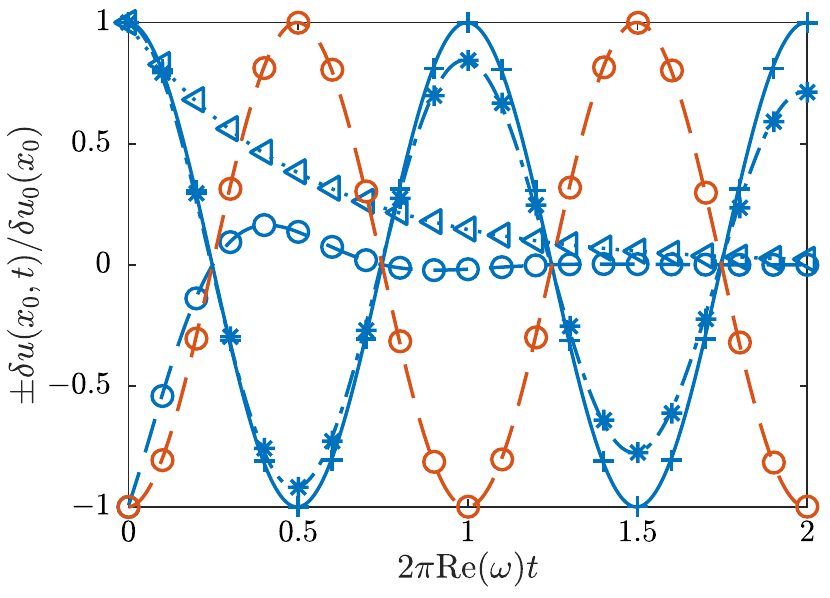}~~~~~~\includegraphics[width=0.46\columnwidth]{\ffold/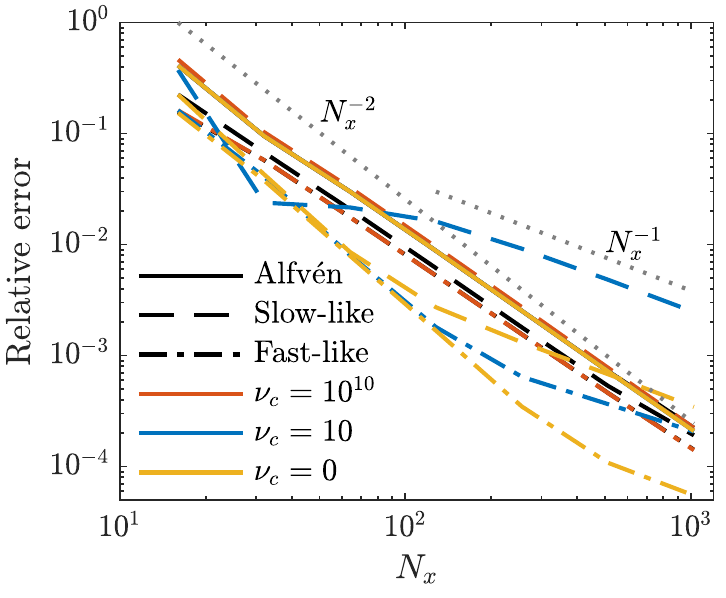}
\caption{Propagation of oblique linear waves within the \textsc{Athena++} CGL Landau-fluid implementation in one spatial dimension. 
Background parameters are $p_{\perp}=p_{\|}=5$, $\rho=1$, $\bm{u}_{0}=0$, and $\bm{B}_{0}=(1,\sqrt{2},0.5)$, with $\bm{k}=2\upi \hat{\bm{x}}$ and $k_{L}=|\bm{k}|$.
The left panel shows the time evolution of $\rho u_{y}(x_{m},t)$, where $x_{m}$ is chosen so that $\rho u_{y}(x_{m},0)$ is the maximum of the sinusoidal initial condition. Results are normalised to the initial amplitude of ${\simeq} 10^{-4}$. The curves show the expected linear solution, 
computed analytically from the dispersion relation~\eqref{eq:entropy.omega}--\eqref{eq:fast.and.slow.om}, while symbols show results from \textsc{Athena++} with $N_{x}=1024$. Blue lines/symbols show $\nu_{\rm c}=10$, with the solid line (plus symbols), dashed line (circle symbols), dot-dashed line (star symbols), and dotted line (triangle symbols) corresponding to the Alfv\'en, slow-like, fast-like, and entropy-like modes, respectively. The red dashed curve (circle symbols) shows the slow mode at  $\nu_{\rm c}=10^{10}$ to illustrate how the system 
reverts to undamped adiabatic MHD  at high $\nu_{\rm c}$. 
The right panel shows the normalised root-mean-squared error in the solution, as measured from the analytic solution,
as a function of spatial resolution $N_{x}$. Line styles and colours are as in the left panel, with the addition 
of yellow curves for collisionless modes ($\nu_{\rm c}=0$) and black for pure CGL (no heat fluxes; $k_{L}\rightarrow\infty$). 
We see almost second-order convergence for most solutions, except for those that are dominated by damping from heat fluxes, in particular the collisionless/weakly collisional slow mode (blue and yellow dashed lines; the former is overdamped $|{\rm Im}(\omega)|>|{\rm Re}(\omega)|$, see left-hand panel).  }
\label{fig: linear waves}
\end{figure}
%%%%%%%%%%%%%%%%%%%%%%%%%%%%%%%%%%

Code testing was carried out with standard problems and methods, such as examining the total-energy conservation, nonlinear Alfv\'en-wave dynamics \citep{Squire2016},
and linear waves. \Cref{fig: linear waves} illustrates the ability of the solver to capture the diverse linear behaviour of the CGL-LF system. 
The tests involve initialising the solver with a chosen linear eigenmode, computed analytically from the linearised system, then 
comparing the frequency and decay rate from the code to the analytic expectations. In the right-hand panel, we show the 
convergence to the analytic solution with spatial resolution for a range of different waves and parameters, scanning from the collisionless system out to
the adiabatic-MHD with $\nu_{\rm c}=10^{10}$. For modes that are only weakly damped, meaning their dynamics are dominated by the 
conservative (CGL) solver, we see almost second-order convergence. With more strongly damped modes, for which the
error becomes dominated by the non-conservative heat flux and/or collisions in the solver, we see first-order convergence 
above some resolution that depends on the mode in question (this behaviour is clearest for the slow-like modes shown with dashed lines, because
these are more strongly damped).

%%%%%%%%%%%%%%%%%%%%%%%%%%

\subsection{Numerical problems and challenges}\label{app:sub: issues }
The CGL system seems to be particularly prone to numerical instability, which hampered efforts 
to design more accurate numerical solvers. As part of the code development effort for this work, 
we tested a wide variety of different numerical options and solvers, all of which
showed some issues in certain situations. Our final choice of the piecewise parabolic method with an HLL Reimman solver 
and the conserved variable $\an$ (see~\cref{eq:u.def}) was made by trading off the various
issues that arose in different numerical tests. 
%For example, using $\mu$ as the eighth conserved variable, rather
%than $\an$, leads to numerical instabilities in regions of small $|\bm{B}|$ that are very difficult to 
%mitigate, despite its advantages for evaluating the heat fluxes and the fact it works well for smaller-amplitude 
%problems (e.g., waves).
Examples include a   pernicious 
numerical instability that acts  on the compressive components of the system at the smallest scales, 
creating flute-like ($k_{\perp}\gg k_{\|}$) grid-scale modes that can grow sufficiently to break up wave solutions 
in certain circumstances\footnote{In particular, they are strongest when  they can directly align with the numerical grid, which 
can occur  when the magnetic field is nearly constant and grid aligned.}; or 
virulent numerical instabilities that arise in regions of small $|\bm{B}|$  when $\mu$ rather than $\an$ is used as the conserved variable.
While we do not fully understand the root causes of these issues, we speculate that
they may be worse than standard in MHD because the system involves two different volumetrically conserved variables ($\rho$ and $\mu$ or $\an$) that  feed back on the
momentum in different ways.

As part of  this development process, we designed a new  Riemann solver for the 
CGL system that treats the Alfv\'enic-branch solutions separately, just like the popular HLLD solver for MHD \citep{Miyoshi2005,Mignone2007}.
This has significantly improved accuracy for problems involving predominantly Alfv\'enic fluctuations.
Although this made it an obvious candidate for our Alfv\'enic turbulence simulations,  the solver was ultimately found to cause issues in our higher-resolution 
turbulence simulations,
even though it worked quite well for many simpler test problems.
This type of behaviour is not uncommon, for example, also occurring  in HLLC and related  hydrodynamic finite-volume  solvers, which can suffer from serious  numerical issues
if shocks develop \citep[see, e.g.,][and references therein]{Simon2019}.
For this reason, for our main simulations, we reverted back to the simplest HLL solver, using parabolic reconstruction to 
reduce the large numerical dissipation that is inherent to this method.
Nonetheless, in case the HLLD solver  proves useful for other problems in future works, we provide a brief description of it below.

\subsubsection{An  HLLD CGL Riemann solver}\label{app:subsub: hlled solver}

Rather than proceeding in the most obvious way, by extending the HLLD method of \citet{Miyoshi2005} to 
the double-adiabatic mode structure, we instead base our solver on the isothermal HLLD solver of \citet{Mignone2007}.
The method thus computes the fluxes of thermodynamic  variables using the simple HLL method \citep{Harten1983}, 
while those of the transverse momentum and magnetic field are obtained using a more accurate computation of the wave structure. 
It enables shear-Alfv\'enic dynamics  (or rotational discontinuities) to be captured accurately, 
while avoiding various very complex and time-consuming nonlinear solves that arise if one attempts to replicate the method of
\citet{Miyoshi2005}\footnote{Indeed, following the logic of \citet{Miyoshi2005} for the CGL system, one finds it
necessary to solve for nearly all inner states of the Riemann fan simultaneously in a complex system of nonlinear equations. 
This leads to a slow and impractical solver.}.

The basis for the solver is to take $\rho$, $m_{x}$, $E$,  $\an$, and their fluxes as constant 
across the Riemann fan. In contrast, the
`Alfv\'enic variables', $m_{y}$, $m_{z}$, $B_{y}$, and $B_{z}$, are allowed to vary across the fan, being separated into 
three states $\bm{U}_{L}^{*}$, $\bm{U}_{c}^{*}$, and $\bm{U}_{R}^{*}$ by the left and right Alfv\'en discontinuities with velocities $S^{*}_{L}$ and $S^{*}_{R}$, as in isothermal MHD \citep{Mignone2007}. We then use the  consistency conditions for a five-state Riemann solver separated by waves $S_{L}$, $S_{L}^{*}$, $S_{R}^{*}$, and $S_{R}$ \citep{Mignone2007,Toro2009}:
\begin{equation}
\frac{(S_{L}^{*}-S_{L})\bm{U}^{*}_{L}+(S_{R}^{*}-S_{L}^{*})\bm{U}^{*}_{c}+(S_{R}-S_{R}^{*})\bm{U}^{*}_{R}}{S_{R}-S_{L}} = \bm{U}^{\rm hll},\label{eq:U.consistency}
\end{equation}
 for the conserved variables $\bm{U}$, and 
\begin{equation}
\frac{(S_{L}^{*-1}-S_{L}^{-1})\bm{F}^{*}_{L}+(S_{R}^{*-1}-S_{L}^{*-1})\bm{F}^{*}_{c}+(S_{R}^{-1}-S_{R}^{*-1})\bm{F}^{*}_{R}}{S_{R}^{-1}-S_{L}^{-1}} = \bm{F}^{\rm hll}\label{eq:F.consistency}
\end{equation}
 for the fluxes $\bm{F}$. Here $\bm{U}^{\rm hll}$ and $\bm{F}^{\rm hll}$ are the HLLE states and fluxes:
 \begin{gather}
\bm{U}^{\rm hll} \doteq \frac{S_{R}\bm{U}_{R} - S_{L}\bm{U}_{L} - \bm{F}_{R} + \bm{F}_{L}}{S_{R}-S_{L}},\label{eq:uhll} \\
\bm{F}^{\rm hll} \doteq \frac{S_{R}^{-1}\bm{F}_{R} - S_{L}^{-1}\bm{F}_{L} - \bm{U}_{R} + \bm{U}_{L}}{S_{R}^{-1}-S_{L}^{-1}}.\label{eq:fhll}
\end{gather}
Applying the assumption $\bm{U}_{L}^{*}=\bm{U}_{c}^{*}=\bm{U}_{R}^{*}=\bm{U}^{*}$ to the non-Alfv\'enic variables gives 
$\bm{U}^{*}=\bm{U}^{\rm hll}$, $\bm{F}^{*}=\bm{F}^{\rm hll}$, as expected. It remains to determine 
$\bm{U}_{L}^{*}$, $\bm{U}_{R}^{*}$, and their fluxes for the Alfv\'enic variables, after which 
$\bm{F}_{c}^{*}$ is determined from~\cref{eq:F.consistency}.

We compute  the jump conditions for the Alfv\'enic variables across the $S_{L}$ and $S_{R}$ waves  following \citet{Mignone2007}:
\begin{gather}
\bm{F}^{*}_{L} =  \bm{F}_{L} + S_{L}(\bm{U}_{L} - \bm{U}_{L}^{*}), \label{eq:fl.star} \\
\bm{F}^{*}_{R} =  \bm{F}_{R} + S_{R}(\bm{U}_{R} - \bm{U}_{R}^{*})\label{eq:fr.star} .
\end{gather}
This leads to
 \begin{gather}
\rho^{*} u_{y\alpha}^{*} = \rho^{*}u_{y\alpha} - B_{x}B_{y\alpha}\frac{u_{x}^{*}\fh_{\alpha}-u_{x\alpha}\fh^{*}+S_{\alpha}(\fh^{*}-\fh_{\alpha})}{(S_{\alpha} - S_{L}^{*})(S_{\alpha}-S_{R}^{*})},\label{eq:my.alpha.star}\\
\rho^{*} u_{z\alpha}^{*} = \rho^{*}u_{z\alpha} - B_{x}B_{z\alpha}\frac{u_{x}^{*}\fh_{\alpha}-u_{x\alpha}\fh^{*}+S_{\alpha}(\fh^{*}-\fh_{\alpha})}{(S_{\alpha} - S_{L}^{*})(S_{\alpha}-S_{R}^{*})},\\
B_{y\alpha}^{*} = \frac{B_{y\alpha}}{\rho^{*}} \frac{\rho_{\alpha}(S_{\alpha} - u_{\alpha})^{2} - B_{x}^{2}\fh_{\alpha}}{(S_{\alpha} - S_{L}^{*})(S_{\alpha}-S_{R}^{*})},\label{eq:by.alpha.star}\\
B_{z\alpha}^{*} = \frac{B_{z\alpha}}{\rho^{*}} \frac{\rho_{\alpha}(S_{\alpha} - u_{\alpha})^{2} - B_{x}^{2}\fh_{\alpha}}{(S_{\alpha} - S_{L}^{*})(S_{\alpha}-S_{R}^{*})}.\label{eq:bz.alpha.star}
\end{gather}
Here $\alpha$ denotes either $L$ or $R$, $\rho^{*} = \rho^{\rm hll}$, $u_{x}^{*}=F_{\rho}^{\rm hll}/\rho^{\rm hll}$, and $\fh^{*}$ is the anisotropy parameter ($\fh\doteq 1+\Dp/B^{2}$) across the fan (see below).
The Alfv\'enic wave speeds, $S_{L,R}^{*}$, are
\begin{equation}
S_{L}^{*} = u_{x}^{*} - \sqrt{\frac{\fh^{*}}{\rho^{*}}}|B_{x}|,\quad S_{R}^{*} = u_{x}^{*} + \sqrt{\frac{\fh^{*}}{\rho^{*}}}|B_{x}|.
\end{equation}
These are fixed by the jump conditions from the left and right  regions to the central regions after asserting that $\fh^{*}$ should be constant (this is in keeping with the
assumption that only Alfv\'enic quantities change inside the fan; see  equations (25)--(28) of \citealt{Mignone2007}). The numerical flux is then computed as 
\begin{equation}
\bm{F} =  \begin{cases} 
\bm{F}_{L}, & 0<S_{L}, \\ 
\bm{F}_{L}^{*}, & S_{L}<0<S_{L}^{*}, \\ 
\bm{F}_{c}^{*}, & S_{L}^{*}<0<S_{R}^{*},\\ 
\bm{F}_{R}^{*}, & S_{R}^{*}<0<S_{R}, \\ 
\bm{F}_{R}, & S_{R}<0,
\end{cases}
\end{equation}
where $\bm{F}_{L}^{*}$ and $\bm{F}_{R}^{*}$ are computed from~\cref{eq:fl.star,eq:fr.star}, and $\bm{F}_{c}^{*}$ from the consistency condition~\eqref{eq:F.consistency}.

The firehose parameter $\fh^{*}$  remains unspecified. Unfortunately, it is  not possible to have $\Dp/B^{2}$, $E$, and $\an$ all constant across the central region for general $B_{y}$ and $B_{z}$ that vary as in~\cref{eq:by.alpha.star,eq:bz.alpha.star}. We are thus forced to consider  $B^{2}$ in
$\fh$ to be some average across the fan. 
There appears to be no way to circumvent this inconsistency, so long as we assume that  non-Alfv\'enic 
variables are constant across then fan, while Alfv\'enic components vary.\footnote{It is worth noting 
that the same inconsistency arises in the isothermal MHD solver of \citet{Mignone2007}, although this is not discussed in his paper. Specifically, 
this solver assumes that, within the fan, the total pressure $p_{T} = c_{\rm s,iso}^{2} \rho + B^{2}/2$ is constant  (this 
is required so that $u_{x}$ is constant), while also taking $\rho$ to be constant. These assumptions are clearly inconsistent if $B^{2}$ varies. 
A similar  inconsistency arises in the  adiabatic MHD solver of \citet{Miyoshi2005} across the $S_{\alpha}^{*}$ waves.}
We are thus left with choosing a suitable average for defining $\fh^{*}$. We have experimented with various possibilities for doing 
this, finding that the most robust  method 
seems to be to use the HLL average to compute $B^{2}$ (i.e., using $B_{y}^{\rm hll}$ and $B_{z}^{\rm hll}$). This method has the
advantage of being straightforward and fast to execute, while also being conceptually consistent with the idea that non-Alfv\'enic variables 
($B^{2}$ in this case)
should be computed using HLL averages. 

\vspace{2mm}\noindent{\textbf{Special cases.}}
There are two special cases that must be dealt with in the solver. The first, which occurs if $S_{\alpha}^{*}\rightarrow S_{\alpha}$, is handled identically to 
 the \citet{Mignone2007} solver by imposing a zero jump across $m_{y}$, $m_{z}$, $B_{y}$, and $B_{z}$. As in isothermal MHD, there is no issue
with the limit $B_{x}\rightarrow 0$, which simply leads to $S_{\alpha}^{*}=u^{*}$ and various simplifications of~\eqref{eq:my.alpha.star}--\eqref{eq:bz.alpha.star}. 

The other special case occurs as $\fh$ approaches 0. For  $\fh<0$, Alfv\'en waves are unstable, and it no longer makes sense to
compute the wave structure in the same way. We thus switch to a standard HLL solver for all variables if $\fh_{L}$, $\fh^{*}$, or $\fh_{R}$ is below some threshold $\fh_{\mathrm{thresh}}$.
In practice, $\fh_{\mathrm{thresh}}\approx 0.1$ seems to give reasonable results 
(recall that Alfv\'enic perturbations propagate very slowly for small $\fh$ anyway). The switch to the HLL solver is carried out using the `anti-diffusion control' method described presently.

\vspace{2mm}\noindent{\textbf{Anti-diffusion control.}}
As mentioned above,  the CGL HLLD method displayed  numerical instabilities, in particular small-scale oscillations in $\fh$, which became 
worse for more complex problems. These can often be mitigated by the `anti-diffusion control' method of  \citet{Simon2019}, which was proposed to 
solve the  `Carbuncle' problem of the hydrodynamic HLLC solver. The method involves continuously switching to the more diffusive HLL solver at the advent 
of grid-scale oscillations. It is derived by rewriting~\cref{eq:fl.star,eq:fr.star} as
\begin{gather}
\bm{F}^{*}_{L} =  \bm{F}^{\rm hll} + S_{L}(\bm{U}^{\rm hll} - \bm{U}_{L}^{*}), \label{eq:fl.star ad} \\
\bm{F}^{*}_{R} =  \bm{F}^{\rm hll} + S_{R}(\bm{U}^{\rm hll}- \bm{U}_{R}^{*})\label{eq:fr.star ad},
\end{gather}
then recognising the second term on each right-hand side as an `anti-diffusive' contribution to the flux (it corrects the HLL flux, making
it less diffusive). Multiplying this contribution by a factor $\omega_{\mathrm{AD}}$ satisfying $0\leq \omega_{\mathrm{AD}}\leq 1$,
gives a method to interpolate between the HLLD and  HLL solvers. As a simple option,  $\omega_{\mathrm{AD}}$ can be chosen as 
$\omega_{\mathrm{AD}} = \exp(-\alpha_{\mathrm{AD}} \max |\Delta \fh|  )$,
where $ \max |\Delta \fh| $ is the maximum value of $|\fh_{L}-\fh_{R}|$ across the interface
and its two neighbors perpendicular to the interface. The free parameter $\alpha_{\mathrm{AD}}$ controls the strength of the 
anti-diffusive effect, with $\alpha_{\mathrm{AD}}\approx 10$ providing a reasonable trade off between accuracy and stability for a range of parameters
 and problems. We also multiply $\omega_{\mathrm{AD}}$ by $H(\fh_{\mathrm{min}}-\fh_{\mathrm{thresh}})$, where $H(x)$ is the Heaviside step function and $\fh_{\mathrm{min}} = \min\{ \fh_{L}, \fh^{*}, \fh_{R}\}$, to switch to the HLL solver as the system approaches the firehose instability condition (see above).
More information and stability analysis 
of the anti-diffusion control method in the HLLC context can be found in \citet{Simon2019}.

\newpage
\bibliographystyle{jpp}
\bibliography{fullbib_formatted}

\end{document}